\documentclass[12pt]{amsart}
\usepackage{amssymb}
\usepackage{amsmath}
\usepackage{amsfonts}
\usepackage{amsthm}
\usepackage{graphicx}
\usepackage{color}
\usepackage[onehalfspacing]{setspace}
\usepackage{hyperref}
\usepackage{multirow}
\usepackage{float}
\usepackage{bm}
\usepackage[flushleft]{threeparttable}
\usepackage{tikz}
\usepackage{booktabs}
\usepackage[round,sectionbib]{natbib}
\usepackage{xr}
\usepackage{booktabs,caption}
\usepackage[flushleft]{threeparttable}
\usepackage{algorithm}
\usepackage{algorithmic}

\def\bY{{\bm Y}}

\def\bU{{\bm U}}

\numberwithin{equation}{section}

\newtheorem{theorem}{Theorem}[section]

\newtheorem{corollary}{Corollary}[section]
\newtheorem{remark}{Remark}[section]

\newtheorem{proposition}{Proposition}[section]

\theoremstyle{definition}
\newtheorem{assumption}{Assumption}[section]

\theoremstyle{definition}

\theoremstyle{definition}

\makeatletter
\def\section{\@startsection{section}{1}
	\z@{1.0\linespacing\@plus\linespacing}{.8\linespacing}{\Large\centering}}

\def\subsection{\@startsection{subsection}{2}
	\z@{.8\linespacing\@plus.7\linespacing}{.7\linespacing}{\large}}

\def\subsubsection{\@startsection{subsubsection}{3}
	\z@{.5\linespacing\@plus.7\linespacing}{-.5em}{\normalfont\bfseries\centering}}
\makeatother

\usepackage{color, hyperref}
\hypersetup{
    bookmarks=true,         % show bookmarks bar?
    unicode=false,          % non-Latin characters
    pdftoolbar=true,        %
    pdfmenubar=true,        %
    pdffitwindow=false,     % window fit to page when opened
    pdfstartview={FitH},    % fits the width of the page to the window
    pdftitle={My title},    % title
    pdfauthor={Author},     % author
    pdfsubject={Subject},   % subject of the document
    pdfcreator={Creator},   % creator of the document
    pdfproducer={Producer}, % producer of the document
    pdfkeywords={keyword1} {key2} {key3}, % list of keywords
    pdfnewwindow=true,      % links in new window
    colorlinks=false,       % false: boxed links; true: colored links
    linkcolor=black,          % color of internal links (change box color with linkbordercolor)
    citecolor=black,        % color of links to bibliography
    filecolor=black,      % color of file links
    urlcolor=black           % color of external links
}

\title{}

\begin{document}
	\vspace*{5ex minus 1ex}
	\begin{center}
		\Large \textsc{Quasi-Bayesian Estimation and Inference \\with Control Functions}
		\bigskip
	\end{center}
	
	%\date{
		%\today%
		%EndExpansion
	%}

	%\vspace*{3ex minus 1ex}
		\begin{center}
			Ruixuan Liu and Zhengfei Yu\\
			\textit{Chinese University of Hong Kong and University of Tsukuba}\\
			\medskip
		\end{center}
	
	\begin{abstract}
		{This paper introduces a quasi-Bayesian method that integrates frequentist nonparametric estimation with Bayesian inference in a two-stage process. Applied to an endogenous discrete choice model, the approach first uses kernel or sieve estimators to estimate the control function nonparametrically, followed by Bayesian methods to estimate the structural parameters. This combination leverages the advantages of both frequentist tractability for nonparametric estimation and Bayesian computational efficiency for complicated structural models. We analyze the asymptotic properties of the resulting quasi-posterior distribution, finding that its mean provides a consistent estimator for the parameters of interest, although its quantiles do not yield valid confidence intervals. However, bootstrapping the quasi-posterior mean accounts for the estimation uncertainty from the first stage, thereby producing asymptotically valid confidence intervals.}
		\bigskip\
		
		%\textcolor{red}{Version edited by ChatGPT: This paper introduces a quasi-Bayesian method that integrates frequentist nonparametric estimation with Bayesian inference in a two-stage process. Applied to an endogenous discrete choice model, the approach first utilizes frequentist techniques to estimate the control function nonparametrically, followed by Bayesian methods to estimate the structural parameters. This combination utilizes the advantages of both frequentist tractability and Bayesian computational efficiency. We analyze the asymptotic properties of the resulting quasi-posterior distribution, finding that its mean provides a consistent estimator for the parameters of interest, although its quantiles do not yield valid confidence intervals. However, bootstrapping the quasi-posterior mean compensates for uncertainties from the first stage, thereby producing asymptotically valid confidence intervals.}

		{\footnotesize \noindent Key words: Bayesian methods, Bernstein--von Mises theorem, endogenous regressors, control function, semiparametric estimation  \bigskip\ }
		
		%{\footnotesize \noindent \textsc{MC2010 Subject Classification: }}
		
	\end{abstract}
	
	\thanks{%The first draft: April 2, 2021.\\
		 Ruixuan Liu, CUHK Business School, Chinese University of Hong Kong, 12 Chak Cheung Street, Cheng Yu Tung Building, Hong Kong, E-mail: \textit{ruixuanliu@cuhk.edu.hk}.\\
		 Zhengfei Yu, Faculty of Humanities and Social Sciences, University of Tsukuba,
		 1-1-1, Tennodai, Tsukuba, Ibaraki  305-8571, Japan. E-mail: \textit{yu.zhengfei.gn@u.tsukuba.ac.jp}.\\
		 We would like to thank Christoph Breunig, Yanqin Fan, Emmanuel Guerre, Hide Ichimura, Shakeeb Khan, Chang-Jin Kim, Yuan Liao, Zhipeng Liao, Essie Maasoumi, Jing Tao, Jun Yu, and Yichong Zhang for helpful comments and discussions. Liu's research is supported by the GRF Number 14503523 from the Research Grants Council Hong Kong. Yu gratefully
acknowledges the support of JSPS KAKENHI Grant Number 21K01419. The usual disclaimer applies.}
	\maketitle

		\section{Introduction}
		The control function approach is commonly used in empirical economic studies to cope with endogenous explanatory variables  \citep{heckman1985control,wooldridge2015control}. This approach typically involves two stages: the first stage projects the endogenous explanatory variables to a set of instruments and other exogenous variables and obtains the residuals; these residuals serve as extra explanatory variables, also known as control functions when estimating the structural equations in the second stage. In this paper, we propose a quasi-Bayesian method that combines a frequentist-type first-stage estimation with some Bayesian approach in the second stage. 
		The study is motivated by structural discrete choice models that apply the control function method to correct for endogeneity bias. In such scenarios, second-stage estimation remains challenging for frequentist methods but can be more conveniently addressed by Bayesian procedures, as the latter transform the difficult integration and/or optimization problem into a sampling problem of posterior distributions. Therefore, the proposed quasi-Bayesian approach offers researchers great flexibility in combining nonparametric frequentist methods with computationally convenient Bayesian approaches. With a clear understanding of the theoretical foundation regarding the asymptotic behavior of the posterior obtained from the second stage--referred to as the quasi-posterior, since it uses frequentist estimates from the first stage--this new procedure can be particularly useful in econometric models characterized by complex likelihood functions and endogeneity issues.
	
	To formalize the idea, we consider independent and identically distributed ($i.i.d.$) observations $\bY^n=(Y_i,i=1,\cdots,n)^\top$. The component $\zeta_0$ represents a vector of unknown functions that one needs to estimate from the first-stage in order to form the control functions. Throughout the paper, we assume that the true $\zeta_0$ is identifiable and can be estimated from the subvector $\bY^n_1=(Y_{1i},i=1,\cdots,n)^\top$ that consists of the observables from the first stage. We denote the corresponding estimator by $\widehat{\zeta}_n$. The second stage postulates a probability density function, $p_{\theta}^n(\cdot;\zeta)$ for the subvector $\bY^n_2=(Y_{2i},i=1,\cdots,n)^\top$ of observables with the unknown parameter $\theta\in\Theta\subset \mathbb{R}^p$, for some integer $p\geq 1$. The unique truth is denoted by $\theta_0$. The Bayesian procedure in the second stage posits a prior distribution $\Pi(\theta)$ and then updates it to the quasi-posterior distribution given the limited-information likelihood\footnote{The current approach is based on the ``limited information" strategy in the sense that the information contained in both stages (or equations) are not simultaneously considered.} in the second stage. Using  Bayes' rule pragmatically, the quasi-posterior of the finite-dimensional parameter $\theta$ is
		\begin{equation}
		\Pi[\theta\in \mathcal{A}|\bY_2^n;\widehat{\zeta}_n]=\frac{\int_{\mathcal{A}} p^n_{\theta}(\bY_2^n;\widehat{\zeta}_n)d\Pi(\theta)} {\int_{\Theta} p^n_{\theta^{\prime}}(\bY_2^n;\widehat{\zeta}_n)d\Pi(\theta^{\prime})},
		\end{equation}
for any given measurable set $\mathcal{A}\subset \Theta$. %\textcolor{red}{I agree that it is better to use $ p^n(\bY_2^n,\theta;\widehat{\zeta}_n)$ rather than $ p^n_{\theta}(\bY_2^n;\widehat{\zeta}_n)$. Because the subscript notation is also used for the partial derivative of the log-likelihood ratio. We need to make the change throughout the proofs too.} 
We refer to the estimation/inferential procedure based on the above quasi-posterior as quasi-Bayesian, because the term $p^n_{\theta}(\bY_2^n;\widehat{\zeta}_n)$ is not a genuine likelihood and it depends on some plug-in estimator $\widehat{\zeta}_n$. Two key questions arise: (1). Is the quasi-posterior distribution of $\theta$ asymptotically Gaussian? (2). What are the center and dispersion of this quasi-posterior distribution? To our best knowledge, such problems have not been investigated in the literature even when the estimator $\widehat{\zeta}_n$ is estimated parametrically. This paper fills this void and allows for general first-stage nonparametric estimation.
		
We investigate the asymptotic properties of this quasi-Bayesian procedure from a frequentist perspective. The main challenge lies in the fact that one encounters a nonstandard posterior distribution that depends on the first-stage estimation. We show that the credible set formed by the quantiles of the quasi-posterior does not achieve correct coverage in large samples, as it fails to account for the estimation uncertainty from the first stage. Although one can not directly rely on the credible set for inference, the quasi-posterior still contains useful information. In particular, we prove that the quasi-Bayesian point estimator, such as the quasi-posterior mean, is asymptotically equivalent to a frequentist two-stage estimator; that is, it is root-$n$ asymptotically normal with the same sandwich-type covariance matrix as its frequentist counterpart. In essence, the quasi-posterior distribution is correctly centered but has incorrect dispersion. We also show that a bootstrap method, when appropriately applied to the quasi-Bayesian approach, yields confidence intervals with the desired coverage probability and is therefore valid for inference. 
	
	 To illustrate our theory, we focus on an endogenous multinomial Probit (MNP) model from \cite{petrin2010control}, which we refer to as the Petrin--Train model in the sequel. The framework is popular in modeling consumer choices over different products, say choosing from $j\in\{0,\cdots, J\}$, with the $0$-th alternative being the null category (non-purchase option). The endogenous variables are usually the prices in the context. To address the endogeneity problem, researchers often construct the Hausman-type price instruments \citep{hausman1994diff}. The first stage involves the estimation of a vector of $(\zeta_j)_{j=1}^J$, as the conditional mean functions of the endogenous variables given the exogenous covariates and instrument variables. Given the estimated  $\widehat{\zeta}_n=(\widehat{\zeta}_{n,j})_{j=1}^J$, we can extract the residuals as the control functions for the second-stage MNP model. In this scenario, the likelihood function $p^n_{\theta}(\bY_2^n;\widehat{\zeta}_n)$ is the product of the conditional choice probabilities that involve complicated multiple integrals. As reviewed in \cite{train2009discrete}, the Bayesian approach enjoys practical advantages over frequentist methods in estimating the MNP-type models. This is also confirmed in our context both with respect to the Monte Carlo simulation as well as the empirical study.
	 
	 Going beyond the Petrin-Train model, our study reflects two generic features of modern econometric models. On one hand, many models used to describe observed data may be so complex that the likelihoods associated with these models are computationally intractable. These analytical difficulties can be alleviated by simulation-based procedures, such as the Bayesian method, which has proven successful in many areas. On the other hand, the control function approach provides an effective solution to deal with the endogeneity problem. This approach treats endogeneity as an omitted variable bias problem, in which the inclusion of estimates of the first-stage errors as additional covariates corrects the inconsistency of the second stage. The idea of combining a frequentist first stage and a Bayesian second stage has been practiced in empirical research on estimating the distribution of students' preferences for public schools \citep{agarwal2018demand}, though lacking a formal theoretical investigation for such a ``hybrid'' or quasi-Bayesian procedure. \cite{agarwal2018demand} also suggest a bootstrap method for inference, which inspires our current proposal. Our study sheds theoretical light on the asymptotic normality of the two-stage quasi-Bayesian estimator and the bootstrap validity therein.
		
 There is a clear advantage of our proposal over the frequentist two-step method or the Bayesian full information paradigm in the current context, because it separates two tasks: one can correct for the endogeneity bias via robust nonparametric first-stage estimation, while tackling the complicated likelihood in the second stage by a Bayesian approach. This separation of tasks allows for considerable algorithmic flexibility. Regarding the frequentist two-step approach, the second stage requires the simulated method of moments \citep{mcfadden1989sme,hajivassiliou1998scores}, which calls for Monte Carlo simulation, in addition to solving hard optimization problems for finding the frequentist point estimators. Given the complicated form of the influence function, it might be a daunting task to compute the correction term or the standard error analytically. When it comes to inference, one has to rely on bootstrap for the frequentist approach too. Referring to the full information Bayesian approach, one would focus on the posterior that is proportional to  
		  $\tilde{p}^n_{\theta,\zeta}(\bY^n)d\Pi(\theta)dQ(\zeta)$, where $ \tilde{p}^n_{\theta,\zeta}(\bY^n)$ is the density function for the joint distribution of $\bm{Y}^n_1$ and $\bm{Y}^n_2$ combining two stages together, and $Q(\cdot)$ is a suitable prior on the functional component $\zeta$. The joint likelihood inevitably requires a full model encompassing both stages, for which the applied researcher may be reluctant to hypothesize such a structure \citep{murphy2002two}. If the researcher does not want to assume a parametric model for $\zeta$, the full-information Bayes must put a nonparametric prior $Q(\zeta)$ on it. Although the non-parametric Bayesian inference is one of the most vibrant research areas in econometric theory, verification of the related Bernstein--von Mises (BvM) theorem either relies on the Gaussian processes priors or explores particular model structures; see \cite{ghosal2017bayesian}. In addition, designing a feasible algorithm for the posteriors remains a skillful task when the joint likelihood functions from both stages are combined. 
		
	 \subsection{Related Literature}	 
	 Studying asymptotic properties of quasi-Bayesian procedures forms an important line of work in econometrics.  Regarding the discrepancy between quasi-Bayesian credible sets and frequentist confidence sets, we highlight a recurring theme when the generalized information identity fails \citep{chernozhukov2003mcmc}. In this case, the local asymptotic normality (LAN) expansion \citep{van1998asymptotic} generates a centering term whose asymptotic variance is of the sandwich form that does not match the (minus) second derivative matrix of the (quasi) log-likelihood function. These phenomena appear in studies of the generalized method of moments (GMM) or the limited information approach \citep{kim2002limited,chernozhukov2003mcmc,chib2018moment}, as well as in misspecified models \citep{kleijn2012mis,muller2013risk,kim2014quasi}. Given the natural link between GMM and two-step estimation \citep{newey1994large}, it is tempting to think that our framework is nested by the works mentioned above. This is not the case. We emphasize that one distinct feature of our problem is that the first-stage estimation is based on a frequentist method and then plugged into the posterior of the second stage. One can view this as approximating the posterior of $\zeta$ using the Dirac delta function. Such feature violates the standard assumptions in proving the BvM theorem. 
	 
Another line of research closely related to this paper is the profile sampler  \citep{lee2005profile,cheng2008profile}. This literature recommends sampling from the profile likelihood $pl_n(\theta)\equiv \sup_{\zeta}p^n_{\theta}(\bY^n;\zeta)$ for semiparametric models indexed by $(\theta,\zeta)$. The infinite-dimensional parameter $\zeta$ therein must be estimated via the profile maximum likelihood so that there is no additional adjustment term. See related discussions on the profiling strategy to eliminate nuisance parameters/functions in the last paragraph on page 1357 of \cite{newey1994var}. Such procedures are not required in our setup. The crux of our analysis is to characterize the first-stage estimation effect on the quasi-posterior. On the technical ground, this paper also refines the proof of \cite{lee2005profile} and \cite{cheng2008profile}: we show the asymptotic negligibility of the posterior outside a shrinking neighborhood of the truth, where the radius depends on the convergence rate of first-stage estimation. Our proof relaxes the convergence rate requirement to match the well-known order of $o(n^{-1/4})$ in general semiparametric models \citep{newey1994var}\footnote{In comparison, the result of Lemma A.1 of \cite{lee2005profile} is stated for some neighborhood with a \textit{fixed} radius, yet later on, this lemma is cited to cover the case with a shrinking radius of the order $o(n^{-1/3})$.}. The proof of our BvM theorem also differs from existing literature from two aspects. First, we need to take into account the first-stage nonparametric estimation when we carefully partition the integral range into the dominating one and negligible parts. Second, despite the semiparametric nature of the problem, we can avoid checking the prior stability condition in the semiparametric BvM literature \citep{ghosal2017bayesian}, because we simply estimate the nonparametric element rather than assigning certain nonparametric prior to it.

\subsection{Organization}	
	The rest of our paper is organized as follows. Section \ref{sec: main results} contains the main theoretical results. This relatively long section is divided into three subsections, in which we introduce necessary assumptions, investigate the asymptotic behavior of the quasi-posterior, and show the validity of bootstrap methods for inference. Section \ref{sec:PT model} applies the general theory to the Petrin--Train model.  We present low-level regularity conditions and state the asymptotic results. Section \ref{sec:simulation} conducts Monte Carlo simulations and provides an empirical illustration. The last section concludes. Proofs of the main results are presented in Appendix A. The supplementary material contains several Appendices. Technical lemmas and their proofs are collected in Appendix B. The formulation of the BvM theorem given parametric first-stage estimation is presented in Appendix C. In Appendix D, we provide details of the influence function for a special case of the Petrin--Train model. Appendix E offers additional simulation evidence comparing our bootstrap proposal with another method.
		
		\subsection{Notations} For a function $f(\cdot)$ of a random vector $Y$ that follows distribution $P$, we use the standard empirical process notations: $P_0f=\int f(y)dP_0(y),\mathbb{P}_n f=n^{-1}\sum_{i=1}^{n}f(Y_i)$, and $\mathbb{G}_n f=n^{1/2}\left(\mathbb{P}_n-P_0\right)f$. We also write $\mathbb{E}_0$ instead of $P_0$ in bounding center stochastic terms. Definitions of the $P_0$-Glivenko-Cantelli class and the $P_0$-Donsker class follow those in \cite{van1996empirical} and \cite{van1998asymptotic}. We denote the bootstrap empirical measure as $\mathbb{P}_n^*$ and the bootstrap empirical process as $\mathbb{G}^*_n=\sqrt{n}(\mathbb{P}_n^*-\mathbb{P}_n)$. For a sequence of random variables $Z^*_n$, we write $Z^*_n=o_{P^*}(1)$ if the law of $Z_n^*$ is governed by the bootstrap law $\mathbb{P}^*$ and if $\mathbb{P}^*(|Z_n^*|>\epsilon)=o_{P_0}(1)$ for any $\epsilon>0$. 
		
		For any $p$-dimensional vector $\nu$, we write $\nu^{\otimes 2}\equiv \nu\nu^\top$. For any $\alpha\in \mathbb{R}$, let $\lfloor \tau\rfloor$ be its integer part. Considering a multi-index $k=(k_1,\cdots,k_d)$, define $k_{.}=k_1+k_2+\cdots+k_d$. For any $L>0,\tau\geq 0$ and non-negative function $L$ on $\mathbb{R}^d$, define the $\tau$-H\"older class with envelope $L$, denoted by $\mathcal{C}^{\tau,L}(\mathbb{R}^d)$ (Section 3 of \cite{ichimura2010semi}) to be the set of all functions $f:\mathbb{R}^d\mapsto\mathbb{R}$ that have finite mixed partial derivatives $D^kf$ of all orders up to $k_{.}\leq \lfloor \tau\rfloor$, 
		\begin{equation}
			\left| (D^kf)(x+y)-(D^k)f(x) \right|\leq L\parallel  y\parallel^{\tau-\lfloor\tau\rfloor}, ~~\forall x,y\in \mathbb{R}^d.
		\end{equation}
		For any real vector $b$, let $\Vert b\Vert_E$ denote the Euclidean norm of $b$. For any $J$-dimensional vector of functions $\bm{h}(x)$ with its $j$-th component equal to $h_j(x)$, we let 
		\begin{equation*}
		\Vert \bm{h}\Vert_{\infty}\equiv \Vert (\Vert h_1\Vert_{\sup},\cdots,\Vert h_J\Vert_{\sup})\Vert_E,
		\end{equation*}
		where $\Vert h_j\Vert_{\sup}\equiv \sup_{x\in\mathcal X} |h_j(x)|$, and $\mathcal{X}$ denotes its support.
 
\section{Main Theoretical Results}\label{sec: main results}	
Bayesian methodology is attractive in its own right. Nonetheless, our problem is fundamentally non-Bayesian, as researchers aim to estimate unknown fixed parameters using the control function approach to correct the endogeneity bias. This motivates us to investigate the large sample behavior of the resulting quasi-posterior under a fixed true probability model $P_0=P_{\theta_0,\zeta_0}$ that generates the observed data. A thorough understanding will provide solid justification for the quasi-Bayesian methods, which can be attractive to non-Bayesian practitioners because of their convenience. Our main use of the quasi-posterior is to interpret it as a frequentist's device, similar to that of a sampling distribution. This is crucial to justify whether the resulting credible set can be used as a confidence set asymptotically, in view of the first stage estimation uncertainty. %Another important perspective of the large sample analysis is to validate the insenstivity of the prior specification, whose influence should be washed off as more one has access to more data.
	 
	\subsection{Assumptions and Discussions}	
	The (limited-information) log-likelihood function is of fundamental importance in the second-stage Bayesian formulation. Throughout the paper, we focus on the case with \textit{i.i.d.} data, so the likelihood function becomes $p_{\theta}^n(\bm{Y}_2^n;\zeta)=\prod_{i=1}^n p_{\theta}(Y_{2i};\zeta)$ and the log-likelihood is denoted by $\ell_n(\theta;\zeta)\equiv\log p_{\theta}^n(\bm{Y}_2^n;\zeta)=\sum_{i=1}^n\log p_{\theta}(Y_{2i};\zeta)$. We define the following first-order derivative with respect to (w.r.t.) the finite-dimensional parameter $\theta$:
	\begin{align}\label{PScore}
		\dot{l}_{\theta}\left(Y_2,\theta;\zeta\right)\equiv \left(\frac{\partial \log p_{\theta}(Y_2;\zeta)}{\partial \beta^\top},\frac{\partial \log p_{\theta}(Y_2;\zeta)}{\partial \eta^\top} \right)^{\top},
	\end{align}
	%\textcolor{magenta}{[It seems that we are using two notations $p_{\theta}(Y_{2};\zeta)$ and $p(Y_2,\theta;\zeta)$ exchangeably. Consider unifying the notation. Also applies to the notations in Assumptions 2.5 and 2.7.] }
where $\theta$ is partitioned into the parameter of interest $\beta$ and the nuisance parameter $\eta$. The metric associated with the parameter $\theta$ is $d_{\Theta}(\cdot,\cdot)$. The second-order derivative is denoted by $\ddot{l}_{\theta}\left(  Y_2,\theta;\zeta\right)$ in the same vein. Thereafter, the score function is $S_{\theta}\left(  \theta,\zeta\right)  =P_0\dot{l}_{\theta}\left(  Y_2,\theta;\zeta\right)$, with the corresponding empirical version as $S_{\theta,n}\left(  \theta,\zeta\right)  =\mathbb{P}_n\dot{l}_{\theta}\left(  Y_2,\theta;\zeta\right)$.
	The negative Hessian matrix $V_0=-P_0\ddot{l}_{\theta}\left(  Y_2,\theta_0;\zeta_0\right)$ is the information matrix for the second-stage likelihood, when $\zeta_0$ is known. The unknown function in the first stage may contain $J$ components, so we write $\zeta_0=(\zeta_{0j})_{j=1}^J$. Consider a map defined by $\Psi(\theta,\zeta)\equiv S_{\theta}(\theta,\zeta)$. 	
	The following pathwise derivative of $\Psi$ plays a key role in determining the effect of the first-stage estimation \citep{newey1994var}:
	\begin{equation*}
		\dot{\Psi}_{\zeta}(\theta_0,\zeta_0)[\zeta-\zeta_0]\equiv\sum_{j=1}^J\dot{\Psi}_{\zeta_j}(\theta_0,\zeta_0)[\zeta_j-\zeta_{0j}],
	\end{equation*}
where each $\dot{\Psi}_{\zeta_j}$ is a bounded linear functional that maps $\zeta_j-\zeta_{0j}$ to the real line. The proper functional space for $\zeta$ is denoted by $\mathcal{G}$ with its pseudo metric $d_G(\cdot,\cdot)$. 
	
We list all regularity conditions followed by some heuristic discussions in order. They are not necessarily the weakest possible assumptions. 

%\textcolor{magenta}{[To be consistent with notation in the proof of Lemma S1, consider change the notation $\mathbb{E}_0$ in Assumption 1 to $\mathbb{P}_0$?]}
\begin{assumption}[Identification]\label{assump:id}
The function $\zeta_0$ is uniquely identified from the first stage. The expected log-likelihood function $\mathbb{E}_0[\log p_{\theta}(Y_2;\zeta_0)]$ has a unique maximum at $\theta_0$ over the parameter space $\Theta$ in the sense that $\sup_{d_{\Theta}(\theta,\theta_0)> \delta}\mathbb{E}_0[\log p_{\theta}(Y_2;\zeta_0)]<\mathbb{E}_0[\log p_{\theta_0}(Y_2;\zeta_0)]$ for any $\delta>0$. Also, $S_{\theta}\left(  \theta_0,\zeta_0\right)  =0$, and the matrix $V_0$ is positive definite with its smallest eigenvalue $\lambda_1>0$.
\end{assumption}

\begin{assumption}[First-stage Estimation]\label{assump:FirstStep}
For some set $\mathcal{G}_n$, it holds that the first-stage estimator $\widehat{\zeta}_n\in\mathcal{G}_n$ with probability 1. Also, $d_G(\widehat{\zeta}_n,\zeta_0)=O_{P_0}(r_n)$, where $r_n=o(n^{-1/4})$ and $nr_n^2\to \infty$. 
\end{assumption}

\begin{assumption}[Frequentist Two-stage Estimation]\label{assump:Freq}
There exists a frequentist-type estimator that approximately maximizes the second-stage likelihood:
\begin{equation}
 \widehat{\theta}_n=\arg\max_{\theta}\ell_n(\theta;\widehat{\zeta}_n)+o_{P_0}(n^{-1/2}).
\end{equation}
In addition, it approximately solves the score equation as follows
	\begin{equation}\label{sample_score}
		S_{\theta,n}\left(\widehat{\theta}_n,\widehat{\zeta}_n\right) 
		=o_{P_0}(n^{-1/2}).
	\end{equation} 
\end{assumption}

\begin{assumption}[Prior]\label{assump:Prior}
	The prior measure $\Pi$ on $\Theta$ is assumed to be a probability measure with a bounded Lebesgue density $\pi$, which is continuous and positive on a neighborhood of the true value $\theta_0$. In addition, $\int |\theta|^a\pi(\theta)d\theta<+\infty$, for some given $a\geq 0$. 
\end{assumption}

%\textcolor{magenta}{[To be consistent with notation in the proof of Lemma S2, consider change the notation $\mathbb{E}_0$ in Assumption \ref{assump:regular} to $\mathbb{P}_0$?]}
\begin{assumption}[Smoothness]\label{assump:regular}
We assume the log-likelihood function satisfies the following restrictions for the set $\mathcal{G}_n$ in Assumption \ref{assump:FirstStep} and for some positive constant terms $C_1,\cdots,C_4$:
	\begin{equation*}
		\sup_{\zeta\in\mathcal{G}_n}\left|\mathbb{E}_0\left[\log p_{\theta_0}(\cdot;\zeta)- \log p_{\theta_0}(\cdot;\zeta_0)\right] \right|\leq C_1 r_n^2,
	\end{equation*}
	and
	\begin{equation*}
		\mathbb{E}_0\left[\log p_{\theta}(\cdot;\zeta)- \log p_{\theta_0}(\cdot;\zeta_0)\right]+e_n(\theta,\zeta)\leq -C_2d^2_{\Theta}(\theta,\theta_0)+C_3d^2_{G}(\zeta,\zeta_0),
	\end{equation*}
	for some $e_n(\cdot,\cdot)$ that satisfies
	\begin{equation*}
		\sup_{d_{\Theta}(\theta,\theta_0)\leq \rho,\zeta\in \mathcal{G}_n}|e_n(\theta,\zeta)|\leq C_4\rho r_n.
	\end{equation*}
\end{assumption}

\begin{assumption}[Differentiability]\label{assump:smooth}
	The function $\mathbb{E}_0[\log p_{\theta}(Y_{2};\zeta)]$ is continuous with respect to $\zeta$ uniformly in $\theta\in \Theta$. There exists a continuous and non-singular matrix $\dot{\Psi}_{\theta}(\theta_0,\zeta_0)$ and a continuous linear functional $\dot{\Psi}_{\zeta}(\theta_0,\zeta_0)$ such that 
	\begin{equation*}
		\parallel\Psi(\theta,\zeta)-\Psi(\theta_0,\zeta_0)-\dot{\Psi}_{\theta}(\theta_0,\zeta_0)(\theta-\theta_0)-\dot\Psi_{\zeta}(\theta_0,\zeta_0)[\zeta-\zeta_0]\parallel\leq o(d_{\Theta}(\theta,\theta_0))+O(d^2_{G}( \zeta,\zeta_0)).
	\end{equation*}
\end{assumption}

\begin{assumption}[Complexity]\label{assump:SE}
(i)	The functional classes $\{\log p_{\theta}(Y_{2};\zeta):\theta\in\Theta,\zeta\in\mathcal{G}_n \}$ and $\left\{\ddot{l}_{\theta}(\cdot,\theta;\zeta):\theta\in\Theta,\zeta\in\mathcal{G}_n \right\}$ belong to $P_0$-Glivenko-Cantelli classes. 
(ii) The functional class $\{\dot{l}_{\theta}(Y_2,\theta_0;\zeta):\zeta\in\mathcal{G}_n\}$ belongs to a $P_0$-Donsker class. The following functional class
		\[
		\{\log p_{\theta}(Y_2;\zeta)-\log p_{\theta_0}(Y_2;\zeta)-(\theta-\theta_0)^\top\dot{l}_{\theta_0}(;\zeta):d_{\Theta}(\theta,\theta_0)\leq Cr_n, \zeta\in\mathcal{G}_n \}
		\]
		divided by each ordinates of $(\theta-\theta_0)$ belongs to a $P_0$-Donsker class with its second moment going to zero with probability approaching one.
		(iii) For some sufficiently small $\delta>0$, there exists $\phi_n(\cdot)$ such that
		\begin{equation*}
			\mathbb{E}_0\left[\sup_{d_{\Theta}(\theta,\theta_0)\leq \rho,\zeta\in \mathcal{G}_n}\mathbb{G}_n \left|\log p_{\theta}(\cdot,\theta;\zeta)- \log p_{\theta}(\cdot,\theta_0;\zeta) \right|\right]\leq \phi_n(\rho),
		\end{equation*}
			for $\delta\geq \rho\geq r_n$, where $\phi_n(\cdot)$ is a sequence of functions defined on $(0,\infty)$ that satisfies: $\rho\mapsto\phi_n(\rho)/\rho^{\gamma}$ is decreasing for some $\gamma<2$; $\rho^{-2}\phi_n(\rho)\leq \sqrt{n}$ for every $n$.
	\end{assumption}
	
\begin{assumption}[Normality]\label{assump:normal} Let $\Delta_{n,0}\equiv \sqrt{n}\left(S_{\theta,n}(\theta_0,\zeta_0)+\dot{\Psi}_{\zeta}(\theta_0,\zeta_0)[\widehat{\zeta}_n-\zeta_0]  \right)$. 
Assume that the following linear representation holds:
	\begin{align*}
		\Delta_{n,0}= \frac{1}{\sqrt{n}}\sum_{i=1}^n \left[\Gamma_2(Y_{2i})+\Gamma_1(Y_{1i})\right]+o_{P_0}(1),
	\end{align*}
and the following asymptotic normality holds: 
	\begin{align*}
	 \frac{1}{\sqrt{n}}\sum_{i=1}^n \left[\Gamma_2(Y_{2i})+\Gamma_1(Y_{1i})\right]\Rightarrow \mathbb{N}(0,\Omega_0),
\end{align*}
for some finite covariance matrix $\Omega_0$.
\end{assumption}
Assumptions \ref{assump:id} to \ref{assump:Freq} are generic, concerning the identifiability of the model parameters, the asymptotics of the first stage nonparametric estimation, as well as the existence of a frequentist two-stage estimator. Considering the convergence rate requirement in Assumption \ref{assump:FirstStep}, we explicitly work with nonparametric first stage estimation. Given this convergence rate, the set $\mathcal{G}_n$ allows us to localize the analysis around the truth related to the $P_0$-Donsker properties in Assumption \ref{assump:SE}. %\textcolor{magenta}{[The previous sentence seems about verifying Assumption 2.7, which is imposed for this section. Do we need to verify Assumption 2.7 in this section?]} 
In addition, one may also impose the shape constraint or perform proper trimming to further regularize the problem. The parametric case is easier and we formulate the necessary changes in the supplementary material. The frequentist two-stage estimator $\widehat{\theta}_n$ in Assumption \ref{assump:Freq} merely serves as a theoretical device, as it will become the centering point in the local asymptotic normal (LAN) expansion \citep{ghosh2002bayesian,chernozhukov2003mcmc,lee2005profile}. In our context of analyzing structural discrete choice models, $\widehat{\theta}_n$ is difficult to compute, which motivates the quasi-Bayesian procedure in the first place. Assumption \ref{assump:Prior} on the prior density is also standard \citep{ghosh2002bayesian,chernozhukov2003mcmc}.
 
Compared with the standard conditions for the BvM theorem \citep{van1998asymptotic,ghosal2017bayesian}, our setting involves two complications due to the presence of the estimated control function. The first task is to ensure that the posterior concentrates in any small neighborhood of the true parameter uniformly over a set $\mathcal{G}_n$ to which the estimated $\widehat{\zeta}_n$ belongs. The second distinction is that one needs an adjustment term in the LAN expansion, which accounts for the first-stage estimation error. Assumption \ref{assump:SE} (i) is needed to ensure that the likelihood ratio $\prod_{i=1}^n[p_{\theta}(Y_{2i};\widehat{\zeta}_n)/p_{\widehat{\theta}_n}(Y_{2i};\widehat{\zeta}_n)]$ decays exponentially fast for $\{\theta:d_{\Theta}(\theta,\theta_0)>\delta\}$ with a fixed radius $\delta>0$ as in Lemma S1. Meanwhile, Assumptions \ref{assump:regular} and \ref{assump:SE} (iii) are used to strengthen the exponential type decay for $\{\theta:Cr_n\leq d_{\Theta}(\theta,\theta_0)\leq\delta\}$, with the rate $r_n$ stated in Assumption \ref{assump:FirstStep}. The differentiability condition in Assumption \ref{assump:smooth} with respect to $\theta$ and $\zeta$ are needed to account for the two stages in our quasi-Bayesian method. Finally, Assumption \ref{assump:normal} is used to establish the quadratic expansion of the log-likelihood over the range $\{\theta: d_{\Theta}(\theta,\theta_0)\leq Cr_n\}$ in Lemma S3 of the supplementary material. Therein we also require Assumption \ref{assump:SE} (ii) to kill various smaller order terms via the stochastic equicontinuity. We also need Assumption \ref{assump:normal}, when studying the frequentist coverage property of the quasi-Bayesian credible set. %\textcolor{magenta}{[How about adding a brief discussion about the roles of Assumption 2.7(i) and (ii)?]} % In addition, we have the following maximal inequality for the 
 	\subsection{Large Sample Behaviors of Quasi-Posteriors}\label{sec: asym_quasi_posterior}
 	 		We develop the asymptotic theory by drawing on two themes. The first is the frequentist two-step estimation and inference \citep{newey1994var,chen2003estimation,ichimura2010semi}. The crux therein is how to characterize the influence of the first-step nonparametric estimation on the remaining parametric components and how to make proper adjustments for inference. This analysis is also central to our study, and we formally show how the quasi-posterior ignores the effect from the first-stage estimation.  The second theme is the asymptotic analysis of Bayesian methods for semiparametric models. Our theoretical findings have not been reported before; they shed new light on the  BvM theorem \citep{van1998asymptotic,ghosal2017bayesian}. For standard semiparametric models using \textit{full information} Bayesian methods, the BvM theorem states that the marginal posterior for the finite-dimensional parameter is approximately a normal distribution centered at the semiparametric efficient estimator. Hence, the point estimators and credible sets are produced by one stroke therein. However, estimation and inference have to be dealt with separately for the quasi-posterior distribution in our setting.
 		
Let $\tilde{\pi}$ be the density function of $h\equiv \sqrt{n}(\theta-\widehat{\theta}_n)$ conditional on the data $\bm{Y}_2^n$ and the first stage estimate $\widehat{\zeta}_n$:
 		\begin{equation*}
 			\tilde{\pi}(h|\bm{Y}_2^n;\widehat{\zeta}_n)=\frac{p^n_{\widehat{\theta}_n+\frac{h}{\sqrt{n}}}(\bm{Y}_2^n|\widehat{\zeta}_n)\pi(\widehat{\theta}_n+\frac{h}{\sqrt{n}})}{\int p^n_{\widehat{\theta}_n+\frac{h}{\sqrt{n}}}(\bm{Y}_2^n|\widehat{\zeta}_n)\pi(\widehat{\theta}_n+\frac{h}{\sqrt{n}})dh}.
 		\end{equation*}
 		Furthermore, the limiting normal density is
 		\begin{equation*}
 			\pi_{\infty}(h)\equiv (2\pi)^{-p/2}|\det V_0|^{1/2}\exp\left\{-\frac{h^\top V_0h}{2}\right\}.
        \end{equation*}
In order to metrize the weak convergence, we consider the following ``total variation of moments'' norm \citep{chernozhukov2003mcmc} for a real-valued function $f$ on $\mathcal{S}$ as
 	\begin{equation*}
 		\Vert f\Vert_{TVM(a)}\equiv \int (1+|h|^{a})|f(h)|dh,
 	\end{equation*}
 	for the choice of $a\geq 0$ in Assumption \ref{assump:Prior}. Let $V_0^{1/2}$ denote the square-root of the positive definite matrix $V_0$.		
 		
		\begin{theorem}[Posterior Measures]\label{thm:BvM}		
		Suppose Assumptions \ref{assump:id} to \ref{assump:normal} hold. Then, we have
		\begin{equation}\label{BvM_result}
	\Vert \tilde{\pi}(h|\bm{Y}_2^n;\widehat{\zeta}_n)-\pi_{\infty}(h)\Vert_{TVM(a)} =o_{P_0}(1).
		\end{equation}
Consequently, the following rescaled sequence of quasi-posterior converges in total variation, that is,
		\begin{equation}\label{ConvTV}
			\sup_{\xi\in \mathbb{R}^p}|\Pi(\sqrt{n}V_0^{1/2}(\theta-\widehat{\theta}_n)\leq \xi|\bY^n_2;\widehat{\zeta}_n)-\Phi_p(\xi)|\to_{P_0} 0.
		\end{equation}
		where $\Phi_p(\cdot)$ denotes the $p$-dimensional standard normal distribution function.
		\end{theorem} 
	Theorem \ref{thm:BvM} may look similar to the standard parametric BvM theorem at first glance. However, the departures are the presence of the first-stage estimation $\widehat{\zeta}_n$ and the centering point $\widehat{\theta}_n$. The latter one is a two-stage frequentist estimator whose asymptotic covariance matrix takes the sandwich form $V_0^{-1}\Omega_0V_0^{-1}$, with $\Omega_0$ corresponding to the variance of $\Gamma_2(Y_{2i})+\Gamma_1(Y_{1i})$ in Assumption \ref{assump:normal}. This has important implication about the frequentist coverage property of the credible set.
	
Let $\mathcal{C}_n(\alpha)$ be the quasi-Bayesian credible set constructed from quantiles of the quasi-posterior such that $\Pi(\theta\in \mathcal{C}_n(\alpha)|\bY^n_2;\widehat{\zeta}_n)=1-\alpha$ for a given nominal level $\alpha\in(0,1)$. An important consequence of Theorem \ref{thm:BvM} is that $\mathcal{C}_n(\alpha)$ does not have the correct asymptotic coverage probability in general, as stated in Corollary \ref{cor:coverage}. Recall the definition of $\Delta_{n,0}$ in Assumption \ref{assump:normal}.
\begin{corollary}\label{cor:coverage} Let $B_n$ be any set that satisfies $(2\pi)^{-p/2}\int_{B_n}e^{-h^\top h/2}dh\to 1-\alpha$ as $n\to\infty$.
Under the same assumptions for Theorem \ref{thm:BvM}, we have 
\begin{equation}\label{ci_cover_general}
\lim_{n\to\infty}	P_0\{\theta_0\in \mathcal{C}_n(\alpha)\}=\lim_{n\to\infty}	P_0\{-V_0^{-1/2}\Delta_{n,0}\in B_n\}.
\end{equation}
If $V_0=\Omega_0$, we have
		\begin{equation}\label{ci_cover_special}
				P_0\{\theta_0\in \mathcal{C}_n(\alpha)\}\to 1-\alpha.
		\end{equation}
	 \end{corollary}
Referring to Corollary \ref{cor:moments}, the variance of the quasi-posterior of $\theta$ is governed by the information matrix $V_0$, which only 
captures the variance of $\Gamma_2(Y_{2i})$. If the nuisance function is known, or we plug in $\zeta_0$ rather than $\widehat{\zeta}_n$, then the Bayesian approach for the second stage would give us
\begin{equation}\label{IFCov}
	\sup_{\xi\in \mathbb{R}^p}|	P_0(\sqrt{n}(V_0)^{1/2}(\theta-\widehat{\theta}_{IF,MLE})\leq \xi|\bm{Y}_2^n)-\Phi_p(\xi)|\to_{P_0} 0.	
\end{equation}
The centering point is the \textit{infeasible MLE} based on $\zeta_0$ such that:
\begin{equation}\label{IFAsymp}
	\sup_{\xi\in \mathbb{R}^p}|	P_0(\sqrt{n}(V_0)^{1/2}(\widehat{\theta}_{IF,MLE}-\theta_0)\leq \xi)-\Phi_p(\xi)|\to_{P_0} 0.	
\end{equation}
Given equations (\ref{IFCov}) and (\ref{IFAsymp}), one can easily show the frequentist validity if $\zeta_0$ is known exactly. For the quasi-Bayesian approach that replaces $\zeta_0$ with $\widehat{\zeta}_n$, its centering point in (\ref{ConvTV}) is the frequentist \textit{two-stage} estimator such that:
\begin{equation}
	\sup_{\xi\in \mathbb{R}^p}|	P_0(\sqrt{n}(V^{-1}_0\Omega_0V_0^{-1})^{-1/2}(\widehat{\theta}_n-\theta_0)\leq \xi)-\Phi_p(\xi)|\to_{P_0} 0.	
\end{equation}
Referring to the statement in (\ref{ConvTV}), we can see that the discrepancy between $\Omega_0$ and $V_0$ prevents the right hand side of (\ref{ci_cover_general}) from converging to the desired coverage probability $1-\alpha$ in general. 

With the additional restriction $V_0=\Omega_0$, which corresponds to the case where the generalized information equality holds \citep{chernozhukov2003mcmc}, $\mathcal{C}_n(\alpha)$ will have the correct coverage probability coincidentally, as (\ref{ci_cover_special}) shows. In our analysis of the endogenous discrete choices model in Section \ref{sec:PT model}, this occurs in the absence of endogeneity.\footnote{Appendix D calculates the influence from the first-stage estimation for the Petrin--Train model with three alternatives, see equations (S.1) to (S.3).} Intuitively, this holds as the likelihood function of the second stage is free from the first stage control variables without the endogeneity. We believe the scenario is rather special, where one does not need the control function approach from the beginning. The general coverage failure of the quasi-Bayesian credible set
is similar to what \cite{kleijn2012mis} found when they studied the Bayesian procedure for misspecified models. Considering the formal decision theory of the interval estimation problem, \cite{muller2013risk} showed that the asymptotic risk of a vanilla posterior is worse than that of a modified posterior using the sandwich covariance matrix. Also, see \cite{kim2014quasi} for the asymptotic theory of the posterior odds ratio for testing hypotheses with the limited-formation Bayesian approach. 

We are interested in whether the mean of this quasi-posterior is a legitimate point estimator for $\theta_0$. For this purpose, we define the quasi-posterior mean and  covariance matrix as follows:
	\begin{equation}\label{MeanVar}
	\widetilde{\theta}_n=\int_{\Theta} \theta d\Pi(\theta|\bY_2^{n};\widehat{\zeta}_n)~~\text{and}~~\widetilde{\mathrm{Var}}_n(\theta)=\int_{\Theta} [\theta-\widetilde{\theta}_n]^{\otimes 2} d\Pi(\theta|\bY_2^{n};\widehat{\zeta}_n).
\end{equation}
	Below, we show that the quasi-posterior mean is asymptotically equivalent to the frequentist two-stage estimator $\widehat{\theta}_n$. However, the quasi-posterior variance is not the same as the asymptotic variance of the frequentist two-stage estimator.
	
	\begin{corollary}\label{cor:moments}
	Assume that the prior for $\theta$ satisfies $\int |\theta|^2d\Pi(\theta)<+\infty$. Then, we have 
		\begin{equation}\label{post_mom}
			\widetilde{\theta}_n=\widehat{\theta}_n+o_{P_0}(n^{-1/2}), ~\text{and}~n^{-1}\widetilde{\mathrm{Var}}_n(\theta)=V^{-1}_{0}+o_{P_0}(1).
		\end{equation}
		As a result of the first equality in (\ref{post_mom}), the quasi-posterior mean is asymptotically normal:
		\begin{equation*}
			\sqrt{n}(\widetilde{\theta}_n-\theta_0)\Rightarrow\mathbb{N}(0,V^{-1}_{0}\Omega_{0}V^{-1}_{0}).
		\end{equation*}
	\end{corollary}

\begin{remark}
In the context of GMM, \cite{chernozhukov2003mcmc} constructed a quasi-likelihood by exponentiating the quadratic criterion function. \cite{chernozhukov2003mcmc} demonstrated that the weighting matrix can be properly chosen so that the generalized information identity holds, and the resulting posterior distribution can be utilized for asymptotically valid confidence sets. When the generalized information equality fails, \cite{chernozhukov2003mcmc} also suggested that the posterior still contains useful information for inference, as the posterior variance is related to the Hessian matrix. If $\Omega_0$ is easier to obtain, one can form the normal confidence interval by plugging in the sandwich-type covariance matrix estimator, which combines the posterior variance and some external estimand for $\Omega_0$. Our motivating example does not belong to this case, as it is not computationally easy to obtain a consistent estimator for $\Omega_0$. This advocates the bootstrap procedure that we examine in the Section \ref{sec:resampling}.
\end{remark}

\subsection{Validity of Bootstrap Inference}\label{sec:resampling}
 %Now, we present the validity of a proper bootstrap inferential procedure that takes into account the first-stage estimation error. 
This section applies nonparametric bootstrap to our quasi-Bayesian method. The resulting inferential procedure yields asymptotically valid confidence intervals and avoids analytical correction for the first stage estimation error needed to be carried out case-by-case. 
 We show that the bootstrap point estimator mimics the asymptotic behavior of the quasi-Bayesian point estimator. The essence of this analysis lies in the bootstrap likelihood with multinomial weights. The technical challenge is to deal with the infinite-dimensional $\zeta$ from the first stage. Our theory also connects the bootstrap procedure suggested by \cite{agarwal2018demand} to a more general context with two-stage semiparametric estimation. In principle, the theory works for other exchangeable weights \citep{van1996empirical}. However, the most convenient approach is Efron's nonparametric bootstrap \citep{efron1979bootstrap}, because it only requires generating a sequence of new data sets by sampling the rows of the original data with replacement and then updating the posterior with the same prior. It is straightforward to compute all bootstrap calculations in parallel.
		
\begin{figure}
			\caption{Quasi-Posterior without and with bootstrap inference: a simulation replication with $J+1=3$, Design I in Section \ref{sec:simulation} with $\theta = \tilde{\beta}=1$, control function $\widehat{\zeta}_n = \left(\hat{v}^\dagger_{i1}, \hat{v}^\dagger_{i2}\right),i=1,\cdots, n$. Solid vertical line in magenta = true value of the parameter, dashed vertical line in black = mean of the distribution.}\label{fig:flowchart}
			\vspace{2mm}
			\begin{tabular}{cc}
				(a). Quasi-posterior & (b). Quasi-posterior with bootstrap\\
				\includegraphics[scale=0.7]{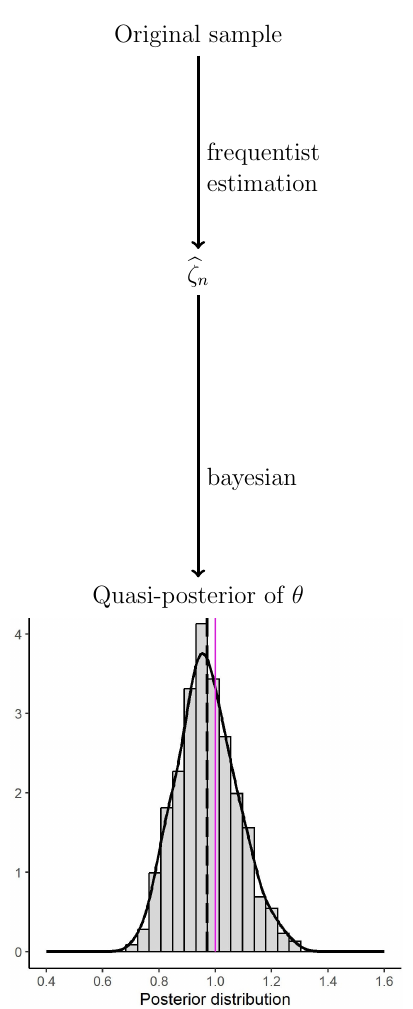}&
				\includegraphics[scale=0.7]{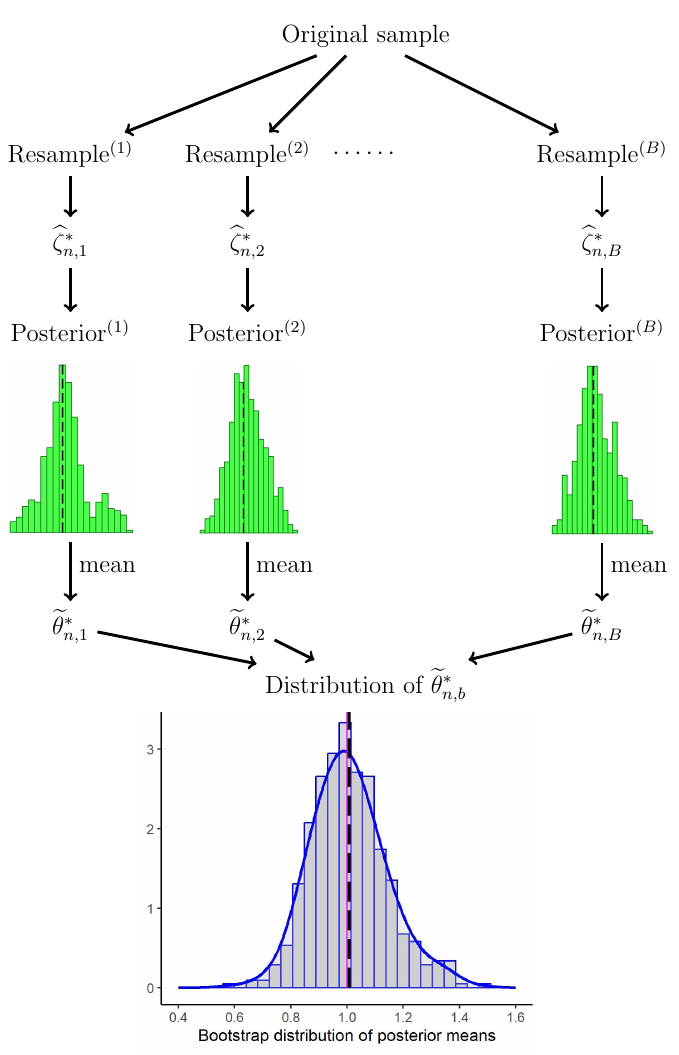}
			\end{tabular}
		\end{figure}

		Let the bootstrap weights $\bm{M}_{n}=\left(  M_{n1},...,M_{nn}\right)^{\top}$ follow the multinomial distribution $Mult\left(  n,\left(n^{-1},...,n^{-1}\right) \right)$ \citep{efron1979bootstrap}. The resulting marginal posterior distribution is given by the Bayes formula:
		\begin{equation}
		\Pi^*[\theta\in \mathcal A|\bY_2^{n};\widehat{\zeta}^*_n]=\frac{\int_{\mathcal A} p^{n*}_{\theta}(\bY_2^n;\widehat{\zeta}^*_n)d\Pi(\theta)} {\int_{\Theta} p^{n*}_{\theta}(\bY_2^n;, \widehat{\zeta}^*_n)d\Pi(\theta)},
		\end{equation}
		where $p^{n*}_{\theta}(\bY_2^n;\zeta)\equiv\prod_{i=1}^n p_{\theta}^{M_{ni}}(Y_{2i};\zeta) $ stands for the bootstrap likelihood function and $\widehat{\zeta}^*_n$ stands for the first-stage bootstrap estimator. The bootstrap quasi-Bayesian point estimator is
		\begin{equation}\label{PointEstBayes}
			\widetilde{\theta}^*_n=\int_{\Theta} \theta d\Pi^*(\theta|\bY_2^{n};\widehat{\zeta}^*_n).
		\end{equation}
		Also, denote by $\widehat{\theta}_n^*$ the bootstrap counterpart of the frequentist two-stage estimator in Assumption \ref{assump:Freq}. Algorithm \ref{algorithm_1} below provides the implementation details.
		\begin{algorithm}[H]
			\caption{Bootstrap inference for Quasi-Bayesian method}
			\label{algorithm_1}
			\begin{algorithmic}
				\STATE \textbf{Input:} Data $\bY^n$, number of bootstrap replications $B$, and number of posterior draws $S$.
				
				\FOR{$b=1,\ldots, B$}
				\STATE Resample data $\bY^n$ with replacement to form the $b$th-bootstrap sample $\bY_b^n$.
				\STATE  (1). Obtain a frequentist estimate $\widehat{\zeta}_{n,b}^*$ from the first-stage data $\bY^n_{1,b}$.
				\STATE (2).  Generate $\left\{\widetilde{\theta}^*_{n,b,s}: s=1,\ldots, S\right\}$ as $S$ draws from the posterior of $\theta$ conditional on $\bY^n_{2,b}$ and $\widehat{\zeta}_{n,b}^*$. 
				\STATE (3). Calculate the bootstrap quasi-Bayesian point estimate $\widetilde{\theta}^*_{n,b} = \frac{1}{S}\sum_{s=1}^S\widetilde{\theta}^*_{n,b,s}$, whose $j$th coordinate is denoted as $\widetilde{\theta}^*_{n,b,j}$. 
				\STATE (4). Calculate the bootstrap quasi-Bayesian standard error $\widetilde{s}^*_{n,b,j}$ for the $j$th coordinate of parameter $\theta_0$ as the $j$-th diagonal element of the quasi-posterior variance matrix $\frac{1}{nS}\sum_{s=1}^S\left(\widetilde{\theta}^*_{n,b,s}-\widetilde{\theta}^*_{n,b}\right)^{\otimes 2}$.
				\ENDFOR
				\STATE \textbf{Output:} $\left\{\widetilde{\theta}^*_{n,b,j},\widetilde{s}^*_{n,b,j} :b=1,\dots,B, j=1,\dots,p\right\}$.
			\end{algorithmic}
		\end{algorithm}
	Figure \ref{fig:flowchart} provides a graphical illustration of Algorithm \ref{algorithm_1} and compares it to the quasi-Bayesian credible set $\mathcal{C}_n(\alpha)$ described in Section \ref{sec: asym_quasi_posterior}. The histogram and density plots in the figure are drawn based on one simulated sample in Section \ref{sec:simulation}. We emphasize that it is crucial to bootstrap the Bayesian point estimator (which is taken to be the quasi-posterior mean) in each bootstrap replication in order to ensure the frequentist validity of the inference procedure. 
		In the supplementary material, we compare our proposal with an alternative procedure that only makes a single draw from the quasi-posterior in each bootstrap sample.
		
Let $\theta_{0,j}$ be the $j$th coordinate of the parameter $\theta_0$. The $1-\alpha$ bootstrap percentile interval for $\theta_{0,j}$ can be formed as 
\begin{equation}\label{bsci_be}
\mathcal{C}^{PC}_{n,j}(\alpha)  =\left[  Q_{n,j}^{\ast}(\alpha/2),  Q_{n,j}^{\ast}(1-\alpha/2) \right], ~\text{for}~ j=1,\cdots,p,
\end{equation}
where $Q^*_{n,j}(a)$ denotes the $a$-th quantile of the bootstrap distribution $\left\{\widetilde{\theta}^*_{n,b,j} :b=1,\ldots,B\right\}$.
Besides the  bootstrap percentile interval, we also consider a percentile-$t$ interval by bootstrapping a $t$-ratio. For the parameter $\theta_j$, calculate the standard error $\widetilde s_{n,j}$ as the square root of the $j$-th diagonal element of the quasi-posterior variance matrix $n^{-1}\widetilde{\mathrm{Var}}_n(\theta)$. Corollary \ref{cor:moments} implies that $\widetilde s_{n,j}\to_{P_0}\sqrt{(V_0^{-1})_{jj}}$.
We introduce the following bootstrap percentile-$t$ interval:
\begin{equation}\label{bsci_ts}
	\mathcal{C}^{PT}_{n,j}(\alpha)  =\left[ \widetilde{\theta}_{n,j}-\widetilde s_{n,j} q_{n,j}^{\ast}(1-\alpha/2),  \widetilde{\theta}_{n,j}+\widetilde s_{n,j} q_{n,j}^{\ast}(\alpha/2) \right], ~\text{for}~ j=1,\cdots,p,
\end{equation}
where $q_{n,j}^{\ast}(a)$ denotes the $a$-th quantile of the bootstrap distribution of the $t$-ratio:
\[
\left\{ (\widetilde{\theta}_{n,b,j}^*-\widehat{\theta}_{n,j})/\widetilde{s}^*_{n,b,j} :b=1,\cdots,B\right\}.
\] %Although this will not make the asymptotic distribution to be the standard Gaussian with unit variance, such a normalization might be helpful in finite samples. 
Although the $t$-ratio studentized by the quasi-posterior standard error would not make its limiting distribution equal to the standard normal, the resulting bootstrap percentile-$t$ interval in (\ref{bsci_ts}) is asymptotic distribution, see Section 10.19 in \cite{hansen2022econometrics} for a related discussion for bootstrapping  frequentist two-stage estimators.\footnote{We thank one anonymous referee for making this suggestion.} 
%Here the $q_{n,j}^{\ast}(\alpha)$ is the $\alpha$-th empirical quantile from the bootstrap draws of the $t$-ratio $(\widehat{\theta}_{n,j}^*-\widehat{\theta}_{n,j})/s^*_{n,j}$, where $s^*_{n,j}$ denotes the bootstrap posterior standard error \textcolor{magenta}{[...the bootstrap distribution of the $t$-ratio $\left\{ (\widetilde{\theta}_{n,b,j}^*-\widehat{\theta}_{n,j})/{s}^*_{n,b,j} :b=1,\ldots,B\right\}$.]}. 
		\begin{theorem}[Bootstrap Consistency]\label{thm:bootstrap}
		In addition to Assumptions \ref{assump:id} to \ref{assump:normal}, we assume that the envelope functions for the functional classes in Parts (i) and (ii) in Assumption (\ref{assump:SE})  have finite $(2+\iota)$ moments, for $\iota>0$.  We have the following equivalence between the bootstrap posterior mean and the bootstrap frequentist-type two-stage estimator: $\widetilde{\theta}^*_n=\widehat{\theta}^*_n+o_{P^*}(n^{-1/2})$.
		Furthermore, we have 
		\begin{equation}
			P_0\left\{\theta_{0j}\in \mathcal{C}^{PC}_{n,j}(\alpha)  \right\}\to 1-\alpha, ~~\text{and} ~~P_0\left\{\theta_{0j}\in \mathcal{C}^{PT}_{n,j}(\alpha)  \right\}\to 1-\alpha,
		\end{equation}
	for $j=1,\cdots,p$.
		\end{theorem}

 \begin{remark}
	%In a different context of analyzing Bayesian variational approximation, 
In the context of Bayesian variational approximation which differs from ours, the bootstrap approach also serves as a useful alternative to restore the valid coverage \citep{blei2017review}. Variational Bayesian inference is a popular approach for approximating the complex posterior density functions through optimization within a given family. The resulting approximate posteriors are generally under-dispersed \citep{blei2017review}. \cite{chen2018bootstrap} developed uncertainty measures for variational inference by using bootstrap procedures. A closer examination reveals that their bootstrap theory essentially follows general bootstrap results on parametric $M$-estimation \citep{van1996empirical}. In contrast, our quasi-Bayesian point estimator requires a subtle analysis of integral of the quasi-posterior distribution. We overcome this challenge by deriving new maximal inequalities for bootstrap likelihood ratios and conducting a detailed LAN expansion of the bootstrap likelihood. To the best of our knowledge, these theoretical developments are new.  
\end{remark}

\section{The Petrin--Train Model}\label{sec:PT model}
We illustrate our theory in Section \ref{sec: main results} by applying it to a class of endogenous discrete choice models originally proposed by \cite{petrin2010control}. We revisit Example 1 in \cite{petrin2010control} with normal latent errors, which does not enforce the independence of irrelevant alternatives (IIA) assumption in Logit models. %The joint normality of the latent errors in both stages naturally leads to the identification strategy by the control function approach. 
The joint normality of the latent errors in both stages facilitates application of the control function approach.
We relax the first-stage specification in \cite{petrin2010control} to allow nonparametric estimation. We henceforth refer to this model as the Petrin-Train model throughout the paper. %Our study provides the theoretical foundation for the quasi-Bayes estimation and inferential tools. To the best of our knowledge, this has not been formally developed. 

As demonstrated below, Petrin-Train model has two features: (I) the conditional choice probability and thus the likelihood function contain complicated multi-dimensional integrals; (II) it involves the pilot estimation of (functional) nuisance parameters. Many econometric models share similar structures. For instance, \cite{agarwal2018demand} estimate the distribution of students' preferences for public schools by first estimating the believed assignment probabilities at various schools using frequentist methods and then conducting Bayesian estimation in the second stage. Their second-stage likelihood involves a multiple integral similar to the MNP, albeit with fairly complicated decision rules in their context. Our analysis of the Petrin-Train model can serve as a template for studying the more complex choices models. 

Let $U_{ij}$ for $j=0,1,\cdots,J$ denote the latent utility of alternative $j$ for individual $i$, with the following linear specification:
\begin{equation}\label{latent_utility_2}
	U_{ij}=X_{ij}^\top\tilde\beta_0+\varepsilon_{ij}, \quad i=1,\cdots,n, j=0,\cdots,J.
\end{equation}
The individual $i$'s choice $C_i\in\{0,1,\cdots,J\}$ is determined by maximizing the utility across $(J+1)$ alternatives. We treat the choice $0$ as the baseline and work with utility differences as follows:
 \begin{equation}\label{MNP_second_stage}
 	U_{ij}-U_{i0}=(X_{ij}-X_{i0})^\top\tilde\beta_0+\varepsilon_{ij}-\varepsilon_{i0}, \quad i=1,\cdots,n, j=0,\cdots,J,
 \end{equation} 
In the sequel, we denote
 \begin{equation}
	U^{\dagger}_{ij}=(X^{\dagger}_{ij})^\top\tilde\beta_0+\varepsilon^{\dagger}_{ij}, \quad i=1,\cdots,n, j=1,\cdots,J,~~~\text{and}~~~U_{i0}^{\dagger}=0,
\end{equation} 
with $U^{\dagger}_{ij}=U_{ij}-U_{i0}$, $X^{\dagger}_{ij}=X_{ij}-X_{i0}$ and $\varepsilon^{\dagger}_{ij}=\varepsilon_{ij}-\varepsilon_{i0}$. Hence, the utility maximization leads to the following decision rule:
\begin{equation}\label{choice}
	C_i=
\begin{cases}
	0, & \text{if}~~\max\{(U_{ij}^{\dagger})_{j=1}^J \}\leq 0 \\
	k, & \text{if}~~\max\{0,(U_{ij}^{\dagger})_{j=1}^J \}=U_{ik}^{\dagger}.
\end{cases}
\end{equation}
%Note that the choice can also be concisely summarized as
%\begin{eqnarray}\label{choice}
%	C_{i} = \sum_{j=1}^J j\mathbb{I}\left\{U^{\dagger}_{ij}>U^{\dagger}_{ik} ~\text{for} ~ k\neq j ~\text{and}~ U^{\dagger}_{ij}>0 \right\}.
%\end{eqnarray}
Among the choice-specific covariates $X^{\dagger}_{ij}$, some components denoted by $X_{ij}^{\dagger e} $ may be correlated with the error term $\varepsilon^{\dagger}_{ij}$. The endogenous regressors $X^{\dagger e}_{ij}$ admit the following representation\footnote{For notation simplicity, we work with a scalar endogenous variable $X_{ij}^{\dagger e}$ so that the range of each function $\zeta_j(\cdot)$ is the real line.}:
\begin{equation}\label{MNP_first_stage_2}
	X^{\dagger e}_{ij}=\zeta_j(Z_{ij})+v^{\dagger}_{ij}, \quad j=1,\cdots,J,
\end{equation}
where $Z_{ij}$ collects the instrumental variables and other exogenous regressors, %that enable the identification
and $v^{\dagger}_{ij}$ stands for the unobservable error that is correlated with $\varepsilon^{\dagger}_{ij}$ %from the structural equation. 
The functions $(\zeta_j)_{j=1}^J$ are completely unspecified, except for some standard smoothness restrictions as detailed later. Following Example 1 of \cite{petrin2010control}, the joint normality of latent errors $\varepsilon^{\dagger}_{ij}$ and $v^{\dagger}_{ij}$  leads to the decomposition of $\varepsilon^{\dagger}_{ij}$:
\begin{equation}\label{MNP_control_func}
	\varepsilon^{\dagger}_{ij}=\mathbb{E}[\varepsilon^{\dagger}_{ij}\mid v^{\dagger}_{ij}]+\epsilon^{\dagger}_{ij}=\lambda_{0}v^{\dagger}_{ij}+\epsilon^{\dagger}_{ij},\quad j=1,\cdots,J,
\end{equation}
where $\lambda_{0}v^{\dagger}_{ij}$ is the control function\footnote{In principle, we can allow for different $\lambda_{0,j}$ over $j=1,\cdots,J$. In this case, the model can be written in the form  of $U^{\dagger}_{ij}=(X^{\dagger}_{ij})^\top\beta_0+ (\tilde{v}_{ij}^\dagger)^{\top}\lambda_0+\epsilon^{\dagger}_{ij}$, by generating proper longer vectors of $\tilde{v}_{ij}^\dagger$ and $\lambda^\top_0=(\lambda_{0,j})_{j=1,\cdots,J}$. See Chapter 6 of \cite{lee2010micro} for the general formulation. For ease of exposition, we do not consider such complications.}, and the the reminder errors $(\epsilon^{\dagger}_{ij})_{j=1}^J$ are also jointly normal. Using (\ref{MNP_control_func}), the latent utility of the second stage can be written as:
\begin{equation}\label{MNP_second_stage_2}
	U^{\dagger}_{ij}=(X^{\dagger}_{ij})^\top\tilde\beta_0+ \lambda_0v^{\dagger}_{ij}+\epsilon^{\dagger}_{ij}, \quad j=1,\cdots,J,~~~\text{and}~~~U_{i0}^{\dagger}=0.
\end{equation}

%Taking difference relative to $U_{i0}$ leads to  
%\begin{equation}\label{MNP_second_stage_diff}
%	U_{ij}-U_{i0}=(X_{ij}-X_{i0})^\top\beta_0+ \lambda_{0}(v_{ij}-v_{i0})+\epsilon_{ij}-\epsilon_{i0}, \quad i=1,\cdots,n, j=1,\cdots,J.
%\end{equation} 
The conditional choice probability for the $j$-th alternative in the second stage involves a multivariate integral:
\begin{eqnarray*}\label{CCP}
P_0\left(C_i=j\mid X_i,Z_i\right)=
\begin{cases}
&\int \mathbb{I}\left\{U^{\dagger}_{ik}\leq 0, \forall k\neq j \right\}dG(\epsilon^{\dagger}_{i1},\cdots,\epsilon^{\dagger}_{iJ}), ~~\text{if}~~j=0,\\
&\int \mathbb{I}\left\{U^{\dagger}_{ik}<U^{\dagger}_{ij},\forall k\neq j, ~\text{and}~ U^{\dagger}_{ij}>0\right\}dG(\epsilon^{\dagger}_{i1},\cdots,\epsilon^{\dagger}_{iJ}), ~~\text{if}~~j\neq0.
\end{cases}
\end{eqnarray*}
The function $G$ is a $J$-dimensional normal distribution with mean zero and covariance matrix $\Sigma_0$. For identification purpose, we assume that $\Sigma_0$ can be parameterized as $\Sigma_0=\Gamma(\eta_0) $ which depends on some given transformation $\Gamma(\cdot)$ and unknown parameter $\eta_0$. For example, one can take $(\Sigma_0)_{jk}=\eta^{|j-k|}_0$ for the $(j,k)$-th element in $\Sigma_0$. We collect the regression coefficients as $\beta^\top_0=(\tilde\beta^\top_0,\lambda_0)$ as the parameter of interest and treat $\eta_0$ as nuisance parameter.

Our quasi-Bayesian method applied to the Petrin--Train model consists of two stages. The first stage estimates the functions $\zeta_j$ nonparametrically (e.g., kernel regression) from equation (\ref{MNP_first_stage_2}) and then obtains the residuals $(\hat v^{\dagger}_{ij})_{j=1}^J$. The second stage corresponds to the MNP model (\ref{MNP_second_stage_2}) with $(v^{\dagger}_{ij})_{j=1}^J$ replaced by the first stage residual estimates. %Due to the analytically intractable conditional choice probability, 
As conditional choice probability $P_0\left(C_i=j\mid X_i,Z_i\right)$ is analytically intractable, Bayesian approach becomes more appealing as it explores the conditional conjugate structure induced by MNP. The posterior of $\beta$ can be drawn using Markov Chain Monte Carlo (MCMC) algorithms coupled with data augmentation techniques \citep{albert1993bayesian,mcculloch1994exact,nobile1998hybrid,imai2005bayesian}. Chapter 5 of \cite{train2009discrete} presents a comprehensive discussion about the advantages of Bayesian methods for MNP-type models. We also refer interested readers to \cite{anceschi2023conjugacy} for a recent review on the contemporary development of various fast and scalable MCMC or deterministic algorithms. We conduct the bootstrap inference following Algorithm \ref{algorithm_1}. Note that the second-stage Bayesian procedure does not require maximization of any function. This reduces the computational burden compared to bootstrapping a frequentist two-stage estimator, whose second stage involves maximization of the simulated likelihood function associated with the MNP model. The computational advantage of the Bayesian second-stage over a frequentist second-stage is amplified when bootstrap resampling is called for valid inference. Our simulation exercise in Section \ref{sec:simulation} illustrates that the quasi-Bayesian method can save the computation time to a large extent.

%The point estimator for $\beta$ can be constructed as the posterior mean. We use bootstrap to construct the confidence interval. Resampling the original data with replacement forms a bootstrap sample. For each bootstrap sample, we apply the first-stage frequentist and the second-stage Bayesian procedures to obtain the bootstrap quasi-posterior mean $\tilde\beta_n^{\ast}$ as in (\ref{PointEstBayes}). Repeatedly drawing the bootstrap sample and calculating the bootstrap point estimator for many times leads to a bootstrap distribution of $\tilde\beta_n^{\ast}$, whose quantiles form the bootstrap confidence intervals described in (\ref{bsci_be}) or (\ref{bsci_ts}). Note that the second-stage Bayesian procedure does not require maximization of any function. This reduces the computational burden compared to bootstrapping a frequentist two-stage estimator, whose second stage involves maximization of the simulated likelihood function associated with the MNP model.

%To spell out the regularity conditions that are sufficient to establish the asymptotic results,
In the following, we present low level conditions that are sufficient to specialize Theorems \ref{thm:BvM} and \ref{thm:bootstrap} to the Petrin--Train model. We introduce notation: denote the collection of vectors across non-baseline choices by $\bU^{\dagger}_i= [U^{\dagger}_{i1},..., U^{\dagger}_{iJ}]^\top$. We write
\begin{eqnarray*}
\bm{W}_i=
\begin{bmatrix} 
W_{i1} & 0 & \cdots & 0 \\
0 & W_{i2} & \cdots & 0\\
\vdots & \vdots & \ddots & \vdots\\
0 &0 &\cdots &W_{iJ}
\end{bmatrix}
~~\text{and}~~\bm{W}_i^\top\beta=
\begin{bmatrix} 
W^\top_{i1}\beta \\
W^\top_{i2}\beta  \\
\vdots \\
W^\top_{iJ}\beta 
\end{bmatrix},
\end{eqnarray*}
where $W_{ij}^\top=((X^{\dagger}_{ij})^\top,v^{\dagger}_{ij})$. Let $R(C_i)$ specify the truncation region associated with the choice variable. If $C_i=0$, $R(C_i)$ consists of the region such that each component of the latent utility is non-positive. For $C_i=j\neq 0$, $R(C_i)$ restricts $U^{\dagger}_{ij}$ to be positive and greater than all other $U^{\dagger}_{ik}, k\neq j$. Referring to the expression (\ref{PScore}) with $\theta=(\beta^\top, \eta^\top)^\top$, the score functions w.r.t. $\beta$ and $\eta$ take the following forms:
\begin{align*}
	\frac{\partial \log p(Y_2,\theta;\zeta)}{\partial \beta^\top}&=\bm{W}_i^\top \Sigma^{-1}\mathbb{E}[\bm{U}^{\dagger}_i-\bm{W}_i^\top\beta|\bm{U}^{\dagger}_i\in R(C_i)],\\
	\frac{\partial \log p(Y_2,\theta;\zeta)}{\partial \eta^\top}&=vec( \Sigma^{-1}\left(\mathbb{I}_{J\times J}-\mathbb{E}[(\bm{U}^{\dagger}_i-\bm{W}_i^\top\beta)^{\otimes 2}|\bm{U}^{\dagger}_i\in R(C_i)]\Sigma^{-1}\right)\Gamma(\eta))\frac{\partial vec(\Gamma(\eta))}{\partial \eta},
\end{align*}
cf. equations (4) and (5) of \cite{hajivassiliou1996simulation}. Clearly, the score functions also involve multivariate integrals, which makes the analytical correction for the first-stage estimation error difficult. In Assumption \ref{MNPAssump:Prior} below, $\mathbb{W}_{J}$ denotes the Wishart distribution for $J\times J$ positive-definite random matrices. 

\begin{assumption}\label{MNPAssump:Identify}
The parameter space $\Theta$ is a compact set, and the true $\theta_0$ is in the interior of $\Theta$. The latent error's covariance matrix $\Gamma_0(\eta_0)$ has its first element $\sigma_1$ normalized to 1, and it is non-singular. The matrix $V_0=-P_0\ddot{l}_{\theta}\left(  Y_2,\theta_0;\zeta_0\right)$ is positive definite.
\end{assumption}
\begin{assumption}\label{MNPAssump:1stStage}
The supports of covariates $X^{\dagger}$ and the control variable $v^{\dagger}$ are bounded. We assume the function $\zeta_{0j}\in \mathcal{C}^{\tau,L}(\mathbb{R}^d)$, for $j=1,\cdots,J$ with $\tau>d/2$. The functions $\zeta_{0j},j=1,\cdots,J$ are uniquely identified from the first stage. 
\end{assumption}
\begin{assumption}\label{MNPAssump:Linear}
	For the first-stage estimator, we assume its convergence rate with respect to the supnorm is $r_n=o(n^{-1/4})$. Furthermore, it has the following linear representation:	
	\begin{equation}
		\widehat{\zeta}_{n,j}(Y_{1i})-\zeta_{0j}(Y_{1i})=\frac{1}{n}\sum_{l=1}^n \phi_{n,j}(Y_{1i})+b_{n,j}(Y_{1i})+R_{n,j}(Y_{1i}), ~~j=1,\cdots,J,
	\end{equation}
	where $\phi_{n,j}(\cdot)$ is a stochastic term that has expectation zero, $b_{n,j}$ is a bias term satisfying $\max_{1\leq j\leq J}\Vert b_{n,j}(\cdot)\Vert_{\sup}=o(n^{-1/2})$ and $\max_{1\leq j\leq J}\Vert R_{n,j}(\cdot)\Vert_{\sup}=o_{P_0}(n^{-1/2})$.
\end{assumption}
\begin{assumption}\label{MNPAssump:Prior}
	Priors for the finite-dimensional parameters follow the Gaussian--Wishart type:
	\begin{equation}
		\tilde\beta\sim \mathbb{N}(\mu_{\beta},V_{\beta}),~~\text{and}~~ \Sigma^{-1}\sim \mathbb{W}_{J}([\rho V_0]^{-1},\rho)\mathbb{I}\{\sigma^2_{1}=1 \},
	\end{equation}
	in which the scalar $\rho$ represents the degree of freedom of the Wishart distribution\footnote{The indicator function on the Wishart prior serves to enforce the identification restriction that the $(1,1)$ element of $\Sigma$ is unity; see \cite{mcculloch2000identified}. This restriction is made for the identification purpose.} and $(\mu_{\beta},V_{\beta})$ specify the mean and variance of the Gaussian distribution. 
\end{assumption}
\begin{assumption}\label{MNPAssump:Moment}
The partial derivatives of $\dot{l}_{\theta}(\cdot,\theta;\zeta)$ with respect to control variables $(v_{ij}^{\dagger})_{j=1}^J$ are uniformly bounded over the support of the covariates and control variables $(W_{ij})_{j=1}^J$. In addition, $\dot{l}_{\theta}(\cdot,\theta;\zeta)$ is locally uniformly $L_2(P_0)$-continuous with respect to $\theta$ and $\zeta$ in the sense that for all small positive $\delta=o(1)$,
	\begin{equation}
		\mathbb{E}\left[\sup_{d_{\Theta}(\theta,\theta_0)<\delta,d_{G}(\zeta,\zeta_0)<\delta}\left|\dot{l}_{\theta}(Y_2,\theta;\zeta)-\dot{l}_{\theta}(Y_2,\theta;\zeta)\right|^2\right]\leq C\delta^2.
	\end{equation}
	Furthermore, the second-order derivative $\left\{\ddot{l}_{\theta}(y_2,\theta;\zeta):\theta\in\Theta,\zeta\in\mathcal{G}_n \right\}$ is upper semi-continuous for almost all $y_2$ and has a integrable envelope function.
\end{assumption}
\begin{assumption}\label{MNPAssump:Inf}
We express the pathwise derivative with respect to the first-stage estimation as
	\begin{equation}
		\dot{\Psi}_{\zeta_j}(\theta_0,\zeta_0)[\bm{h}]=\int \bm{h}(\cdot)g_j(\cdot,\theta)dP_0,
	\end{equation}
for some function $g_j(\cdot,\theta)$ differentiable w.r.t. $\theta$, where $j=1,\cdots,J$. We further assume that $\mathbb{E}_0\left[\frac{\partial g_j(\cdot,\theta_0)}{\partial \theta}|Z_{ij}\right]$ has finite fourth-order moment for $j=1,\cdots,J$.
\end{assumption}

The following proposition specializes Theorems \ref{thm:BvM} and \ref{thm:bootstrap} to the Petrin--Train Model. It establishes the asymptotic normality of the quasi-Bayesian point estimator and the asymptotic validity of the bootstrap confidence intervals.
\begin{proposition}\label{Prop:Example}
Under Assumptions \ref{MNPAssump:Identify} to \ref{MNPAssump:Inf}, the quasi-posterior mean is asymptotically normal:
\begin{equation*}
	\sqrt{n}(\widetilde{\theta}_n-\theta_0)\Rightarrow\mathbb{N}(0,V^{-1}_{0}\Omega_{0}V^{-1}_{0}).
\end{equation*}
Moreover, the bootstrap confidence intervals in (\ref{bsci_be}) and (\ref{bsci_ts}) are asymptotically valid.
\end{proposition}
We provide heuristic discussions about how Assumptions \ref{MNPAssump:Identify} to \ref{MNPAssump:Inf} imply the high-level assumptions in Section \ref{sec: main results}. We separate the discussion for the first and the second stages.

Regarding the first stage, we need to properly control the complexity of the underlying functional class which contains the $\zeta_0$. Herein, we work with the H\"older class. The desired Glivenko-Cantelli or Donsker properties are satisfied with sufficient smoothness restrictions (relative to the dimensionality of first-stage regressors) in Assumption \ref{MNPAssump:1stStage}. For nonparametric estimation in the first stage, one may use kernel smoothing estimators, including the local constant or local polynomial estimators in \cite{chen2003estimation,ichimura2010semi}, or sieve estimators in \cite{chen2007sieve}. For example, if we estimate $\zeta_{0j}$ by local constant kernel estimator with the kernel function $K(\cdot)$ and bandwidth $h^{d_z}$, the linear representation in Assumption \ref{MNPAssump:Linear} is
\begin{equation}
	\phi_{n,j}(Y_{1l})=\frac{1}{f_Z(Z_{lj})nh^{d_z}}\sum_{i=1}^nX_{lj}^{\dagger e}K\left(\frac{Z_{ij}-Z_{lj}}{h}\right).
\end{equation}
The control of bias $b_{n,j}$ or remainder terms $R_{n,j}$ are also standard; see Equation (3.14) and the discussion in \cite{ichimura2010semi}. 

 Referring to the second-stage likelihood structure, the high-level conditions mainly rely on the differentiability conditions, which are indeed satisfied for the MNP. Assumption \ref{MNPAssump:Moment} provides more primitive conditions for showing the Glivenko-Cantelli and Donsker properties in Assumption \ref{assump:SE}. 
 In Section S3 of the supplementary material, we provide detailed expressions of $g_j(\cdot,\theta)$ in Assumption \ref{MNPAssump:Inf} in the case of three alternatives ($J+1=3$), which capture the influence from first-stage estimation. Indeed, the required moment or smoothness condition is easily satisfied under mild regularity conditions, such as the compact support of covariates. Nonetheless, the expression itself is lengthy, which again advocates bootstrap for inference. Given the expression of $g_j(\cdot,\theta)$, the term $\Gamma_1(\cdot)$ appearing in the influence function in Assumption \ref{assump:normal} can be written as $\Gamma_1(\cdot)=\sum_{j=1}^J v^{\dagger}_{ij}\mathbb{E}_0[\partial g_j(\cdot,\theta_0)/\partial \theta|Z_{ij}]$.\footnote{This is a special case of the expression (3.15) from \cite{ichimura2010semi}, because the conditional mean regression in the first stage does not depend on $\theta$ from the second stage.} Regarding the prior choice in Assumption \ref{MNPAssump:Prior}, one can also impose the restriction on the eigenvalue rather than the first diagonal element as in \cite{imai2005bayesian}. Examples of other proper normalizations for the unrestricted or restricted covariance can be found in \citet[Chapter 5.2]{train2009discrete}. 
 
Next, we draws readers' attention to several important restrictions of our current exposition. Namely, our asymptotic results are built on the point identification, smooth criterion functions in the second stage, and the \textit{i.i.d.} data structure.
\begin{remark}
	Throughout the paper, we maintain the point identification assumption of all finite-dimensional parameters in the MNP model. This is also assumed in the classical literature on frequentist estimators \citep{mcfadden1989sme,pakes1989simulation}. As noted by \cite{keane1992note}, the identification in the MNP model can be tenuous even if formal identification conditions are satisfied. \cite{keane1992note} showed through simulation results that the sample likelihood function can be flat to the parameters in the covariance matrix of latent error terms. In response, there are several proposals by further restricting the covariance structure via the trace restriction \citep{burgette2012trace}, the independence assumption \citep{johndrow2013diagonal} or the variance reparameterization \citep{munkin2023note}. Another possibility is to explicitly allowing for the robustness to the partial identification in the spirit of \cite{chen2018MC}. It is certainly interesting and challenging to extend our approach to the direction of \cite{chen2018MC} and develop a valid inferential procedure under partial identification. 
\end{remark}
	
	\begin{remark}
		The asymptotic normality established here crucially depends on the smoothness of the log-likelihood function. %If one works with a general non-smooth criterion function, 
		Our results do not cover non-smooth criterion functions such as the indicator function in the maximum score estimation \citep{manski1975mse}, non-standard asymptotics including the cubic-root rate and Chernoff type limiting distribution are expected. In the context of modeling multinomial choices, the maximum score are often used to relax the parametric error assumptions. Such examples can be found in \cite{fox2007multi} and \cite{chen2014maximum}, where the second stage estimation of the discrete choice model only invokes the conditional median restriction, after controling the endogeneity or including generated regressors from the first stage estimation. It is demonstrated by \cite{jun2015laplace} that Bayesian method brings significant computational convenience compared with the brutal-force optimization of the discontinuous function, nonetheless with the aforementioned nonstandared asymptotics. Note that \cite{chen2014maximum} worked with one special case that the first stage nonparametric regression converges faster than the cubic-root rate, so it does not affect the limiting distribution of the second stage. A thorough investigation of the analogous quasi-Bayesian inference is beyond the scope of the current work and will be pursued elsewhere. 
	\end{remark}

\begin{remark}
Throughout the paper, we have worked exclusively with the \textit{i.i.d.} data.	Another interesting direction is macroeconometrics\footnote{In the earlier literature on the empirical macroeconomics, a two-step procedure is examined by \cite{murphy2002two} in which the unobservables, such as the rational expectations of future forcing variables, are imputed by an auxiliary econometric model.} that combines macro and micro level data to address the heterogeneity issue. Recently, \cite{chang2024hetero} develop a state-space model that stacks macroeconomic aggregates and a cross-sectional density. A novel two-step estimation procedure is proposed by \cite{chang2024hetero} in which the cross-sectional density is first estimated by the sieve approach and the state-space model is then fitted by the Bayesian algorithm. We plan to develop the corresponding BvM theorem in the future work that properly accounts for the delicate functional data feature and time series dependence structure in \cite{chang2024hetero}. We anticipate that it is a highly non-trivial task to design the proper bootstrap procedure in this case.
\end{remark}

\section{Monte Carlo Simulation}\label{sec:simulation}
We conduct Monte Carlo simulations for the Petrin--Train Model. Simulation results confirm that the quasi-Bayesian credible set does not have desirable coverage probabilities, and this problem can be solved by using bootstrapping the quasi-posterior mean or $t$-ratio. Our simulation design is a multinomial choice model with three or four choices and one endogenous regressor for each choice. For each alternative, the latent utility level takes the following form:
			\begin{equation}\label{simu_latent_utility}
				U_{ij}=\tilde\beta X^e_{ij}+\varepsilon_{ij},  \quad j=0,1,\cdots,J.
			\end{equation}
The true value of $\tilde\beta$ is $1$. The endogenous regressor $X^e_{ij}$ depends on $\xi_{ij}$ that is excluded from equation (\ref{simu_latent_utility}):
			\begin{equation}\label{simu_MNP_first_stage}
				X^e_{ij}=\tau(\xi_{ij})+v_{ij},  \quad j=0,1,\cdots,J.
			\end{equation}
Observables $\xi_{i0},.., \xi_{iJ} $ and error terms $v_{i0},..., v_{iJ}$ in the first stage (\ref{simu_MNP_first_stage}) are independent standard normal. The error term $\varepsilon_{ij}$ in the latent utility (\ref{simu_latent_utility}) is generated by $\varepsilon_{ij}=\lambda v_{ij}+\epsilon_{ij}$ for $j=0,1,\cdots,J$, where $\epsilon_{ij}$ are jointly normal with means equal to zero, and variances $\sigma_{0}^2=1$, $\sigma_{j}^2= \sigma^2= 0.75$ for $j=1,\cdots J$. In the case of three choices ($J=2$), the correlation coefficient for any pair of $\epsilon_{ij}$ is $corr(\epsilon_{ij},\epsilon_{is})=\rho$ for $j\neq s$. In the case of four choices ($J=3$), $corr(\epsilon_{i0},\epsilon_{i1})=corr(\epsilon_{i2},\epsilon_{i3})=\rho$, and $corr(\epsilon_{ij},\epsilon_{is})=0$ for other $(j,s)$ pairs. We set $\rho=\sigma/2$ and $\lambda=0.6$. The latent utility relative to the null category becomes
\begin{equation}\label{simu_latent_utility_diff}
	U^{\dagger}_{ij}=\tilde\beta X^{\dagger e}_{ij}+\varepsilon^{\dagger}_{ij},  \quad j=1,\cdots,J,
\end{equation}
and the endogenous regressors can be written as
	\begin{equation}\label{simu_MNP_first_stage_diff}
	X^{\dagger e}_{ij}=\tau(\xi_{ij})-\tau(\xi_{i0})+v^{\dagger}_{ij}=\zeta_j(Z_{ij})+v^{\dagger}_{ij},  \quad j=1,\cdots,J,
\end{equation}
with the instrument $Z_{ij}=(\xi_{ij},\xi_{i0})$ and the conditional mean function $\zeta_j(Z_{ij})=\tau(\xi_{ij})-\tau(\xi_{i0})$. Substituting the control function $v_{ij}$ into the (differenced) utility equation (\ref{simu_latent_utility_diff}) leads to
\begin{equation}\label{simu_latent_utility_diff_controlled}
	U^{\dagger}_{ij}=\tilde\beta X^{\dagger e}_{ij}+ \lambda v_{ij}^{\dagger} + \epsilon^{\dagger}_{ij},  \quad j=1,\cdots,J,
\end{equation}
where $\epsilon^{\dagger}_{ij}= \epsilon_{ij}-\epsilon_{i0}$ is independent of $X^{\dagger e}_{ij}$ and $v_{ij}^{\dagger}$.

We consider two designs for the functional form for $\tau(\cdot)$ in (\ref{simu_MNP_first_stage}):  Design I with $\tau(z)= 0.9z + 0.9z^2 + \ln(z+1)^2$ and Design II with $\tau(z)=0.9z+0.9z^2+\exp(0.9z)$.  
%Our pilot estimate $\hat{\tau}$ for $\tau$ explores the specification for endogenous variable at the level in (\ref{simu_MNP_first_stage}), where we use a local constant regression with the bandwidth chosen by leave-one-out cross-validation. This step is implemented by the R package $\mathtt{np}$. Then we take the control functions $\hat \zeta(Z_{ij})=\hat\tau(\xi_{ij})-\hat\tau(\xi_{i0})$. Alternatively, one can carry out the first-stage estimation by estimating an additive model for the conditional mean of the differenced endogenous variable $X^{\dagger e}_{ij}$, conditional on $\xi_{ij}$ and $\xi_{i0}$. Controlling for the estimated $\hat v^{\dagger}_{ij}=X^{\dagger e}_{ij}-\hat \zeta(Z_{ij})$, we draw the posterior of $\tilde\beta$ from the MNP model in the second stage, using Gibbs sampler with data augmentation \citep{imai2005bayesian}. This step uses the R package $\mathtt{MNP}$ \citep{imai2005mnp}. 
Our first stage estimates the model (\ref{simu_MNP_first_stage}) using a kernel regression\footnote{Alternatively, if one proceeds with the differenced endogenous variable in equation (\ref{simu_MNP_first_stage_diff}) directly, one can estimate a nonparametric additive model with regressors $(\xi_{ij},\xi_{i0})$.} with the bandwidth chosen by leave-one-out cross-validation and obtains the residual $\widehat v_{ij}$. This step is implemented by the R package $\mathtt{np}$ \citep{hayfield2008np}. Our second stage estimates the coefficient $\beta=(\tilde\beta,\lambda)^\top$ in the MNP model (\ref{simu_latent_utility_diff_controlled}), after substituting $ v_{ij}^{\dagger}$ by its estimator $\widehat v^{\dagger}_{ij}$ obtained in the first stage. %The second stage can be conducted by applying a Bayesian MNP model. 
The Bayesian second stage draws the quasi-posterior of $\beta$ using a Gibbs sampler with data augmentation \citep{imai2005bayesian}, which is implemented by the R package $\mathtt{MNP}$ \citep{imai2005mnp}.

Table \ref{table:CR} presents the empirical coverage probabilities and the average lengths of the confidence intervals for the scalar parameter $\tilde\beta$ produced by different inference procedures: \textbf{QB} stands for the quasi-Bayesian credible set $\mathcal{C}_n(\alpha)$ discussed in Corollary 2. \textbf{QBB} and \textbf{QBB-$t$} represent the quasi-Bayesian bootstrap percentile and percentile-$t$ intervals given by Algorithm \ref{algorithm_1}, (2.12) and (2.13).
%\textbf{QBB$_1$}:a variant of QBB as described in Algorithm [] in which for each bootstrap sample we obtain one posterior draw instead of the posterior mean. 
For comparison, we also consider a frequentist two-stage method that has the same first stage as our quasi-Bayesian method, but employs simulated maximum likelihood estimation for the second stage (\ref{simu_latent_utility_diff_controlled}), which is implemented by the R package $\mathtt{mlogit}$ \citep{croissant2020estimation}.\footnote{Setting the argument ``$\mathtt{probit = TRUE}$'' when calling the function $\mathtt{mlogit}$ estimates a multinomial probit model.}
The resulting frequentist bootstrap percentile (or percentile-$t$) interval is denoted as \textbf{F2B} (or \textbf{F2B-$t$}). The sample size is $1000$. The number of bootstrap replications is $500$, and the number of simulation replications is $1000$. Table \ref{table:CR} reveals that the quasi-Bayesian credible interval (QB) systematically under-covers the true parameter $\tilde\beta$ in all cases. Bootstrap quasi-Bayesian intervals (QBB and QBB-$t$) significantly improve the coverage performance and yield coverage probabilities close to nominal levels. Since QBB-$t$ is studentized using the posterior standard error that does not account for the first stage uncertainty, it is not expected to have better coverage performance than QBB -- a fact also confirmed by simulation evidence. We also observe that bootstrap confidence intervals QBB and QBB-$t$ are longer than the credible interval QB, as bootstrap resampling incorporates estimation uncertainty from the first stage. 

As Table \ref{table:CR} shows, the frequentist bootstrap confidence intervals F2B and F2B-$t$--especially the former--also have good coverage performance in finite samples. The second stage of the frequentist approach uses a simulated maximum likelihood estimator and is computationally much slower than the Bayesian second stage. For example, to construct a confidence interval in the case of three alternatives ($J+1=3$) and Design I, quasi-Bayesian bootstrap intervals QBB and QBB-$t$ on average take about $130$ seconds per simulation replication, while the frequentist counterparts F2B and F2B-$t$ take more than $4400$ seconds. When it comes to four alternatives ($J+1=4$) and Design I, QBB and QBB-$t$ take about $170$ seconds to compute whereas F2B and F2B-$t$ take about $8700$ seconds. This confirms the computational advantage of using a Bayesian second stage to handle complicated likelihood functions derived from structural models, especially when the bootstrap is used for inference.
All codes are run using R version 4.4.1 (2024-06-14) on a desktop machine (3.8GHz 8-Core Intel Core i7, 40GB Memory) with macOS system (x86 64-apple-darwin20).
When simulating the posterior, the initial $2000$ Gibbs draws are discarded, and the following $2000$ draws are stored. 
%Overall, for the Petrin–Train model considered in our simulation exercise, the quasi-Bayesian approach coupled with bootstrap forms a computationally much more efficient inference procedure than the frequentist counterpart.

\begin{figure} 
\caption{Coverage probability of quasi-Bayesian credible intervals (QB, black, dashed) and quasi-Bayesian bootstrap confidence intervals(QBB, blue, solid),  with the prior variance for $\tilde\beta$  being $100$, $1000$ and $\infty$, and the nominal coverage probability $(1-\alpha)$ being $0.90$ ($\bullet$), $0.95$ ($\blacksquare$), and $0.99$ ($\blacktriangle$); \# of choices $=3$, $n=1000$.}\label{figure: CR_diff_prior}
\vspace{2mm}
\begin{tabular}{cc}
 Design I &  Design II\\
\includegraphics[scale=0.55]{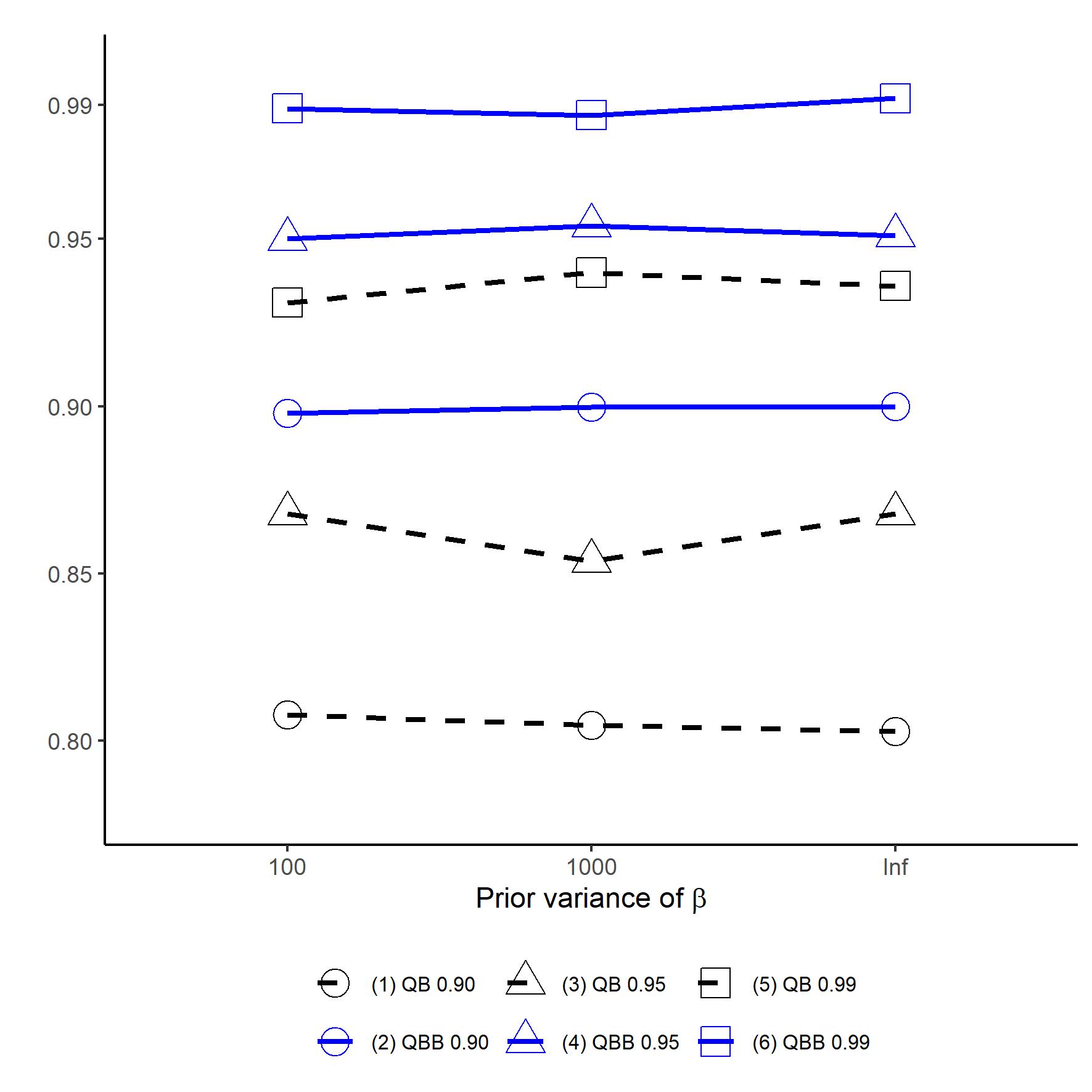}&
\includegraphics[scale=0.55]{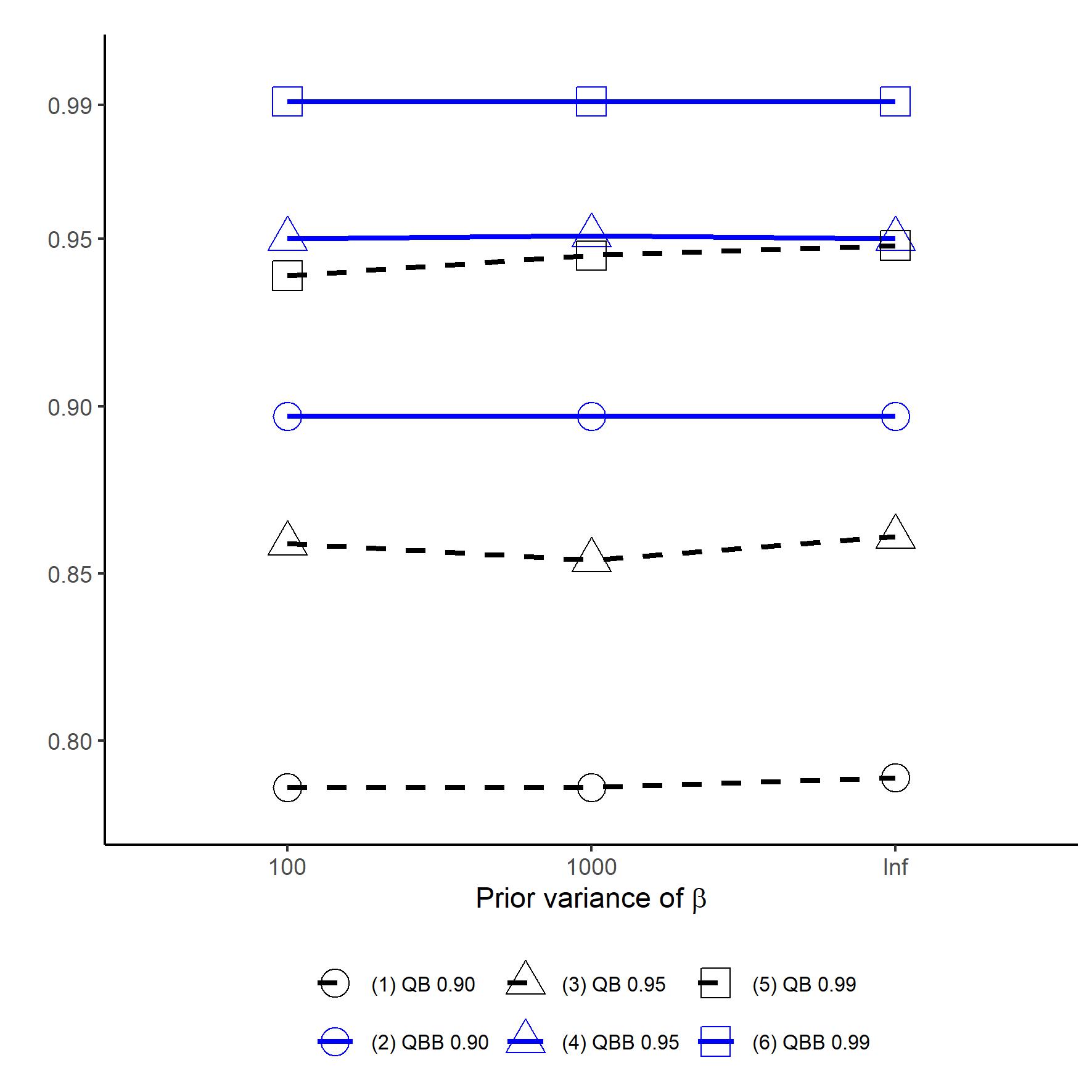}\\
\end{tabular}
\end{figure}

The quasi-Bayesian method in Table \ref{table:CR} imposes a non-informative prior distribution on the parameter $\tilde\beta$. Figure \ref{figure: CR_diff_prior} illustrates how the coverage performance is affected by the variance of the prior distribution. We consider the prior variance for $\tilde\beta$ equal to $100$, $1,000$, and $+\infty$ ($+\infty$ corresponds to the non-informative prior used in Table \ref{table:CR}). The prior mean for $\tilde\beta$ is set as zero in all cases. Figure \ref{figure: CR_diff_prior} shows that the under-coverage of QB and the improvement achieved by QBB are robust to the variance of the prior.

%In sum, our simulation findings suggest that the under-coverage of the quasi-Bayesian credible interval is ubiquitous for different first-stage relationships, number of choices, and prior variance values. On the other hand, our recommended bootstrap procedure successfully restores the coverage probabilities to the nominal levels. 
\begin{table}
\caption{Finite sample performance of different inference methods (confidence intervals) for the coefficient $\tilde\beta$, $J+1$= number of choices, $n=1000$, nominal coverage $1-\alpha=0.90$, $0.95$, and $0.99$.}\label{table:CR}
\begin{tabular}{cclcccccccccccccc}\toprule
\multicolumn{1}{c}{\multirow{1}{*}{}}&\multicolumn{1}{c}{\multirow{1}{*}{}}&\multicolumn{1}{c}{\multirow{1}{*}{}}&\multicolumn{1}{c}{}& \multicolumn{3}{c}{Coverage probability}&\multicolumn{1}{c}{}&\multicolumn{3}{c}{Average length}\\
\cline{5-7}\cline{9-11}
\multicolumn{1}{c}{\multirow{1}{*}{$J+1$}}&\multicolumn{1}{c}{\multirow{1}{*}{Design}}&\multicolumn{1}{c}{Methods}&\multicolumn{1}{c}{\multirow{1}{*}{}}& \multicolumn{1}{c}{$0.90$}&\multicolumn{1}{c}{$0.95$}&\multicolumn{1}{c}{$0.99$}&\multicolumn{1}{c}{}&\multicolumn{1}{c}{$0.90$}&\multicolumn{1}{c}{$0.95$}&\multicolumn{1}{c}{$0.99$}\\
\toprule
3 & I & QB && 0.803 & 0.868 & 0.936 && 0.325 & 0.389 & 0.499\\
& & QBB && 0.900 & 0.951 & 0.992 && 0.414 & 0.511 & 0.738\\
%& & QBB-hb && 0.862 & 0.901 & 0.958 && 0.414 & 0.511 & 0.738\\
& & QBB-$t$ && 0.872 & 0.930 & 0.983 && 0.429 & 0.522 & 0.729\\
%& & QBB$_1$ && 0.961 & 0.984 & 0.997 && 0.527 & 0.657 & 0.953\\
& & F2B && 0.915 & 0.961 & 0.986 && 0.361 & 0.438 & 0.592\\
& & F2B-$t$ &&  0.870 & 0.923 & 0.979 && 0.333 & 0.396 & 0.514\\
\cline{2-11}
 & II & QB && 0.789 & 0.861 & 0.948 && 0.303 & 0.362 & 0.467\\
& & QBB && 0.897 & 0.950 & 0.991 && 0.391 & 0.480 & 0.676\\
%& & QBB-hb && 0.876 & 0.933 & 0.978 && 0.391 & 0.480 & 0.676 \\
& & QBB-$t$ && 0.864 & 0.934 & 0.986 && 0.389 & 0.470 & 0.639\\
%& & QBB$_1$ && 0.967 & 0.988 & 0.998 && 0.496 & 0.617 & 0.888\\
& & F2B && 0.904 & 0.950 & 0.987 && 0.335 & 0.404 & 0.541\\
& & F2B-$t$ && 0.860 & 0.922 & 0.979 && 0.304 & 0.361 & 0.467 \\
\midrule
4 & I & QB && 0.802 & 0.865 & 0.926 && 0.381 & 0.461 & 0.573\\
& & QBB &&  0.915 & 0.958 & 0.992 && 0.482 & 0.602 & 0.886\\
%& & QBB-hb && 0.864 & 0.911 & 0.982 && 0.482 & 0.602 & 0.886\\
& & QBB-$t$ && 0.901 & 0.952 & 0.993 && 0.546 & 0.673 & 0.963 \\
%& & QBB$_1$ && 0.960 & 0.986 & 0.998 && 0.615 & 0.792 & 1.196 \\
& & F2B && 0.925 & 0.963 & 0.992 && 0.484 & 0.597 & 0.847\\
& & F2B-$t$ &&  0.833 & 0.899 & 0.958 &&
 0.430 & 0.514 & 0.673\\
\cline{2-11}
 & II & QB && 0.833 & 0.889 & 0.952 && 0.371 & 0.442 & 0.545\\
& & QBB && 0.939 & 0.971 & 0.997 && 0.483 & 0.606 & 0.896\\
%& & QBB-hb && 0.904 & 0.957 & 0.990 && 0.483 & 0.606 & 0.896\\
& & QBB-$t$ && 0.924 & 0.966 & 0.996 && 0.519 & 0.639 & 0.916\\
%& & QBB$_1$ && 0.974 & 0.993 & 0.999 && 0.617 & 0.808 & 1.217\\
& & F2B && 0.922 & 0.959 & 0.990 && 0.444 & 0.541 & 0.750\\
& & F2B-$t$ && 0.851 & 0.913 & 0.966 && 0.390 & 0.464 & 0.604 \\
\bottomrule
\end{tabular}
\end{table}

\section{Empirical Application}
We apply the quasi-Bayesian approach to investigate U.S. firms' incorporation decisions, initially studied by \cite{eldar2020regulatory}. This research concerns about how the corporate law at the state level affects firms' choice of locations to incorporate their business, in particular, whether firms prefer to incorporate in states where the corporate laws are more friendly to share-holders or more protective of managers. In the U.S, two states dominate the incorporation market: in year 2013, $63.9\%$ of firms were incorporated in Delaware, and $8.5\%$ in Nevada. Interestingly, these two dominating states have quite different legal characteristics: Delaware's laws are more market-oriented and facilitate takeover activities, while Nevada's laws provide more protection for managers' interests. In the dataset, each state's legal characteristics are measured by three indices: \textit{ATS} counts the number of anti-takeover statutes, resulting in a score from $0$ to $5$; \textit{DIR} (or \textit{OFF}) measures the extent to which state laws protect directors (or officers) from liability by providing them exemption and indemnification. The \textit{DIR} index takes a value from $0.5$ to $6$ and the \textit{OFF} index ranges from $0.5$ to $5$. We use the data in year 2013, which contains 2,922 firms. Let $i$ denote the firm and $j$ denote the incorporation location, we consider an MNP model with three alternatives: Delaware ($j=1$), Nevada ($j=2$), and other states ($j=0$). A firm's latent utility is expressed as:
\begin{eqnarray}\label{empirical:latent utility}
U_{ij} &=& \tilde\beta_1 Law_j  + \tilde\beta_2(Law_j\times Inst.~Owner_i) + \tilde\beta_3(Law_j\times Small_i) + \varepsilon_{ij},
\end{eqnarray}
where we consider three specifications for the regressor $Law_j$ for the location alternative $j$: $Law_j=ATS_j$, $DIR_j,$ or $OFF_j$, respectively. Other regressors are firm-level characteristics and are the same for all specifications: $Inst.~Owner_i$ represents the share of institutional ownership of firm $i$, and $Small_i$ is a dummy variable for small-sized firms. 
The concern about endogeneity arises from the variable $Inst.Owner$. Following \cite{eldar2020regulatory}, we use a dummy instrumental variable $S\&P500$ which indicates whether the firm is included in the S\&P 500 index. Inclusion in the S\&P 500 index is positively correlated with a firm's institutional ownership. On the other hand, the S\&P 500 is mainly decided by the market's view about the firm's representativeness and thus does not depend on the firm's decision.  The first-stage regression model is
\begin{eqnarray}\label{empirical:1st stage}
Inst.~Owner_i &=& \tau( S\& P 500, Small_i)+ \nu_i, 
\end{eqnarray}
where $\tau$ is an unknown function. The control function takes the form $\widehat{v}_{ij}=ATS_j\times \hat{\nu}_i$. Our first stage estimates the function $\tau(\cdot)$ in (\ref{empirical:1st stage}) by a local-constant kernel regression where the bandwidths are selected by the leave-one-out cross-validation. Including $\widehat{v}_{ij}$ as an additional regressor, the Bayesian second stage draws the quasi-posteriors of $\tilde\beta_1$, $\tilde\beta_2$ and $\tilde\beta_3$, upon which the point estimates and 
$90\%$ bootstrap confidence intervals QBB and QBB-$t$ are constructed in the same way as the simulation exercise in Section \ref{sec:simulation}. Results are presented in Table \ref{table:incorporation} with the percentile and percentile-$t$ intervals reported in brackets and parentheses, respectively. For comparison, Table \ref{table:incorporation} also includes the frequentist two-stage estimator and its bootstrap intervals F2B and F2B-$t$.

The negative estimates of $\tilde\beta_1$ in Table \ref{table:incorporation} suggest that firms on average prefer less legal restrictions on the takeover and less protections on directors/officers, which explains why a majority of firms choose to incorporate in Delaware which has market oriented laws friendly to shareholders.
On the other hand, a positive estimate of $\tilde\beta_3$ suggests that a firm's preference about corporate laws varies with respect to its size. Small firms are more likely to welcome the protectionist laws that restrict takeover activities and protects directors/offices from liability. This explains why Nevada, with $ATS=5$, occupies a notable market share for incorporation. Our results align with the findings of \cite{eldar2020regulatory}, who estimated a multinomial Logit model. In Table \ref{table:incorporation}, the quasi-Bayesian and frequentist two-stage approaches generate very similar point estimates and confidence intervals, but the former is about 50 times faster. For one specification of the law variable,  QBB and QBB-$t$ cost about 20 minutes while F2B and F2B-$t$ cost 16 hours.

\begin{table}
\begin{threeparttable}
\caption{Quasi-Bayesian and frequentist estimates for incorporation decision,  $90\%$ bootstrap percentile interval QBB in brackets and percentile-$t$ interval QBB-$t$ in parentheses, $n=2,922$. Three specifications: Spec I with the $Law$ variable = $ATS$, Spec II with the $Law$ variable = $DIR$, and Spec III with the $Law$ variable = $OFF$.}
\label{table:incorporation}
\begin{tabular}{llccccc}\toprule
&&& Quasi-Bayesian && Frequentist &  \\
\toprule
\multicolumn{1}{l}{Spec I}\\
\cline{1-1}
&$ATS$ &&  -0.437  &  & -0.419 &\\
&&& [-0.789, -0.012] &  & [-0.792, -0.026] \\
&&& (-0.845, -0.082) &  & (-0.763, -0.061) \\
 &$ATS\times Inst.~Owner.$ && 0.274  & & 0.250  \\
&&& [-0.383, 0.827] &  & [-0.365, 0.816] \\
&&&  (-0.279, 0.905) & & (-0.309, 0.788)\\
&$ATS\times Small$ && 0.344 && 0.329 \\
&&& [0.007, 0.624] & & [0.017, 0.619]  \\
&&&  (0.067, 0.674) &  & (0.055, 0.601)\\
\midrule
\multicolumn{1}{l}{Spec II}\\
\cline{1-1}
&$DIR$ && -0.782 &  & -0.821\\
&&& [-1.492, -0.078] & & [-1.512, -0.091]\\
&&&  (-1.464, -0.073) & & (-1.452, -0.169)\\
 &$DIR\times Inst.~Owner.$ && 0.405 &&  0.471\\
&&& [-0.742, 1.518] && [-0.683, 1.524]\\
&&& (-0.692, 1.467) && (-0.570, 1.450)\\
&$DIR\times Small$ && 0.569 && 0.602 \\
&&& [0.016, 1.123] && [0.032, 1.149]\\
&&& (0.008, 1.107) && (0.088, 1.100)\\
\midrule
\multicolumn{1}{l}{Spec III}\\
\cline{1-1}
&$OFF$ && -0.770 && -0.785\\
&&& [-1.433, -0.052] && [-1.438, -0.093]\\
&&& (-1.407, -0.133) && (-1.379, -0.170)\\
 &$OFF\times Inst.~Owner.$ && 0.421 && 0.444\\
&&& [-0.682, 1.446] && [-0.645, 1.437]\\
&&& (-0.567, 1.412)  && (-0.533, 1.363)\\
&$OFF\times Small$ && 0.551 && 0.565\\
&&& [-0.009, 1.075] && [0.025, 1.081]\\
&&& (0.054, 1.046)  && (0.079, 1.031)\\
\bottomrule
\end{tabular}
\end{threeparttable}
\end{table}

\section{Conclusion}        
     Our research highlights two key aspects of modern econometric models. First, the complexity of these models leads to analytically intractable likelihoods, which can be more conveniently dealt with by the Bayesian method. Second, the control function approach effectively handles the endogeneity problem by including the first-stage residuals as additional covariates, rectifying the second stage's inconsistency. Our study extensively investigates this novel quasi-Bayesian method. Given the widespread use of the Bayesian approach in contemporary econometrics and machine learning, our method provides practitioners with enhanced flexibility to integrate cutting-edge algorithms from various estimation stages. This methodology can be beneficial for other structural models when the underlying asymptotic theory is well understood. %Besides the Petrin-Train model, one can also apply our method to the consumer search model developed by \cite{chen2017sequential}, where one can control the endogeneity via the control function approach and then adopt the Bayesian approach to handle the complicated likelihood due to the sequential search. 

\section{Appendix A: Proofs of Main Results}
This appendix collects proofs for Theorems \ref{thm:BvM}, \ref{thm:bootstrap}, Corollary \ref{cor:coverage}, \ref{cor:moments}, and Proposition \ref{Prop:Example}.
Our proof of (\ref{BvM_result}) in Theorems \ref{thm:BvM} is patterned in line with generic arguments in the proof of Theorem 1.4.2 from \cite{ghosh2002bayesian} or Theorem 1 from \cite{chernozhukov2003mcmc}. The crux is to address the first-stage nonparametric estimator $\widehat{\zeta}_n$. This is reflected in a different choice of splitting ranges of the integration and a more delicate expansion of the (quasi-)log-likelihood ratio in our proof.

\begin{proof}[Proof of Theorem \ref{thm:BvM}]
 Consider the quasi-posterior distribution of $h\equiv \sqrt{n}(\theta-\widehat{\theta}_n)$:
\begin{align*}
	\tilde{\pi}(h|\bm{Y}_2^n;\widehat{\zeta}_n)=&\frac{\prod_{i=1}^np_{\widehat{\theta}_n+\frac{h}{\sqrt{n}}}(Y_{2i}|\widehat{\zeta}_n)\pi(\widehat{\theta}_n+\frac{h}{\sqrt{n}})}{\int \prod_{i=1}^np_{\widehat{\theta}_n+\frac{h}{\sqrt{n}}}(Y_{2i}|\widehat{\zeta}_n)\pi(\widehat{\theta}_n+\frac{h}{\sqrt{n}})dh}\\
	=&\frac{\prod_{i=1}^n\left[p_{\widehat{\theta}_n+\frac{h}{\sqrt{n}}}(Y_{2i}|\widehat{\zeta}_n)/p_{\widehat{\theta}_n}(Y_{2i}|\widehat{\zeta}_n)\right]\pi(\widehat{\theta}_n+\frac{h}{\sqrt{n}})}{\int \prod_{i=1}^n\left[p_{\widehat{\theta}_n+\frac{h}{\sqrt{n}}}(Y_{2i}|\widehat{\zeta}_n)/p_{\widehat{\theta}_n}(Y_{2i}|\widehat{\zeta}_n)\right]\pi(\widehat{\theta}_n+\frac{h}{\sqrt{n}})dh}.
\end{align*}
Denote its denominator as 
\begin{equation}
		D_n\equiv\int \prod_{i=1}^n\left[\frac{p_{\widehat{\theta}_n+\frac{h}{\sqrt{n}}}(Y_{2i}|\widehat{\zeta}_n)}{p_{\widehat{\theta}_n}(Y_{2i}|\widehat{\zeta}_n)}\right]\pi(\widehat{\theta}_n+\frac{h}{\sqrt{n}})dh.
\end{equation}
Recall that the limiting normal probability density function is as follows:
\begin{equation*}
	\pi_{\infty}(h)\equiv (2\pi)^{-p/2}|\det V_0|^{1/2}\exp\left\{-\frac{h^\top V_0h}{2}\right\}.
\end{equation*}
Note that the total variation norm contains the weighting function $(1+ |h|^a)$. Nonetheless, it suffices to bound the following
\begin{align}
&\int |h|^a \left|\tilde{\pi}(h|\bm{Y}_2^n;\widehat{\zeta}_n)-\pi_{\infty}(h)\right|dh \notag \\
\leq&D_n^{-1}\underbrace{\int |h|^a \left|\prod_{i=1}^n \left[\frac{p_{\widehat{\theta}_n+\frac{h}{\sqrt{n}}}(Y_{2i}|\widehat{\zeta}_n)}{p_{\widehat{\theta}_n}(Y_{2i}|\widehat{\zeta}_n)}\right]\pi(\widehat{\theta}_n+\frac{h}{\sqrt{n}})-e^{-\frac{h^\top V_0h}{2}}\pi(\theta_0)\right|dh}_{N_{n1}}\notag \\
+&D_n^{-1}\underbrace{\int |h|^a e^{-\frac{h^\top V_0h}{2}}\left(\pi(\theta_0)-D_n(2\pi)^{-p/2}\left|\det V_0 \right|^{1/2}\right)dh}_{N_{n2}},\label{tv_ub}
\end{align} 
because we can take $a=0$ if the above term converges to zero in probability.

It is sufficient to show that
\begin{equation}\label{Numerator}
	N_{n1}\equiv\int |h|^a \left|\prod_{i=1}^n \left[\frac{p_{\widehat{\theta}_n+\frac{h}{\sqrt{n}}}(Y_{2i}|\widehat{\zeta}_n)}{p_{\widehat{\theta}_n}(Y_{2i}|\widehat{\zeta}_n)}\right]\pi(\widehat{\theta}_n+\frac{h}{\sqrt{n}})-e^{-\frac{h^\top V_0h}{2}}\pi(\theta_0)\right|dh=o_{P_0}(1).
\end{equation}
Because if (\ref{Numerator}) holds, taking $a=0$ yields
\begin{equation*}
	\int \left|\prod_{i=1}^n \left[\frac{p_{\widehat{\theta}_n+\frac{h}{\sqrt{n}}}(Y_{2i}|\widehat{\zeta}_n)}{p_{\widehat{\theta}_n}(Y_{2i}|\widehat{\zeta}_n)}\right]\pi(\widehat{\theta}_n+\frac{h}{\sqrt{n}})-e^{-\frac{h^\top V_0h}{2}}\pi(\theta_0)\right|dh=o_{P_0}(1),
\end{equation*}
which leads to
\begin{equation}\label{C_n_converge}
	D_n=(2\pi)^{p/2}\left|\det V_0 \right|^{-1/2}\pi(\theta_0)+o_{P_0}(1).
\end{equation}
Also note that $D_n=O_{P_0}(1)$. The convergence in (\ref{C_n_converge}) then implies
\begin{equation*}
	N_{n2}=\left|D_n(2\pi)^{-p/2}|\det V_0|^{1/2}-\pi(\theta_0)\right|\int_{\mathbb{R}^p}|h|^{a}\exp\left\{-\frac{h^\top V_0h}{2}\right\}dh=o_{P_0}(1),
\end{equation*}
which makes the upper bound in (\ref{tv_ub}) for the total variation norm satisfy $(N_{n1}+N_{n2})/D_n=o_{P_0}(1)$. This completes the proof of (\ref{BvM_result}).

To show (\ref{Numerator}), we split its integration range into three mutually exclusive areas:
\begin{itemize}
	\item $H_{1,n}\equiv \{h:|h|\leq C\sqrt{n}r_n\}$;
	\item  $H_{2,n}\equiv \{h:C r_n\sqrt{n}<|h|\leq \delta\sqrt{n}\}$;
	\item  $H_{3,n}\equiv \{h:|h|>\delta\sqrt{n}\}$;
\end{itemize}
for a large positive constant $C$ and a small $\delta>0$. Here we need to choose $\delta$ to be sufficiently small so that the requirement of Assumption \ref{assump:SE} (iii) is satisfied. Note that
\begin{equation*}
	\int_{H_{2,n}\cup H_{3,n}} \exp\left(-\frac{h^\top V_0h}{2}\right)\pi(\theta_0)dh=o(1),
\end{equation*}
given the fact that the smallest eigenvalue of $V_0$ is positive and $r_n\sqrt{n}\to\infty$ from Assumptions \ref{assump:id} and \ref{assump:FirstStep}. 
We further define 
\begin{equation}\label{Delta_nr}
	\Xi_n^r\equiv \sup_{d_{\Theta}(\theta,\theta_0)\geq r}\left\{\frac{1}{n}\sum_{i=1}^n[\log p_{\theta}(Y_{2i};\widehat{\zeta}_n)-\log p_{\widehat{\theta}_n}(Y_{2i};\widehat{\zeta}_n)]\right\}.
\end{equation}

First consider the integration over the outer range $H_{3,n}$. By Lemma S1, for any small $\delta>0$ there exits a positive constant $c>0$, s.t. %\textcolor{magenta}{[Lemma \ref{lemma:outer} is stated for a proper choice $c$, which corresponds to $\epsilon$. So here can we have it for any $\epsilon>0$?]} \textcolor{red}{We should apply the lemma for any $\delta>0$, there exits such $c>0$}, we have
\begin{equation}
	\int_{H_{3,n}} |h|^a \prod_{i=1}^n \frac{p_{\widehat{\theta}_n+\frac{h}{\sqrt{n}}}(Y_{2i}|\widehat{\zeta}_n)}{p_{\widehat{\theta}_n}(Y_{2i}|\widehat{\zeta}_n)}\pi(h)dh\leq \sqrt{n}^{a+1}e^{-nc}\int_{\theta\in\Theta}|\theta|^{a}\pi(\theta)d\theta,~w.p.a.1. 
\end{equation}
Therein, we make use of Assumptions \ref{assump:id}, \ref{assump:FirstStep}, \ref{assump:smooth} and \ref{assump:SE} (i). The right hand side of the above inequality is of $o_{P_0}(1)$. 
 
Then consider the integration over the middle range $H_{2,n}$. By Assumptions \ref{assump:FirstStep}, \ref{assump:Freq}, \ref{assump:regular} and \ref{assump:SE}(iii), we apply Lemma S2 to get
\begin{equation*}
	\int_{H_{2,n}} |h|^a \prod_{i=1}^n \frac{p_{\widehat{\theta}_n+\frac{h}{\sqrt{n}}}(Y_{2i}|\widehat{\zeta}_n)}{p_{\widehat{\theta}_n}(Y_{2i}|\widehat{\zeta}_n)}\pi(h)dh\leq \sqrt{n}^{a+1}e^{-cnr_n^2}\int_{\theta\in\Theta}|\theta|^{a}\pi(\theta)d\theta, 
\end{equation*}
for a positive $c$ and a large enough constant $C$, $w.p.a.1$. The right hand side of the above inequality is also of $o_{P_0}(1)$.

Lastly, we consider the inner range $H_{1,n}$. We first switch from $\pi(\theta_0)$ to $\pi(\widehat{\theta}_n+h/\sqrt{n})$ by noting that $\sup_{h\in H_{1,n}}\limsup\left|\widehat{\theta}_n+\frac{h}{\sqrt{n}}-\theta_0\right|=o_{P_0}(1)$, which implies
\begin{equation*}
	\int_{H_{1,n}} |h|^a \exp\left(-\frac{h^\top V_0h}{2}\right)\left|\pi(\widehat{\theta}_n+\frac{h}{\sqrt{n}})-\pi(\theta_0)\right|dh=o_{P_0}(1),
\end{equation*}
by the continuity of the prior density $\pi(\cdot)$ in Assumption \ref{assump:Prior} and the dominated convergence theorem.
The remaining analysis is about
\begin{align*}
	&\int_{H_{1,n}} |h|^a \left|\prod_{i=1}^n \left[\frac{p_{\widehat{\theta}_n+\frac{h}{\sqrt{n}}}(Y_{2i}|\widehat{\zeta}_n)}{p_{\widehat{\theta}_n}(Y_{2i}|\widehat{\zeta}_n)}\right]-e^{-\frac{h^\top V_0h}{2}}\right|\pi(\widehat{\theta}_n+\frac{h}{\sqrt{n}})dh\\
	&=\int_{H_{1,n}}|h|^a e^{-\frac{1}{2}h^\top V_0h}\left|e^{\sum_{i=1}^n \log p_{\widehat{\theta}_n+\frac{h}{\sqrt{n}}}(Y_{2i}|\widehat{\zeta}_n)-\log p_{\widehat{\theta}_n}(Y_{2i}|\widehat{\zeta}_n)  +\frac{1}{2}h^\top V_0h}-1\right|\pi(\widehat{\theta}_n+\frac{h}{\sqrt{n}})dh\\
	&\leq A_n\times B_n,
\end{align*}
where
\begin{align*}
	A_n&\equiv\int_{\mathcal{H}_{1,n}}|h|^a\exp\left\{-\frac{h^\top V_0h}{4} \right\}\pi(\widehat{\theta}_n+\frac{h} {\sqrt{n}})dh,\\
	B_n&\equiv\sup_{h\in H_{1,n}}\left[e^{-\frac{1}{4}h^\top V_0h }\left|\exp\left\{\sum_{i=1}^n \log p_{\widehat{\theta}_n+\frac{h}{\sqrt{n}}}(Y_{2i}|\widehat{\zeta}_n)-\log p_{\widehat{\theta}_n}(Y_{2i}|\widehat{\zeta}_n)  +\frac{1}{2}h^\top V_0h\right\}-1\right|\right].
\end{align*}
It is clear that $A_n$ is stochastically bounded due to the boundedness of the prior density. Given Assumptions \ref{assump:id}, \ref{assump:smooth}, \ref{assump:SE} and \ref{assump:normal}, Lemma S4 shows that $B_n=o_{P_0}(1)$. Then we obtain $A_n\times B_n=o_{P_0}(1)$, which completes the proof of (\ref{Numerator}). The convergence in the norm $TVM(a)$ stated in (\ref{BvM_result}) implies the convergence of the distribution in the total variation as stated in (\ref{ConvTV}); see \cite{ghosh2002bayesian}. 
\end{proof}

Next, we prove the coverage property of the resulting quasi-Bayesian credible set, given the results in (\ref{BvM_result}) and (\ref{ConvTV}). We denote the $p$-dimensional standard normal measure of a generic set $A$ by 
\begin{equation*}
\mathbb{N}(A)\equiv (2\pi)^{-p/2}\int_{A}e^{-h^\top h/2}dh. 
\end{equation*}
Recall the definition of $\Delta_{n,0}$ in Assumption \ref{assump:normal}. 
\begin{proof}[Proof of Corollary \ref{cor:coverage}]
Consider the quasi-Bayesian credible set $\mathcal{C}_{n}(\alpha)$, which satisfies $\Pi(\theta\in \mathcal{C}_{n}(\alpha)|\bY_2^n,\widehat{\zeta}_n)=1-\alpha$. Applying (\ref{BvM_result}) in Theorem \ref{thm:BvM} with $a=0$ and using the relationship (2.1) in Appendix B of the supplementary material, we have 
\begin{equation*}
\mathbb{N}(V_0^{1/2}n^{1/2}(\mathcal{C}_{n}(\alpha)-\theta_0-n^{-1/2}V_0^{-1}\Delta_{n,0}))\to_{P_0} 1-\alpha.
\end{equation*}
Thus, $\mathcal{C}_{n}(\alpha)=\theta_0+n^{-1/2}V_0^{-1}\Delta_{n,0}+n^{-1/2}V_0^{-1/2}B_n$, where $B_n$ is a set that satisfies $\mathbb{N}(B_n)\to_{P_0} 1-\alpha$, as $n\to\infty$. Therefore, the frequentist coverage of the Bayesian credible set is 
\begin{align*}
\lim_{n\to\infty}	P_0\{\theta_0\in \mathcal{C}_{n}(\alpha)\}&=\lim_{n\to\infty}	P_0\{\theta_0\in  \theta_0+n^{-1/2}V_0^{-1}\Delta_{n,0}+n^{-1/2}V_0^{-1/2}B_n\}\\
&=\lim_{n\to\infty}	P_0\{-V_0^{-1/2}\Delta_{n,0}\in B_n\}.
\end{align*}
The coverage probability of the set $\mathcal{C}_{n}(\alpha)$ does not converge to $1-\alpha$ in general, unless the matrix $V_0$ is equal to $\Omega_0$. 
\end{proof}

\begin{proof}[Proof of Corollary \ref{cor:moments}]
	The convergence in the total variation of moments norm in (\ref{BvM_result}) with $a=1$ implies that
	\begin{equation*}
	\int h\left(\tilde{\pi}(h|\bm{Y}_2^n;\widehat{\zeta}_n)-\pi_{\infty}(h) \right)dh\to_{P_0}0.
\end{equation*}
As the limiting normal distribution is centered around zero, i.e., $\int h \pi_{\infty}(h) dh=0$, we have
\begin{equation*}
	\int h \tilde{\pi}(h|\bm{Y}_2^n;\widehat{\zeta}_n)dh\to_{P_0}0.
\end{equation*}
Translating the above result to the quasi-posterior mean, we note that
\begin{align*}
	\widetilde{\theta}_n\equiv\int \theta d\Pi_n(\theta|\bm{Y_2};\widehat{\zeta}_n)
	=\int\left(\widehat{\theta}_n+\frac{h}{\sqrt{n}}\right)\tilde{\pi}(h|\bm{Y}_2^n;\widehat{\zeta}_n)dh
	=\widehat{\theta}_n+o_{P_0}(n^{-1/2}),
\end{align*}
which leads to the desired result. The proof of the posterior variance follows along similar lines involving the second-order moment. It is straightforward and hence omitted.
\end{proof}

\begin{proof}[Proof of Theorem \ref{thm:bootstrap}]
Lemma S5 in the supplementary material proves the convergence in total variation norm for the bootstrap posterior density function of $h\equiv \sqrt{n}(\theta-\widehat{\theta}_n)$. This implies the asymptotic equivalence of the mean of the bootstrap quasi-posterior $\widetilde{\theta}^{*}_n$ and the bootstrap frequentist-type two-stage estimator $\widehat{\theta}_n^*$, i.e., $\sqrt{n}\left(\widetilde{\theta}_n^{*}-\widehat{\theta}^*_n\right)=o_{P^*}(1)$, following arguments similar to the proof of Corollary \ref{cor:moments}. For the frequentist two-stage estimator $\widehat{\theta}_n$, we have:
\begin{equation}\label{FreqCov}
	\sup_{\xi\in \mathbb{R}^p}|	P_0(\sqrt{n}(V^{-1}_0\Omega_{0}V^{-1}_0)^{-1/2}(\widehat{\theta}_n-\theta_0)\leq \xi)-\Phi_p(\xi)|\to_{P_0} 0,	
\end{equation}
and its bootstrap analog has:
\begin{equation}\label{BootCov}
	\sup_{\xi\in \mathbb{R}^p}|	P^*(\sqrt{n}(V^{-1}_0\Omega_{0}V^{-1}_0)^{-1/2}(\widehat{\theta}^*_n-\widehat{\theta}_n)\leq \xi)-\Phi_p(\xi)|\to_{P^*} 0.
\end{equation}
We denote the cumulative distribution of $\mathbb{N}(0,\omega_j)$ as $H_j(\cdot)$, where $\omega_j\equiv(V^{-1}_0\Omega_{0}V^{-1}_0)_{jj}$. Meanwhile, we let $H^{-1}_{j}(\alpha)$ denote the $\alpha$-th quantile of this disbribution function.

The conditional weak convergence in (\ref{BootCov}) implies the corresponding convergence of the bootstrap quantiles as
\begin{equation*}
	\sqrt{n}(Q^{\ast}_{n,j}(\alpha)-\widehat{\theta}_{n,j})\to_{P_0}H_{j}^{-1}(\alpha).
\end{equation*}
Therefore, the percentile interval $\mathcal{C}^{PC}_{n,j}(\alpha)$ satisfies
\begin{align*}
	P_0\{\theta_{0j}\in \mathcal{C}^{PC}_{n,j}(\alpha) \}&=\Pr\{Q^{\ast}_{n,j}(\alpha/2)\leq \theta_{0j}\leq Q^{\ast}_{n,j}(1-\alpha/2) \}\\
	&=	P_0\{-\sqrt{n}(Q^{\ast}_{n,j}(\alpha/2)-\widehat{\theta}_{n,j})\geq \sqrt{n}(\widehat{\theta}_{n,j}-\theta_{0j})\geq -\sqrt{n}(Q^{\ast}_{n,j}(1-\alpha/2)-\widehat{\theta}_{n,j}) \}\\
	&\to H_j(-H^{-1}_j(\alpha/2))-H_j(-H^{-1}_j(1-\alpha/2))\\
	&=H_j(H^{-1}_j(1-\alpha/2))-H_j(H^{-1}_j(\alpha/2))=1-\alpha,
\end{align*}
where the last equality uses the symmetry of the normal distribution $\mathbb{N}(0,\omega_j)$. 

For the percentile-$t$ interval $\mathcal{C}^{PT}_{n,j}(\alpha)$, note that $\widetilde s_{n,j}=\sqrt{(V_0^{-1})_{jj}}+o_{P_0}(1)$ and its bootstrap counterpart $\widetilde s^*_{n,j}=\sqrt{(V_0^{-1})_{jj}}+o_{P^*}(1)$. As a result, the bootstrap $t$-statistic
\begin{equation*}
	T^*_{n,j}=\frac{\sqrt{n}(\widetilde{\theta}_{n,j}^*-\widehat{\theta}_{n,j})}{\widetilde s^*_{n,j}}
\end{equation*}
converges in distribution to $\mathbb{N}(0,\omega_j/\nu_j)$ conditional on $\bm{Y}^{n}$, where $\nu_j=(V_0^{-1})_{jj}$. This is also the same limiting distribution of $T_{n,j}=\frac{\sqrt{n}(\widetilde{\theta}_{n,j}-\theta_{0j})}{\widetilde s_{n,j}}$. 
We denote the cumulative distribution of $\mathbb{N}(0,\omega_j/\nu_j)$ and its quantile function by $\tilde{H}_j(\cdot)$ and $\tilde{H}^{-1}_j(\cdot)$, respectively. Therefore, $\mathcal{C}^{PT}_{n,j}(\alpha) $ satisfies
\begin{align*}
	P_0\{\theta_{0j}\in \mathcal{C}^{PT}_{n,j}(\alpha) \}&=P_0\{\widetilde{\theta}_{n,j}-\widetilde s_{n,j}q^{\ast}_{n,j}(1-\alpha/2)\leq \theta_{0j}\leq \widetilde{\theta}_{n,j}+ \widetilde s_{n,j}q^{\ast}_{n,j}(\alpha/2) \}\\
	&=	P_0\{q^{\ast}_{n,j}(\alpha/2)\leq T_{n,j}\leq q^{\ast}_{n,j}(1-\alpha/2) \}\\
	&\to \tilde{H}_j(\tilde{H}^{-1}_j(1-\alpha/2))-\tilde{H}_j(\tilde{H}^{-1}_j(\alpha/2))=1-\alpha.
\end{align*}
This completes the proof.
\end{proof}

\begin{proof}[Proof of Proposition \ref{Prop:Example}]
We verify the high-level Assumptions \ref{assump:id} to \ref{assump:normal} that lead to our Theorems \ref{thm:BvM} and \ref{thm:bootstrap}. Regarding Assumption \ref{assump:id}, we have a well-separated maximum point, due to the identification and the compactness of the parameter space in Assumption \ref{MNPAssump:Identify}, as well as the continuity of the log-likelihood function \citep{newey1994large}. Assumption \ref{MNPAssump:Linear} also satisfies the convergence rate condition in Assumption \ref{assump:FirstStep}. Referring to the frequentist estimator $\widehat{\theta}_n$, one can take the simulated score estimator by \cite{hajivassiliou1998scores} in the second stage. The prior specification given in Assumption \ref{MNPAssump:Prior} satisfies the requirement in Assumption \ref{assump:Prior}. The normality of the error term generates sufficient smoothness of the conditional choice probabilities, which satisfy Assumption \ref{assump:smooth}. Therein, the function $e_n(\cdot,\cdot)$ in Assumption \ref{assump:regular} can be taken as the cross product term $(\theta-\theta_0)^\top \mathbb{E}_0[\ddot{l}_{\theta,\zeta}][\zeta-\zeta_0]$ as in Lemma C.2 of \cite{chen2014maximum}. The smoothness of MNP implies a stronger notion of differentiability than what is required in Assumption \ref{assump:smooth}, i.e., the likelihood is Frech\'et differentiable w.r.t. $\zeta$ \citep{ichimura2010semi}. 
Next we verify Assumption \ref{assump:SE}. For Assumption \ref{assump:SE}(i), let $\mathcal{P}\equiv\{\log p_{\theta}(Y_{2};\zeta):\theta\in\Theta,\zeta\in\mathcal{G}_n \}$. Our Assumption \ref{MNPAssump:1stStage} on the H\"older class implies the $P_0$-Glivenko-Cantelli property of $\mathcal{P}$. Given the smoothness requirement in Assumption \ref{MNPAssump:1stStage}, we can utilize the Lipschitz continuity and apply Theorem 2.7.11 and Theorem 2.7.1 in \cite{van1996empirical} to bound its bracketing entropy by
\begin{align}\label{EntropyBounds}
	\log N_{\left[  {}\right]  }(\epsilon,\mathcal{P},\left\Vert \cdot\right\Vert _{2})\lesssim\log N_{\left[  {}\right]  }(\epsilon/2C,\Theta,\left\Vert \cdot\right\Vert _{2})+\log N_{\left[  {}\right]  }(\rho/2C,\mathcal{G}_n,\left\Vert \cdot\right\Vert _{2})\lesssim p\log(\epsilon^{-1})+ \epsilon^{-d/\tau};
\end{align}
see Appdendix B in the supplementary material. To check the $P_0$-Donsker property in Assumption \ref{assump:SE}(ii), we follow the route in Example 19.7 from \cite{van1998asymptotic} given our Assumption \ref{MNPAssump:Moment}, along with the restriction that $\tau>d/2$ in Assumption \ref{MNPAssump:1stStage}. Other properties, such as the functional class of the second derivative $\ddot l_{\theta}$ being $P_0$-Glivenko-Cantelli, can be checked along similar lines, utilizing the smoothness of the MNP in the second stage, as well as the entropy bound for the H\"older class from the first stage in (\ref{EntropyBounds}); see \cite{mcfadden1989sme} and \cite{pakes1989simulation}. The score functions of $\beta$ and $\eta$ (equations (14) and (15) in \cite{hajivassiliou1998scores}) have finite second-order moments given the bounded support of covariates and control variables. Given the smoothness and boundedness of covariates suppport, we also have the envelope functions being bounded. Thus, the slightly stronger assumption used in the bootstrap part (Theorem \ref{thm:bootstrap}) is also satisfied. When it comes to Assumption \ref{assump:SE}(iii), one can apply Lemma 2.14.3 in \cite{van1996empirical} to obtain $\phi_n(\rho)=\rho^{1-\frac{d}{2\tau}}$, which satisfy the restrictions under the maintained assumption $\tau>d/2$. By equation (3.15) of \cite{ichimura2010semi}, the influence function of the first-stage estimation defined by $\Gamma_1(Y_{1i})=\sum_{j=1}^Jv^{\dagger}_{ij}\mathbb{E}_0\left[\frac{\partial g_j(\cdot,\theta_0)}{\partial \theta}|Z_{ij}\right]$. The asymptotic normality in Assumption \ref{assump:normal} follows from the Lindeberg-L\'evy central limit theorem under Assumption \ref{MNPAssump:Inf} and Cauchy-Shwarz inequality. 
\end{proof}

\bibliographystyle{econometrica}
\bibliography{Bayes_bib}

\end{document}

\vspace*{5ex minus 1ex}
	\begin{center}
		\Large \textsc{Supplementary Material for ``Quasi-Bayesian Estimation and Inference with Control Functions''}
		\bigskip
	\end{center}
	
	%\date{
		%\today%
		%EndExpansion
	%}

	%\vspace*{3ex minus 1ex}
		\begin{center}
			Ruixuan Liu and Zhengfei Yu\\
			\textit{Chinese University of Hong Kong and University of Tsukuba}\\
			\medskip
		\end{center}

	\thanks{}
	\maketitle
		
The supplementary material contains several Appendices. Appendix B collects technical lemmas and their proofs. Appendix C presents the BvM theorem and its proof when the first stage is modeled and estimated parametrically. Appendix D provides details of the influence function for a special case of the Petrin--Train model. Appendix E offers additional simulation evidence comparing our bootstrap proposal with another method.

\section{Appendix B: Proofs of Technical Lemmas}\label{supp:lemmas}
Appendix B proves technical lemmas used in the proof of Theorems 2.1 and 2.2. For simplicity, we state the lemmas under all of our maintained Assumptions 2.1 to 2.8. From the proofs, it is clear which specific conditions are actually invoked.  For two sequences of real numbers $a_n$ and $b_n$, we write $a_n\lesssim b_n$ if $\limsup(a_n/b_n)<\infty$. 
Our assumptions imply Conditions (i)-(iv) in Corollary 1 of \cite{nan2013general}. Thus, the frequentist two-stage estimator satisfies the following expansion:
\begin{align}
	(\widehat{\theta}_n-\theta_0) 			&=-V_0^{-1}\left(S_{\theta,n}(\theta_0,\zeta_0)+\dot{\Psi}_{\zeta}(\theta_0,\zeta_0)[\widehat{\zeta}_n-\zeta_0]  \right)+o_{P_0}(n^{-1/2})\notag \\
	&\equiv -V_0^{-1}\frac{1}{\sqrt{n}}\sum_{i=1}^n \left[\Gamma_2(Y_{2i})+\Gamma_1(Y_{1i})\right]+o_{P_0}(n^{-1/2}),\label{freq_estimator_linearize}
\end{align}
which implies that $\widehat{\theta}_n$ is root-$n$ consistent and asymptotically normal. 
Recall the definition of $\Xi_n^r$ as follows. 
\begin{equation}
	\Xi_n^r\equiv \sup_{d_{\Theta}(\theta,\theta_0)\geq r}\left\{\frac{1}{n}\sum_{i=1}^n[\log p_{\theta}(Y_{2i};\widehat{\zeta}_n)-\log p_{\widehat{\theta}_n}(Y_{2i};\widehat{\zeta}_n)]\right\}.
\end{equation}
Recall that the bracketing number $N_{\left[  {}\right]  }\left(
\epsilon,\mathcal{F},\left\Vert \cdot\right\Vert _{2}\right)  $ for
a functional class $\mathcal{F}$ is defined to be the minimum of $m$ such that $\exists$
$f_{1}^{L},f_{1}^{U},\dots,f_{m}^{L},f_{m}^{U}$ for $\forall f\in\mathcal{F}$,
$f_{j}^{L}\leq f\leq f_{j}^{U}$ for some $j$, and $\left\Vert f_{j}^{U}%
-f_{j}^{L}\right\Vert _{2}\leq\epsilon$ \citep{van1996empirical}. For the empirical process, we write $\Vert \mathbb{G}_n f\Vert_{\mathcal{F}}:=\sup_{f\in\mathcal{F}}\{|\mathbb{G}_nf|:f\in\mathcal{F} \}$. Similar notation is also used for its bootstrap version $\Vert \mathbb{G}^*_n f\Vert_{\mathcal{F}}$
\begin{lemma}\label{lemma:outer}
	Under Assumptions 2.1 to 2.8, for any $\delta>0$, we have the following inequality holds with probability approaching one,
	\begin{equation}\label{LikelihoodRatio}
		\sup_{d_{\Theta}(\theta,\theta_0)> \delta}\mathbb{P}_n[\log p_{\theta}(Y_{2i};\widehat{\zeta}_n)-\log p_{\widehat{\theta}_n}(Y_{2i};\widehat{\zeta}_n)] \leq -c,
	\end{equation}
	for some positive constant term $c$.
\end{lemma}
\begin{proof}
For any $\theta$ s.t. $d_{\Theta}(\theta,\theta_0)> \delta$, we start with the following decomposition:
	\begin{align*}
		&\mathbb{P}_n[\log p_{\theta}(Y_{2i};\widehat{\zeta}_n)-\log p_{\widehat{\theta}_n}(Y_{2i};\widehat{\zeta}_n)]\\
		=&\mathbb{P}_n[\log p_{\theta}(Y_{2i};\widehat{\zeta}_n)-\log p_{\theta_0}(Y_{2i};\zeta_0)]+\mathbb{P}_n[\log p_{\theta_0}(Y_{2i};\zeta_0)-\log p_{\widehat{\theta}_n}(Y_{2i};\widehat{\zeta}_n)].
	\end{align*}

	The second term above can be shown to be $o_{P_0}(1)$ following the standard consistency argument for the two-stage estimation given the identification in Assumption 2.1 and $P_0$-Glivenko-Cantelli property in Assumption 2.7; see \cite{chen2007sieve}. Therefore, we focus on the first term and further decompose it as follows:
	\begin{align*}
		&\mathbb{P}_n[\log p_{\theta}(Y_{2i};\widehat{\zeta}_n)-\log p_{\theta_0}(Y_{2i};\zeta_0)]\\
		=&(\mathbb{P}_n-P_0)[\log p_{\theta}(Y_{2i};\widehat{\zeta}_n)]-(\mathbb{P}_n-P_0)[\log p_{\theta_0}(Y_{2i};\zeta_0)]\\
		+&P_0[\log p_{\theta}(Y_{2i};\widehat{\zeta}_n)-\log p_{\theta}(Y_{2i};\zeta_0)]+P_0[\log p_{\theta}(Y_{2i};\zeta_0)-\log p_{\theta_0}(Y_{2i};\zeta_0)].
	\end{align*}
	By the $P_0$-Glivenko-Cantelli property, i.e., $\sup_{\theta\in\Theta,\zeta\in\mathcal{G}_n}\left|(\mathbb{P}_n-P_0)[\log p_{\theta}(Y_{2i};\zeta)] \right|=o_{P_0}(1)$, we have the first two terms on the right hand side of the above equality converging to zero in probability uniformly for any $\theta\in\Theta$. When it comes to the third term, we utilize the uniform (regarding $\theta$) continuity of the criterion function with respect to the first stage nuisance function in our Assumption 2.6, so that $P_0[\log p_{\theta}(Y_{2i};\widehat{\zeta}_n)-\log p_{\theta}(Y_{2i};\zeta_0)]=o_{P_0}(1)$. For any given $\delta>0$, Assumption 2.1 implies that the last term satisfies
	\begin{align*}
		\sup_{d_{\Theta}(\theta,\theta_0)> \delta}P_0\left[\log p_{\theta}(\cdot;\zeta_0)- \log p_{\theta_0}(\cdot;\zeta_0)\right]\leq -c,
	\end{align*}
	with a proper choice of constant $c>0$, which concludes the proof.
\end{proof}

\begin{lemma}\label{lemma:MaximalIneq}
	
Under Assumptions 2.1 to 2.8, for the decreasing sequence $r_n=o(n^{-1/4})$ and $nr^2_n\to\infty$, we have
	\begin{equation}
		\sup_{C r_n\leq d_{\Theta}(\theta,\theta_0)\leq \delta}\mathbb{P}_n[\log p_{\theta}(Y_{2i};\widehat{\zeta}_n)-\log p_{\widehat{\theta}_n}(Y_{2i};\widehat{\zeta}_n)]\leq -cr_n^2
	\end{equation}
with probability approaching one, for a given positive $c$ and a large enough positive constant $C$.
\end{lemma}
\begin{proof}[Proof of Lemma \ref{lemma:MaximalIneq}]	
	Because $\widehat{\theta}_n$ is root-$n$ consistent and $nr^2_n\to\infty$, we have
	\begin{equation*}
		\{\theta:d_{\Theta}(\theta,\widehat{\theta}_n)\geq C r_n \}\subset \{\theta:d_{\Theta}(\theta,\theta_0)\geq C^\prime r_n \},
	\end{equation*}
	for a proper choice of the constant term $C^\prime$. Therefore, we obtain
	\begin{align*}
		&\{\sup_{\theta:d_{\Theta}(\theta,\theta_0)\geq C^\prime r_n}\mathbb{P}_n[\log p_{\theta}(Y_2;\widehat{\zeta}_n)-\log p_{\widehat{\theta}_n}(Y_2;\widehat{\zeta}_n)] \geq -cr^2_n\}\\
		&\subset \{\sup_{\theta:d_{\Theta}(\theta,\theta_0)\geq C^\prime r_n}\mathbb{P}_n[\log p_{\theta}(Y_2;\widehat{\zeta}_n)-\log p_{\theta_0}(Y_2;\widehat{\zeta}_n)] \geq -cr^2_n\},
	\end{align*}
	where the inclusion follows from $\mathbb{P}_n[\log p_{\widehat{\theta}_n}(Y_2;\widehat{\zeta}_n)]\geq \mathbb{P}_n[\log p_{\theta_0}(Y_2;\widehat{\zeta}_n)]$ by the definition of $\widehat{\theta}_n$ from Assumption 2.3. We can further restrict our attention to the set $\{d_G(\zeta,\zeta_0)\leq Cr_n\}$, as its complement is asymptotically negligible by Assumption 2.2. It suffices to examine 
	\begin{align}\label{PEvent}
		\left\{\sup_{d_{\Theta}(\theta,\theta_0)\geq C^\prime r_n,d_G(\zeta,\zeta_0)\leq Cr_n}\mathbb{P}_n[\log p_{\theta}(Y_2;\zeta)-\log p_{\theta_0}(Y_2;\zeta)] \geq -cr^2_n\right\}
	\end{align}
	Thanks to our Lemma \ref{lemma:outer}, we can localize on the set $\Theta_n:=\{\theta:d_{\Theta}(\theta,\theta_0)\leq \delta \}\bigcap\{\theta:d_{\Theta}(\theta,\theta_0)\geq C^\prime r_n\}$ for a given small $\delta$. This localized set will be partitioned into the shells for $\rho=C^\prime r_n$:
	\begin{align*}
		S_{n,j,M}=\left\{(\theta,\zeta)\in \Theta_n\times \mathcal{G}_n:2^{j-1}\rho<d_{\Theta}(\theta,\theta_0)\leq 2^j\rho, d_G(\zeta,\zeta_0)\leq2^{-M}d_{\Theta}(\theta,\theta_0) \right\}.
	\end{align*}
	In the slicing argument, we sum over the shells that $2^j\rho<\delta$. For the $j$-th set involved, we have 
	\begin{align}\label{SliceBound}
		&P_0(\log p_\theta(Y_2;\zeta)-\log p_{\theta_0}(Y_2;\zeta)) \notag \\
		=&P_0(\log p_\theta(Y_2;\zeta)-\log p_{\theta_0}(Y_2;\zeta_0))+P_0(\log p_{\theta_0}(Y_2;\zeta_0)-\log p_{\theta_0}(Y_2;\zeta)) \notag\\
		\leq& P_0(\log p_\theta(Y_2;\zeta)-\log p_{\theta_0}(Y_2;\zeta_0))+e_n(\theta,\zeta)+C_1r_n^2 \notag\\
		\lesssim&-d_{\Theta}^2(\theta,\theta_0)+d^2_G(\zeta,\zeta_0)+C_2 2^j\rho r_n+C_1r_n^2 \lesssim-(1-2^{-2M})d_{\Theta}^2(\theta,\theta_0)\lesssim -2^{2j-2}\rho^2,
	\end{align}
	under Assumption 2.5. In other words, we have
	\begin{equation*}
		\inf_{	(\theta,\zeta)\in S_{n,j,M}}	-P_0(\log p_\theta(Y_2;\zeta)-\log p_{\theta_0}(Y_2;\zeta))\geq 2^{2j-2}\rho^2.
	\end{equation*}
	Thus, the probability of (\ref{PEvent}) can be bounded by
	\begin{align*}
		&\sum_{j\geq M, 2^j\rho_n\leq \delta} 	P_0\left(\sup_{(\theta,\zeta)\in S_{n,j,M}} \mathbb{P}_n[\log p_\theta(Y_2;\zeta)-\log p_{\theta_0}(Y_2;\zeta)]\geq -c_1r_n^2\right)\\
		&\leq\sum_{j\geq M} 	P_0\left(\sup_{(\theta,\zeta)\in S_{n,j,M}} \mathbb{G}_n[\log p_{\theta}(Y_2;\zeta)-\log p_{\theta_0}(Y_2;\zeta)]\geq C2^{2j-2}\sqrt{n}\rho^2\right)	\\
		&\lesssim\sum_{j\geq M} \frac{1}{2^{2j}\sqrt{n}\rho^2}\mathbb{E}_0\sup_{(\theta,\zeta)\in S_{n,j,M}} \left|\mathbb{G}_n[\log p_{\theta}(Y_2;\zeta)-\log p_{\theta_0}(Y_2;\zeta)]\right|\\
		&\lesssim\sum_{j\geq M} \frac{1}{2^{2j}\sqrt{n}\rho^2}\phi_n(2^j\rho)\leq \sum_{j\geq M} \frac{1}{2^{2j}\sqrt{n}\rho^2}2^{j\gamma}\phi_n(\rho)\lesssim\sum_{j\geq M}2^{j(\gamma-2)}.
	\end{align*}
	In the first inequality, we add/subtract $P_0[\log p_\theta(\cdot;\zeta)-\log p_{\theta_0}(\cdot;\zeta)]$ and apply (\ref{SliceBound}). The fourth inequality makes use of the monotonicity of the mapping $\delta\mapsto\phi_n(\delta)/\delta^{\gamma}$ for some $\gamma<2$, as stated in Assumption 2.7(iii), so that we have $\phi_n(2^j\rho)\leq 2^{j\gamma}\phi_n(\rho)$. The final inequality follows from $\phi_n(\rho)\leq \sqrt{n}\rho^{2}$ for every $n$. The desired result follows by choosing a large enough $M$, which is induced by a large $C$.
	%Therefore, we can restrict to our attention to the set $\{\theta:d_{\Theta}(\theta,\theta_0)\geq 2^M(r_n+d_G(\zeta,\zeta_0))\}$. 
\end{proof}

\begin{lemma}\label{lemma:local_expan}
	Under Assumptions 2.1 to 2.8, for any $\theta_n$ converging to $\theta_0$ such that $d_{\Theta}(\theta_n,\theta_0)\leq Cr_n$ w.p.a.1, one has
	\begin{equation}
		\ell_n(\theta_n;\widehat{\zeta}_n)-\ell_n(\widehat{\theta}_n;\widehat{\zeta}_n)=-\frac{1}{2}(\theta_n-\widehat{\theta}_n)^\top V_0(\theta_n-\widehat{\theta}_n)+o_{P_0}(1+\sqrt{n}d_{\Theta}(\theta_n,\theta_0))^2.
	\end{equation}
\end{lemma}
\begin{proof}
	Recall that $\Delta_{n,0}\equiv \sqrt{n}\left(S_{\theta,n}(\theta_0,\zeta_0)+\dot{\Psi}_{\zeta}(\theta_0,\zeta_0)[\widehat{\zeta}_n-\zeta_0]  \right)$. We start with the following string of equations:
	\begin{align*}
		&\ell_n(\theta_n;\widehat{\zeta}_n)-\ell_n(\theta_0;\widehat{\zeta}_n)\\
		&=n\mathbb{P}_n[\log p_{\theta_n}(Y_2;\widehat{\zeta}_n)-\log p_{\theta_0}(Y_2;\widehat{\zeta}_n)]\\
		&=\sqrt{n}\mathbb{G}_n[\log p_{\theta_n}(Y_2;\widehat{\zeta}_n)-\log p_{\theta_0}(Y_2;\widehat{\zeta}_n)]+nP_0[\log p_{\theta_n}(Y_2;\widehat{\zeta}_n)-\log p_{\theta_0}(Y_2;\widehat{\zeta}_n)]\\
		&=\sqrt{n}\mathbb{G}_n\left[\log p_{\theta_n}(Y_2;\widehat{\zeta}_n)-\log p_{\theta_0}(Y_2;\widehat{\zeta}_n)-(\theta_n-\theta_0)^\top\dot{l}_{\theta}(Y_2,\theta_0;\widehat{\zeta}_n) \right]	\\
		&\quad + \sqrt{n}(\theta_n-\theta_0)^\top\mathbb{G}_{n}\left[\dot{l}_{\theta}(Y_2,\theta_0;\widehat{\zeta}_n)-\dot{l}_{\theta}(Y_2,\theta_0;\zeta_0) \right]\\
		&	+\sqrt{n}(\theta_n-\theta_0)^\top\mathbb{G}_{n}\left[\dot{l}_{\theta}(Y_2,\theta_0;\zeta_0) \right]+nP_0\left[\log p_{\theta_n}(Y_2;\widehat{\zeta}_n)-\log p_{\theta_0}(Y_2;\widehat{\zeta}_n)\right]
	\end{align*}
	By the $P_0$-Donsker property in our Assumption 2.7, we have
	\begin{equation}
		\sup_{d_{\Theta}(\theta_n,\theta_0)\leq Cr_n}	\left|\mathbb{G}_{n}\left[\log p_{\theta_n}(Y_2;\widehat{\zeta}_n)-\log p_{\theta_0}(Y_2;\widehat{\zeta}_n)-(\theta_n-\theta_0)^\top\dot{l}_{\theta}(Y_2,\theta_0;\widehat{\zeta}_n) \right]\right|=o_{P_0}(d_{\Theta}(\theta_n,\theta_0)),
	\end{equation}
	and
	\begin{equation}
		\mathbb{G}_{n}\left[\dot{l}_{\theta}(Y_2,\theta_0;\widehat{\zeta}_n)-\dot{l}_{\theta}(Y_2,\theta_0;\zeta_0) \right]=o_{P_0}(n^{-1/2}).
	\end{equation}
	In addition, the smoothness in Assumption 2.6 leads to
	%\begin{footnotesize}
	\begin{align*}
		&\sup_{d_{\Theta}(\theta_n,\theta_0)\leq Cr_n}	\biggl\vert	nP_0\left[\log p_{\theta_n}(Y_2;\widehat{\zeta}_n)-\log p_{\theta_0}(Y_2;\widehat{\zeta}_n) \right] \\
		&\quad\quad\quad\quad\quad -\frac{1}{2}(\theta_n-\theta_0)^{\top}V_0(\theta_n-\theta_0)-(\theta_n-\theta_0)^\top\dot{\Psi}_{\zeta}(\theta_0,\zeta_0)[\widehat{\zeta}_n-\zeta_0]\biggl\vert\\
		& =o_{P_0}(1+\sqrt{n}d_{\Theta}(\theta_n,\theta_0))^2, \notag
	\end{align*}
	%\end{footnotesize}
	where $r_n=o(n^{-1/4})$. Combining the previous steps, we have
	%\begin{footnotesize}
	\begin{align*}
		\ell_n(\theta_n;\widehat{\zeta}_n)-\ell_n(\theta_0;\widehat{\zeta}_n)&=\sqrt{n}(\theta_n-\theta_0)^{\top}\Delta_{n,0}-\frac{n}{2}(\theta_n-\theta_0)^\top V_0(\theta_n-\theta_0)\\
		& \quad+o_{P_0}(1+\sqrt{n}d_{\Theta}(\theta_n,\theta_0))^2\\
		\ell_n(\widehat{\theta}_n;\widehat{\zeta}_n)-\ell_n(\theta_0;\widehat{\zeta}_n)&=\frac{n}{2}(\widehat{\theta}_n-\theta_0)^\top V_0(\widehat{\theta}_n-\theta_0)+o_{P_0}(1+\sqrt{n}d_{\Theta}(\widehat{\theta}_n,\theta_0))^2.
	\end{align*}
	%\end{footnotesize}
	Now we take the difference of the above two equations and utilize the linear representation of $\widehat{\theta}_n$ to get
	%\begin{footnotesize}
	\begin{align*}
		\ell_n(\theta_n;\widehat{\zeta}_n)-\ell_n(\widehat{\theta}_n;\widehat{\zeta}_n)&=n(\theta_n-\theta_0)^{\top}V_0(\widehat{\theta}_n-\theta_0)
		-\frac{n}{2}(\theta_n-\theta_0)^\top V_0(\theta_n-\theta_0)\\
		&-\frac{n}{2}(\widehat{\theta}_n-\theta_0)^\top V_0(\widehat{\theta}_n-\theta_0)+o_{P_0}(1+\sqrt{n}d_{\Theta}(\widehat{\theta}_n,\theta_0))^2\\
		&=-\frac{1}{2}(\theta_n-\widehat{\theta}_n)^\top V_0(\theta_n-\widehat{\theta}_n)+o_{P_0}(1+\sqrt{n}d_{\Theta}(\theta_n,\theta_0))^2.
	\end{align*}
	%\end{footnotesize}
	The last step follows from completing the square of $(\theta_n-\widehat{\theta}_n)^\top V_0(\theta_n-\widehat{\theta}_n)$.
\end{proof}

\begin{lemma}\label{lemma:Bn}
	Let
	\begin{equation*}
		\tilde{B}_n(\theta)\equiv e^{-\frac{n}{4}(\theta-\widehat{\theta}_n)^\top V_0(\theta-\widehat{\theta}_n) }\left|e^{\sum_{i=1}^n \log p_{\theta}(Y_{2i}|\widehat{\zeta}_n)-\log p_{\widehat{\theta}_n}(Y_{2i}|\widehat{\zeta}_n)  +\frac{n}{2}(\theta-\widehat{\theta}_n)^\top V_0(\theta-\widehat{\theta}_n)}-1\right|.
	\end{equation*}
	Under Assumptions 2.1 to 2.8, we have $B_n\equiv \sup_{d_{\Theta}(\theta,\theta_0)\leq C r_n}\tilde{B}_n(\theta)=o_{P_0}(1)$.
\end{lemma}
\begin{proof}
	For any sequence $\theta_n$ such that $d_{\Theta}(\theta_n,\widehat{\theta}_n)\leq Cr_n$ for all $n$ and some fixed $C<\infty$, we have
	\begin{align*}
		\tilde{B}_n(\theta_n)\leq \exp\left\{-\frac{n}{4} (\theta_n-\widehat{\theta}_n)^\top V_0(\theta_n-\widehat{\theta}_n)\right\}\times\left|\exp\{o_{P_0}(1+\sqrt{n}d_{\Theta}(\theta_n,\theta_0))^2 \}-1 \right|,
	\end{align*}
	where the remainder term satisfies
	\begin{align*}
		o_{P_0}(1+\sqrt{n}d_{\Theta}(\theta_n,\theta_0))^2
		\leq & o_{P_0}(1)[(1+\sqrt{n}d_{\Theta}(\theta_n,\widehat{\theta}_n))^2+nd_{\Theta}(\theta_0,\widehat{\theta}_n)^2]\\
		= & o_{P_0}(1)(1+\sqrt{n}d_{\Theta}(\theta_n,\widehat{\theta}_n))^2,
	\end{align*}
	due to Lemma \ref{lemma:local_expan}. Note that $|e^x-1|\leq e^{|x|}$ for any $x$, we can proceed as follows for a given large positive constant $C$:
	\begin{align*}
		\tilde{B}_n(\theta_n)&\leq \exp\left\{-\frac{n}{4} (\theta_n-\widehat{\theta}_n)^\top V_0(\theta_n-\widehat{\theta}_n)\right\}\left|\exp\{o_{P_0}(1)\}-1 \right|\times \mathbb{I}\{\sqrt{n}d_{\Theta}(\theta_n,\widehat{\theta}_n)\leq C \}\\
		&+\exp\left\{-\frac{n}{4} (\theta_n-\widehat{\theta}_n)^\top V_0(\theta_n-\widehat{\theta}_n)+o_{P_0}(1)(1+\sqrt{n}d(\theta_n,\theta_0))^2\right\}\times \mathbb{I}\{\sqrt{n}d_{\Theta}(\theta_n,\widehat{\theta}_n)\geq C \}\\
		&\leq o_{P_0}(1)+\exp\left\{ \frac{C^2}{4}\lambda_1(1-o_{P_0}(1))\right\}\to\exp\{-\lambda_1C^2/4\},
	\end{align*}
	where $\lambda_1$ is the smallest eigenvalue of $V_0$ which is strictly positive by our Assumption 2.1. Therefore, $B_n=o_{P_0}(1)$, as $C$ can be made arbitrarily large.
\end{proof}

\begin{lemma}\label{lemma:BootstrapDensity}
	Let the bootstrap posterior density function of $h\equiv \sqrt{n}(\theta-\widehat{\theta}^*_n)$ be
	\begin{equation}
		\tilde{\pi}^*(h|\bm{Y}_2^{n};\widehat{\zeta}^*_n)\equiv\frac{\prod_{i=1}^np^{M_{ni}}_{\widehat{\theta}^*_n+\frac{h}{\sqrt{n}}}(\bm{Y}_2^n|\widehat{\zeta}^*_n)\pi(\widehat{\theta}^*_n+\frac{h}{\sqrt{n}})}{\int \prod_{i=1}^np^{M_{ni}}_{\widehat{\theta}^*_n+\frac{h}{\sqrt{n}}}(\bm{Y}_2^n|\widehat{\zeta}^*_n)\pi(\widehat{\theta}^*_n+\frac{h}{\sqrt{n}})dh}.
	\end{equation}	
	In addition to Assumptions 2.1 to 2.8, we assume that the envelope functions for the functional classes in Parts (i) and (ii) in Assumption 2.7 have finite $(2+\iota)$ moments, for $\iota>0$. Then, we have the following convergence in total variation norm:
	\begin{equation}
		\Vert \tilde{\pi}^*(h|\bm{Y}_2^{n};\widehat{\zeta}^*_n)-\pi_{\infty}(h)\Vert_{TVM(a)} =o_{P^*}(1).
	\end{equation}
\end{lemma}
\begin{proof}
	The bootstrap version follows along similar lines of the proof of our Theorem 2.1. We highlight two main modifications needed herein. 
	
	First, we show the posterior mass outside an shrinking neighborhood around the truth is negligible conditional on the observed data with probability approaching one. Specifically, we need to characterize the order of the following term
	\begin{equation*}
		\sup_{\theta:\delta \geq d_{\Theta}(\theta,\widehat{\theta}^*_n)\geq C r_n}\mathbb{P}^*_n[\log p_{\theta}(Y_2;\widehat{\zeta}^*_n)-\log p_{\widehat{\theta}_n^*}(Y_2;\widehat{\zeta}^*_n)].
	\end{equation*}
	A key step in the chaining argument is to use a proper maximal inequality related to the bootstrap empirical process. For this purpose, we resort to Corollary 1 from \cite{han2019heavy} to obtain
	\begin{equation}\label{WeightedRate}
		\mathbb{E}^*_0\sup_{(\theta,\zeta)\in S_{n,j,M}} \left|\mathbb{G}^*_n[\log p_{\theta}(Y_2;\zeta)-\log p_{\theta_0}(Y_2;\zeta)]\right|\leq C\phi_n(2^j\rho).
	\end{equation}
	
	Second, when we carry out an expansion of the bootstrap log-likelihood, we need to verify the following stochastic equicontinuity as 
	\begin{equation}
		\mathbb{G}^*_{n}\left[\dot{l}_{\theta}(Y_2,\theta_0;\widehat{\zeta}^*_n)-\dot{l}_{\theta}(Y_2,\theta_0;\zeta_0) \right]=o_{P^*}(n^{-1/2}).
	\end{equation}
	Denote $\mathcal{F}\equiv\{f=\dot{l}_{\theta}(\cdot,\theta_0;\zeta)-\dot{l}_{\theta}(\cdot,\theta_0;\zeta_0),\zeta\in\mathcal{G}_n\}$. We apply the multiplier inequality in Lemma 3.6.7 of \cite{van1996empirical}, which states for any $1\leq n_0\leq n$
	\begin{equation}\label{BootVW}
		\mathbb{E}_{YM}\parallel \mathbb{G}^*_nf \parallel_{\mathcal{F}} \lesssim 2(n_0-1)\mathbb{E}_{Y}|F|\frac{1}{\sqrt{n}}\mathbb{E}_{M}\left|\max_{i}M_{ni}\right|+2\mathbb{E}_MM^2_{n1} \left| \max_{n_0\leq k\leq n}\mathbb{E}_{Y}\parallel\mathbb{G}_{k}f\parallel_{\mathcal{F}} \right|,
	\end{equation}
	where $F$ is the corresponding envelope function for the functional class $\mathcal{F}$, and the subscript on the expectation signifies the source of randomness. 
	
	For the multinomial weights, we have $\mathbb{E}|M_{ni}|^2<\infty$. We start with applying the Jensen's inequality to obtain
	\begin{align*}
		\left(\mathbb{E}_{M}\left|\max_{1\leq i\leq n}\frac{M_{ni}}{\sqrt{n}}\right|\right)^2\leq \mathbb{E}_{M}\left|\max_{1\leq i\leq n}\frac{M^2_{ni}}{n}\right|
		\leq\frac{m^2}{n}+\frac{1}{n}\sum_{i=1}^n\mathbb{E}_{M}\left[M^2_{ni}\mathbb{I}\{M_{ni}>m\}\right],
	\end{align*}
	which goes to $0$. Thus, the first term in (\ref{BootVW}) is of smaller order as $\frac{1}{\sqrt{n}}\mathbb{E}_{M}\left|\max_{i}M_{ni}\right|=o(1)$. Regarding the term $\max_{n_0\leq k\leq n}\mathbb{E}_{Y}\parallel\mathbb{G}_{k}f\parallel_{\mathcal{F}} $, it sufficies to bound $\mathbb{E}_{Y}\parallel\mathbb{G}_{n}f\parallel_{\mathcal{F}}$ by Lemma 4.3 of \cite{praestgaard1993weight}. Because of the stochastic equicontinuity of the empirical process and the Hoffmann-Jorgensen inequality, the first-order moment version also holds, i.e., $\mathbb{E}_{Y}\parallel\mathbb{G}_{n}f\parallel_{\mathcal{F}}=o_{P_0}(1)$. In sum, this makes the second term of (\ref{BootVW}) negligible, given that the original empirical process also satisfies the $P_0$-Donsker property.
\end{proof}

\section{Appendix C: Two-stage Parametric Models}
Our general theory presented in Section 2 does not cover the parametric (either frequentist or Bayesian) first-stage estimation with the usual root-$n$ rate of convergence per se. As we demonstrate herein, the parametric first-stage estimation significantly simplifies the technical proof of the BvM theorem. We do not need to rely on the maximal inequality via the empirical process theory to handle the nonparamtric first-stage estimation. Instead, one can simply take a third-order Taylor expansion of the log-likelihood function and proceed with some elementary bounds. We need to work with the third-order derivative $\dddot{l}_{\theta}\left(  Y_2,\theta;\zeta\right)$. $\dddot{l}_{\theta}\left(  Y_2,\theta;\zeta\right)$ is a tensor in the sense that $h^\top \otimes \dddot{l}_{\theta}\left(  Y_2,\theta;\zeta\right)\otimes h$ is a $p$-dimensional vector whose $i$-th element equals $h^\top \otimes(\partial^2/\partial \theta\partial\theta^\top) \dot{l}_{\theta}\left(  Y_2,\theta;\zeta\right)\otimes h$ \citep{cheng2008profile}. For readers' convenience, we state the high-level assumptions with proper modifications tailored to the parametric first-stage estimation.
\begin{assumption}[Identification]\label{assump:Pid}
	The finite dimensional parameter $\zeta_0$ is uniquely identified from the first stage, and it belongs to the interior of the parameter space $\mathcal{G}$. The expected log-likelihood function $\mathbb{E}_0[\log p_{\theta}(Y_2;\zeta_0)]$ has a unique maximum at $\theta_0$ over the parameter space $\Theta$ in the sense that $\sup_{d_{\Theta}(\theta,\theta_0)> \delta}\mathbb{E}_0[\log p_{\theta}(Y_2;\zeta_0)]<\mathbb{E}_0[\log p_{\theta_0}(Y_2;\zeta_0)]$ for any $\delta>0$. Also, $S_{\theta}\left(  \theta_0,\zeta_0\right)  =0$, and the matrix $V_0$ is positive definite with its smallest eigenvalue $\lambda_1>0$.
\end{assumption}

\begin{assumption}[First-stage Estimation]\label{assump:PFirstStep}
	The first-stage estimator $\widehat{\zeta}_n$ satisfies that $d_G(\widehat{\zeta}_n,\zeta_0)=O_{P_0}(r_n)$, where $r_n=O(n^{-1/2})$. 
\end{assumption}

\begin{assumption}[Frequentist Two-stage Estimation]\label{assump:PFreq}
	There exists a frequentist-type estimator that approximately maximizes the second-stage likelihood:
	\begin{equation}
		\widehat{\theta}_n=\arg\max_{\theta}\ell_n(\theta;\widehat{\zeta}_n)+o_{P_0}(n^{-1/2}).
	\end{equation}
	In addition, it approximately solves the score equation as follows
	\begin{equation}\label{sample_score}
		S_{\theta,n}\left(\widehat{\theta}_n,\widehat{\zeta}_n\right) 
		=o_{P_0}(n^{-1/2}).
	\end{equation} 
\end{assumption}

\begin{assumption}[Prior]\label{assump:PPrior}
	The prior measure $\Pi$ on $\Theta$ is assumed to be a probability measure with a bounded Lebesgue density $\pi$, which is continuous and positive on a neighborhood of the true value $\theta_0$. In addition, $\int |\theta|^a\pi(\theta)d\theta<+\infty$, for $a\geq 0$.
\end{assumption}

%\begin{assumption}[Smoothness]\label{assump:Pregular}
%	We assume the log-likelihood function satisfies the following restrictions for some positive constant terms $C_1,\cdots,C_4$:
%	\begin{equation*}
%		\sup_{\zeta\in\mathcal{G}_n}\left|\mathbb{E}_0\left[\log p_{\theta_0}(\cdot;\zeta)- \log p_{\theta_0}(\cdot;\zeta_0)\right] \right|\leq C_1 r_n^2,
%	\end{equation*}
%	and
%	\begin{equation*}
%		\mathbb{E}_0\left[\log p_{\theta}(\cdot;\zeta)- \log p_{\theta_0}(\cdot;\zeta_0)\right]+e_n(\theta,\zeta)\leq -C_2d^2_{\Theta}(\theta,\theta_0)+C_3d^2_{G}(\zeta,\zeta_0),
%	\end{equation*}
%	for some $e_n(\cdot,\cdot)$ that satisfies
%	\begin{equation*}
%		\sup_{d_{\Theta}(\theta,\theta_0)\leq \rho,\zeta\in \mathcal{G}_n}|e_n(\theta,\zeta)|\leq C_4\rho r_n.
%	\end{equation*}
%\end{assumption}
\begin{assumption}[Differentiability]\label{assump:Psmooth}
	The function $\mathbb{E}_0[\log p_{\theta}(Y_{2i};\zeta)]$ is continuous with respect to $\zeta$ uniformly in $\theta\in \Theta$. There exists a non-singular matrix $\dot{\Psi}_{\theta}(\theta_0,\zeta_0)$ and a matrix $\dot{\Psi}_{\zeta}(\theta_0,\zeta_0)$ such that 
	\begin{equation*}
		\parallel\Psi(\theta,\zeta)-\Psi(\theta_0,\zeta_0)-\dot{\Psi}_{\theta}(\theta_0,\zeta_0)(\theta-\theta_0)-\dot\Psi_{\zeta}(\theta_0,\zeta_0)(\zeta-\zeta_0)\parallel\leq o(d_{\Theta}(\theta,\theta_0)+d_{G}( \zeta,\zeta_0)).
	\end{equation*}
\end{assumption}
\begin{assumption}[Complexity]\label{assump:PSE}
	(i)	The functional classes $\{\log p_{\theta}(Y_{2i};\zeta):\theta\in\Theta,\zeta\in\mathcal{G} \}$ and $\left\{\ddot{l}_{\theta}(\cdot,\theta;\zeta):\theta\in\Theta,\zeta\in\mathcal{G} \right\}$ belong to $P_0$-Glivenko-Cantelli classes. 
		(ii) We assume that the third-order derivative $\dddot{l}_{\theta}\left(  Y_2,\theta;\zeta\right)$ exists and is continuous with respect to both $\theta$ and $\zeta$. In addition, for a local neighborhood around the truth $(\theta_0,\zeta_0)$, each element in the tensor has an envelope function that has a finite first order moment.
\end{assumption}
\begin{assumption}[Normality]\label{assump:Pnormal} Let $\Delta_{n,0}\equiv \sqrt{n}\left(S_{\theta,n}(\theta_0,\zeta_0)+\dot{\Psi}_{\zeta}(\theta_0,\zeta_0)(\widehat{\zeta}_n-\zeta_0)  \right)$. 
	Assume that the following linear representation holds:
	\begin{align*}
		\Delta_{n,0}= \frac{1}{\sqrt{n}}\sum_{i=1}^n \left[\Gamma_2(Y_{2i})+\Gamma_1(Y_{1i})\right]+o_{P_0}(1).
	\end{align*}
Also, the following asymptotic normality holds:
	\begin{align*}
		\frac{1}{\sqrt{n}}\sum_{i=1}^n \left[\Gamma_2(Y_{2i})+\Gamma_1(Y_{1i})\right]\Rightarrow \mathbb{N}(0,\Omega_0),
	\end{align*}
	for some finite covariance matrix $\Omega_0$.
\end{assumption}

In comparison with the statement in the main text, we can drop Assumption 2.7 (ii) regarding the Donsker properties needed in the quadratic approximation to the log-likelihood involving nonparametric components. Likewise, Assumptions 2.5 and 2.7 (iii) can also be removed, which are mainly used in the maximal inequality related to the log-likelihood ratio. In addition, one can also modify the Assumptions 2.6 and 2.8 related to the pathwise derivative to the standard derivative with respect to the unknown parameter from the first-stage estimation.

\begin{proposition}
Suppose Assumptions \ref{assump:Pid} to \ref{assump:Pnormal} hold. Then, we have
\begin{equation}\label{BvM_Parametric}
	\Vert \tilde{\pi}(h|\bm{Y}_2^n;\widehat{\zeta}_n)-\pi_{\infty}(h)\Vert_{TVM(a)} =o_{P_0}(1).
\end{equation}	
In addition, the conclusions in Corollary 2.1 hold.
\end{proposition}

\begin{proof}
	In the proof of the BvM theorem, the way that we split the integral can be modified as follows:
	\begin{itemize}
		\item $H_{1,n}\equiv \{h:|h|\leq C\log\sqrt{n}\}$;
		\item  $H_{2,n}\equiv \{h:C\log\sqrt{n}<|h|\leq \delta\sqrt{n}\}$;
		\item  $H_{3,n}\equiv \{h:|h|>\delta\sqrt{n}\}$.
	\end{itemize}
For this purpose, recall that $\ell_n(\theta;\widehat\zeta_n)=\sum_{i=1}^n p_{\theta}(Y_{2i}|\widehat\zeta_n)$. We take the third order derivative with respect to $\theta$, while $\zeta$ evaluated at $\widehat\zeta_n$ as $\ell_n^{(3)}(\theta)$. Let $V_n=-\mathbb{P}_n\ddot{l}_{\theta}\left(  Y_2,\widehat\theta_n;\widehat\zeta_n\right)$. 

It proves more convenient to first show that
\begin{equation}
	\int |h|^a \left|\prod_{i=1}^n \left[\frac{p_{\widehat{\theta}_n+\frac{h}{\sqrt{n}}}(Y_{2i}|\widehat{\zeta}_n)}{p_{\widehat{\theta}_n}(Y_{2i}|\widehat{\zeta}_n)}\right]\pi(\widehat{\theta}_n+\frac{h}{\sqrt{n}})-e^{-\frac{h^\top V_nh}{2}}\pi(\widehat\theta_n)\right|dh=o_{P_0}(1),
\end{equation}
and then one can employ standard argument, including the uniform LLN for the second order derivative of log-likelihood and continuity of the prior to get
 \begin{equation}
 	N_{n1}\equiv\int |h|^a \left|\prod_{i=1}^n \left[\frac{p_{\widehat{\theta}_n+\frac{h}{\sqrt{n}}}(Y_{2i}|\widehat{\zeta}_n)}{p_{\widehat{\theta}_n}(Y_{2i}|\widehat{\zeta}_n)}\right]\pi(\widehat{\theta}_n+\frac{h}{\sqrt{n}})-e^{-\frac{h^\top V_0h}{2}}\pi(\theta_0)\right|dh=o_{P_0}(1).
 \end{equation}

Note that for the outer range $H_{3,n}$, we can upper bound
\begin{align*}
&	\int_{H_{3,n}} |h|^a \left|\prod_{i=1}^n \left[\frac{p_{\widehat{\theta}_n+\frac{h}{\sqrt{n}}}(Y_{2i}|\widehat{\zeta}_n)}{p_{\widehat{\theta}_n}(Y_{2i}|\widehat{\zeta}_n)}\right]\pi(\widehat{\theta}_n+\frac{h}{\sqrt{n}})-e^{-\frac{h^\top V_nh}{2}}\pi(\widehat\theta_n)\right|dh\\
\leq	&	\int_{H_{3,n}}  |h|^a  \exp\left(\ell_n(\widehat{\theta}_n+\frac{h}{\sqrt{n}};\widehat{\zeta}_n)- \ell_n(\widehat{\theta}_n;\widehat{\zeta}_n)\right)\pi(\widehat{\theta}_n+\frac{h}{\sqrt{n}})+\int_{H_{3,n}}  e^{-\frac{h^\top V_nh}{2}}\pi(\widehat\theta_n)dh.
\end{align*}
The first integral converges to zero by Lemma \ref{lemma:outer} for any $\delta>0$, which requires Assumption \ref{assump:Pid}, \ref{assump:PFirstStep}, \ref{assump:Psmooth}, and \ref{assump:PSE} (i). The second goes to zero by the Gaussian tail inequality.

Considering the range $H_{1,n}$ and $H_{2,n}$, we apply a third-order Taylor expansion of the likelihood ratio as follows
\begin{align*}
	\prod_{i=1}^n \left[\frac{p_{\widehat{\theta}_n+\frac{h}{\sqrt{n}}}(Y_{2i}|\widehat{\zeta}_n)}{p_{\widehat{\theta}_n}(Y_{2i}|\widehat{\zeta}_n)}\right]
	=\exp\left(\ell_n(\widehat{\theta}_n+\frac{h}{\sqrt{n}};\widehat{\zeta}_n)- \ell_n(\widehat{\theta}_n;\widehat{\zeta}_n)\right)
	=\exp\left(\frac{1}{2}h^\top V_n h+R_n(h)\right),
\end{align*}
given Assumption \ref{assump:PFreq}. Then, we can work with 
	\begin{align*}
	&\int_{H_{1,n}}|h|^a\left|\pi\left(\widehat{\theta}_n+\frac{h}{\sqrt n}\right)e^{\frac{h^\top V_nh}{2}+R_n(h)}-\pi\left(\widehat{\theta}_n\right)e^{-\frac{h^\top V_nh}{2}}\right|dh\\
	\leq&		\int_{H_{1,n}}|h|^a\pi\left(\widehat{\theta}_n+\frac{h}{\sqrt n}\right)|e^{-\frac{h^\top V_nh}{2}+R_n(h)}-e^{-\frac{h^\top V_nh}{2}}|dh\\
	+&	\int_{H_{1,n}}|h|^a\left|\pi\left(\widehat{\theta}_n+\frac{h}{\sqrt n}\right)-\pi\left(\widehat{\theta}_n\right)\right|e^{-\frac{h^\top V_nh}{2}}dh.
\end{align*}
The second term goes to zero by the continuity of $\pi(\cdot)$ in Assumption \ref{assump:PPrior} and the dominated convergence theorem. The first term can be bounded as follows
\begin{align*}
&\int_{H_{1,n}}|h|^a\pi\left(\widehat{\theta}_n+\frac{h}{\sqrt n}\right)e^{-\frac{h^\top V_nh}{2}}|e^{R_n(h)}-1|dh\\
\leq&\int_{H_{1,n}}|h|^a\pi\left(\widehat{\theta}_n+\frac{h}{\sqrt n}\right)e^{-\frac{h^\top V_nh}{2}}e^{R_n(h)}|R_n(h)|dh\\
\leq&\sup_{h\in H_{1,n}} |h|^aR_n(h)e^{R_n(h)}\pi\left(\widehat{\theta}_n+\frac{h}{\sqrt n}\right)\int_{H_{1,n}}e^{-\frac{h^\top V_nh}{2}}dh\to 0,
\end{align*}
because $\sup_{h\in H_{1,n}}R_n(h)\lesssim (\log \sqrt{n})^3/n$.

For the middle range $H_{2,n}$, we start with the following inequality
	\begin{align*}
		&\int_{H_{2,n}}|h|^a\left|\pi\left(\widehat{\theta}_n+\frac{h}{\sqrt n}\right)e^{-\frac{h^\top V_nh}{2}+R_n(h)}-\pi\left(\widehat{\theta}_n\right)e^{-\frac{h^\top V_nh}{2}}\right|dh\\
		\leq&		\int_{H_{2,n}}|h|^a\pi\left(\widehat{\theta}_n+\frac{h}{\sqrt n}\right)e^{-\frac{h^\top V_nh}{2}+R_n}dh+	\int_{H_{2,n}}|h|^a\pi\left(\widehat{\theta}_n\right)e^{-\frac{h^\top V_nh}{2}}dh.
	\end{align*}
	For the first term, because $h\in H_{2,n}$, we have $|h|/\sqrt{n}<\delta$, thus the remainder term is
	\begin{equation*}
	\sup_{h\in H_{2,n}}	|R_n(h)|\leq \frac{\delta}{6}\frac{1}{n}|h^\top\otimes \ell_n^{(3)}(\theta^\prime)\otimes h|.
	\end{equation*}
	Because of the integrability of the envelope function on the third-order derivatives in Assumption \ref{assump:PSE} (ii), we have $\sup_{\theta^\prime\in (\theta_0-\delta,\theta_0+\delta)}\frac{1}{n}|\ell_n^{(3)}(\theta^\prime)|=O_{P_0}(1)$. Thus, we can choose a small enough $\delta>0$ to ensure that
	\begin{equation*}
		P_0\left\{|R_n(h)|<\frac{h^\top V_nh}{4}, \forall h\in H_{2,n} \right\}>1-\epsilon,
	\end{equation*} 
	i.e.,
	\begin{equation*}
		P_0\left\{R_n(h)-\frac{h^\top V_nh}{2}<-\frac{h^\top V_nh}{2}, \forall h\in H_{2,n} \right\}>1-\epsilon,
	\end{equation*} 
	for any small $\epsilon$ and large enough $n$. Therefore, we can proceed with the following bound:
	\begin{align*}
		\int_{H_{2,n}}|h|^a\pi\left(\widehat{\theta}_n+\frac{h}{\sqrt n}\right)e^{-\frac{h^\top V_nh}{2}+R_n(h)}dh\leq \sup_{h\in H_{2,n}}\pi\left(\widehat{\theta}_n+\frac{h}{\sqrt n}\right) \int_{H_{2,n}}|h|^ae^{-h^\top V_nh/4}dh\to0.
	\end{align*}
	Recall $\theta$ is a $p$-dimensional vector. For the second term, we make use of the range of the integral to get
	\begin{align*}
		\int_{H_{2,n}}|h|^a\pi\left(\widehat{\theta}_n\right)e^{-\frac{h^\top V_nh}{2}}dh\leq 2\pi(\widehat{\theta}_n)e^{-\frac{c\log\sqrt{n}}{2}}\left[\delta\sqrt{n}-c\log\sqrt{n}\right]^{p}(\delta\sqrt{n})^a\lesssim\pi(\widehat{\theta}_n)\frac{n^{(p+a)/2}}{n^{c/4}},
	\end{align*}
which converges to zero	for a large enough $c$. Finally, the conclusion about the coverage property of the quasi-Bayesian credible set follows from the same proof in the main text, given Assumption \ref{assump:Psmooth} and \ref{assump:Pnormal}. 
\end{proof}

\section{Appendix D: Calculation of the Petrin-Train Model}\label{sec:PTDetail}
Appendix D explicitly calculates the influence from the first-stage nonparametric estimation for the Petrin--Train model with three alternatives. It provides the expression for the pathwise derivative in Assumption 3.6, which helps to verify Assumption 3.6 for that special case.
%We present the explicit correction term for a special case of the Petrin--Train model. 
Recall that we need to obtain
\begin{equation*}
	\dot{\Psi}_{\zeta}(\theta_0,\zeta_0)[\zeta-\zeta_0]\equiv \sum_{j=1}^J\dot{\Psi}_{\zeta_j}(\theta_0,\zeta_0)[\zeta_j-\zeta_{0j}],
\end{equation*}
where $\Psi(\theta,\zeta)\equiv P_0\dot{l}_{\theta}\left(  Y_2,\theta;\zeta\right)$.

For notation-wise simplicity, we focus on the case with three alternatives ($J+1=3$), in accordance with the original paper of \cite{hausman1978probit}. Also, see Section 2 of \cite{munkin2023note}. Recall that we normalize the baseline utility $U_{i0}\equiv 0$. The covariance matrix of the error terms $(\epsilon_{i1}^{\dagger},\epsilon_{i2}^{\dagger})$ is specified as follows:
\begin{equation*}
	\Sigma=
	\begin{bmatrix} 
		1 & \sigma_{12} \\
		\sigma_{12} & \sigma_2^2
	\end{bmatrix},
\end{equation*}
where the variance of $\epsilon^{\dagger}_{i1}$ is normalized to be one for the identification purpose. For a pair of bivariate normal random variables with zero mean, unit variance, and correlation coefficient $\rho$, we write the joint cumulative distribution function as $\Phi(u_1,u_2;\rho)$. We denote the partial derivatives of this function by
\begin{equation*}
	\Phi_{k}(u_1,u_2;\rho)\equiv\frac{\partial}{\partial u_k}\Phi(u_1,u_2;\rho),~~k=1,2,~~\text{and}~~\Phi_{\rho}(u_1,u_2;\rho)\equiv\frac{\partial}{\partial \rho}\Phi(u_1,u_2;\rho).
\end{equation*}
In addition, we write the cross derivatives by $\Phi_{kl}(u_1,u_2;\rho)\equiv\frac{\partial}{\partial u_l}\Phi_k(u_1,u_2;\rho),~~k,l=1,2$, and $\Phi_{\rho l}(u_1,u_2;\rho)\equiv\frac{\partial}{\partial u_l}\Phi_{\rho}(u_1,u_2;\rho)$ for $l=1,2$.

We obtain the conditional choice probabilities as follows 
\begin{align*}
	&P_0\left(C_i=0\mid W_i\right)=\Phi\left(-W_{i1}^\top\beta_0,-W_{i2}^\top\beta_0/\sigma_2;\frac{\sigma_{12}}{\sigma_2} \right),\\
	&P_0\left(C_i=1\mid W_i\right)=\Phi\left(\frac{(W_{i1}-W_{i2})^\top\beta_0}{\sqrt{1+\sigma_2^2-2\sigma_{12}}},W_{i1}^\top\beta_0;\frac{1-\sigma_{12}}{\sqrt{1+\sigma_2^2-2\sigma_{12}}} \right),\\
	&P_0\left(C_i=2\mid W_i\right)=\Phi\left(\frac{(W_{i2}-W_{i1})^\top\beta_0}{\sqrt{1+\sigma_2^2-2\sigma_{12}}},W_{i2}^\top\beta_0/\sigma_2;\frac{\sigma_2^2-\sigma_{12}}{\sigma_2\sqrt{1+\sigma_2^2-2\sigma_{12}}} \right).
\end{align*}
We proceed with additional simplification where $\sigma_2=1$ and $\sigma_{12}=0$. In this case, there is no additional parametric nuisance parameter in the second stage, so that we take $\theta_0=\beta_0$ in the sequel. We work with the following choice probabilities: 
\begin{align*}
	&P_0\left(C_i=0\mid W_i\right)=\Phi\left(-W_{i1}^\top\theta_0,-W_{i2}^\top\theta_0;0 \right),\\
	&P_0\left(C_i=1\mid W_i\right)=\Phi\left(\frac{(W_{i1}-W_{i2})^\top\theta_0}{\sqrt{2}},W_{i1}^\top\theta_0;\frac{1}{\sqrt{2}} \right),\\
	&P_0\left(C_i=2\mid W_i\right)=\Phi\left(\frac{(W_{i2}-W_{i1})^\top\theta_0}{\sqrt{2}},W_{i2}^\top\theta_0;\frac{1}{\sqrt{2}} \right).
\end{align*}
Then, we get
\begin{footnotesize}
	\begin{align*}
	\left.	\frac{\partial}{\partial \theta}P_0\left(C_i=0\mid W_i\right)\right\rvert_{\theta=\theta_0}&=	-\Phi_1\left(-W_{i1}^\top\theta_0,-W_{i2}^\top\theta_0;0 \right)W_{1i}-\Phi_2\left(-W_{i1}^\top\theta_0,-W_{i2}^\top\theta_0;0 \right)W_{i2},\\
	\left.	\frac{\partial}{\partial \theta_0}P_0\left(C_i=1\mid W_i\right)\right\rvert_{\theta=\theta_0}&=\Phi_1\left(\frac{(W_{i1}-W_{i2})^\top\theta_0}{\sqrt{2}},W_{i1}^\top\theta_0;\frac{1}{\sqrt{2}} \right)\frac{W_{i1}-W_{i2}}{\sqrt{2}}+\Phi_2\left(\frac{(W_{i1}-W_{i2})^\top\theta_0}{\sqrt{2}},W_{i1}^\top\theta_0;\frac{1}{\sqrt{2}} \right)W_{i1},\\
	\left.	\frac{\partial}{\partial \theta_0}	P_0\left(C_i=2\mid W_i\right)\right\rvert_{\theta=\theta_0}&=\Phi_1\left(\frac{(W_{i2}-W_{i1})^\top\theta_0}{\sqrt{2}},W_{i2}^\top\theta_0;\frac{1}{\sqrt{2}} \right)\frac{W_{i2}-W_{i1}}{\sqrt{2}}+\Phi_2\left(\frac{(W_{i2}-W_{i1})^\top\theta_0}{\sqrt{2}},W_{i2}^\top\theta_0;\frac{1}{\sqrt{2}} \right)W_{i2}.
	\end{align*}
\end{footnotesize}

The pathwise derivatives of conditional choice probabilities with respect to the first stage estimation are
\begin{footnotesize}
	\begin{align*}
		\dot{P}_0\left(C_i=0\mid W_i\right)[\widehat{\zeta}-\zeta_0]&=\Phi_1\left(-W_{i1}^\top\theta_0,-W_{i2}^\top\theta_0;0 \right)\lambda_{0}(\widehat{\zeta}_1-\zeta_1)+\Phi_2\left(-W_{i1}^\top\theta_0,-W_{i2}^\top\theta_0;0 \right)\lambda_{0}(\widehat{\zeta}_2-\zeta_2),\\
		\dot{P}_0\left(C_i=1\mid W_i\right)[\widehat{\zeta}-\zeta_0]&=-\Phi_1\left(\frac{(W_{i1}-W_{i2})^\top\theta_0}{\sqrt{2}},W_{i1}^\top\theta_0;\frac{1}{\sqrt{2}} \right)\frac{\lambda_{0}}{\sqrt{2}}(\widehat{\zeta}_1-\zeta_1-\widehat{\zeta}_2+\zeta_2)\\
		-&\Phi_2\left(\frac{(W_{i1}-W_{i2})^\top\theta_0}{\sqrt{2}},W_{i1}^\top\theta_0;\frac{1}{\sqrt{2}} \right)\lambda_{0}(\widehat{\zeta}_1-\zeta_1),\\
		\dot{P}_0\left(C_i=2\mid W_i\right)&=-\Phi_1\left(\frac{(W_{i2}-W_{i1})^\top\theta_0}{\sqrt{2}},W_{i2}^\top\theta_0;\frac{1}{\sqrt{2}} \right)\frac{\lambda_{0}}{\sqrt{2}}(\widehat{\zeta}_2-\zeta_2-\widehat{\zeta}_1+\zeta_1)\\
		-& \Phi_2\left(\frac{(W_{i2}-W_{i1})^\top\theta_0}{\sqrt{2}},W_{i2}^\top\theta_0;\frac{1}{\sqrt{2}} \right)\lambda_{0}(\widehat{\zeta}_2-\zeta_2).
	\end{align*}
\end{footnotesize}
Next, we further differentiate with respect to the parameter in the second stage. Given that the differentiation commutes with the expectation (integral) sign, this leads to the $\dot{\Psi}_{\zeta}(\cdot)$ that we are looking for. \footnote{Alternatively, one could first compute the derivative w.r.t. $\theta$ and then the path-wise derivative w.r.t. $\zeta$. There are extra terms which have zero expectation under this approach. In sum, they all lead to the same result after taking the expectation as we currently present.} By the chain rule, we get
	
\begin{eqnarray}\label{1st_influ_0}
&&	\left.\frac{\partial}{\partial \theta}\dot{P}_0\left(C_i=0\mid W_i\right)[\widehat{\zeta}-\zeta_0]\right\rvert_{\theta=\theta_0}\notag\\
	&=&	-\Phi_{11}\left(-W_{i1}^\top\theta_0,-W_{i2}^\top\theta_0;0 \right)W_{i1}\lambda_{0}(\widehat{\zeta}_1-\zeta_1)\notag\\
		&& -\Phi_{21}\left(-W_{i1}^\top\theta_0,-W_{i2}^\top\theta_0;0 \right)W_{i1}\lambda_{0}(\widehat{\zeta}_2-\zeta_2)\notag\\
		&& -\Phi_{12}\left(-W_{i1}^\top\theta_0,-W_{i2}^\top\theta_0;0 \right)W_{i2}\lambda_{0}(\widehat{\zeta}_1-\zeta_1)\notag\\
		&& -\Phi_{22}\left(-W_{i1}^\top\theta_0,-W_{i2}^\top\theta_0;0 \right)W_{i2}\lambda_{0}(\widehat{\zeta}_2-\zeta_2).
\end{eqnarray}
	
Similarly, we can obtain
	\begin{eqnarray}\label{1st_influ_1}
		&& \left.\frac{\partial}{\partial \theta}\dot{P}_0\left(C_i=1\mid W_i\right)[\widehat{\zeta}-\zeta_0]\right\rvert_{\theta=\theta_0}\notag \\
		&=& -\Phi_{11}\left(\frac{(W_{i1}-W_{i2})^\top\theta_0}{\sqrt{2}},W_{i1}^\top\theta_0;\frac{1}{\sqrt{2}} \right)\frac{W_{i1}-W_{i2}}{2}\lambda_{0}(\widehat{\zeta}_1-\zeta_1-\widehat{\zeta}_2+\zeta_2)\notag \\
		&& -\Phi_{12}\left(\frac{(W_{i1}-W_{i2})^\top\theta_0}{\sqrt{2}},W_{i1}^\top\theta_0;\frac{1}{\sqrt{2}} \right)\frac{W_{i1}}{\sqrt{2}}\lambda_{0}(\widehat{\zeta}_1-\zeta_1)\notag \\
		&& -\Phi_{21}\left(\frac{(W_{i1}-W_{i2})^\top\theta_0}{\sqrt{2}},W_{i1}^\top\theta_0;\frac{1}{\sqrt{2}} \right)\frac{W_{i1}-W_{i2}}{\sqrt{2}}\lambda_{0}(\widehat{\zeta}_1-\zeta_1-\widehat{\zeta}_2+\zeta_2)\notag \\
		&& -\Phi_{22}\left(\frac{(W_{i1}-W_{i2})^\top\theta_0}{\sqrt{2}},W_{i1}^\top\theta_0;\frac{1}{\sqrt{2}} \right)W_{i1}\lambda_{0}(\widehat{\zeta}_1-\zeta_1),
		\end{eqnarray}
and

	\begin{eqnarray}\label{1st_influ_2}
		&& \left.\frac{\partial}{\partial \theta}\dot{P}_0\left(C_i=2\mid W_i\right)[\widehat{\zeta}-\zeta_0]\right\rvert_{\theta=\theta_0}\notag\\
		& = & -\Phi_{11}\left(\frac{(W_{i2}-W_{i1})^\top\theta_0}{\sqrt{2}},W_{2i}^\top\theta_0;\frac{1}{\sqrt{2}} \right)\frac{W_{i2}-W_{i1}}{2}\lambda_0(\widehat{\zeta}_2-\zeta_2-\widehat{\zeta}_1+\zeta_1)\notag\\
		&& -\Phi_{12}\left(\frac{(W_{i2}-W_{i1})^\top\theta_0}{\sqrt{2}},W_{i2}^\top\theta_0;\frac{1}{\sqrt{2}} \right)\frac{W_{i2}}{\sqrt{2}}\lambda_{0}(\widehat{\zeta}_2-\zeta_2)\notag\\
		&& -\Phi_{21}\left(\frac{(W_{i2}-W_{i1})^\top\theta_0}{\sqrt{2}},W_{i2}^\top\theta_0;\frac{1}{\sqrt{2}} \right)\frac{W_{i2}-W_{i1}}{\sqrt{2}}\lambda_{0}(\widehat{\zeta}_2-\zeta_2-\widehat{\zeta}_1+\zeta_1)\notag\\
		&& -\Phi_{22}\left(\frac{(W_{i2}-W_{i1})^\top\theta_0}{\sqrt{2}},W_{i2}^\top\theta_0;\frac{1}{\sqrt{2}} \right)W_{i2}\lambda_{0}(\widehat{\zeta}_2-\zeta_2).
	\end{eqnarray}

The pathwise derivative with respect to the first stage functional components is obtained by taking the expectation of the following term:
\begin{small}
\begin{align*}
	\sum_{j=0}^2\mathbb{I}\{C_i=j\}\left[ \frac{\partial \dot{P}_0\left(C_i=j\mid W_i\right)/\partial \theta^\top[\widehat{\zeta}-\zeta_0]}{P_0\left(C_i=j\mid W_i\right)}-\frac{\partial P_0\left(C_i=j\mid W_i\right)/\partial \theta^\top \dot{P}_0\left(C_i=j\mid W_i\right)[\widehat{\zeta}-\zeta_0]}{P^2_0\left(C_i=j\mid W_i\right)} \right].
\end{align*}
\end{small}
It is clear that the influence from the first-stage estimation is null in the absence of the endogeneity, i.e., when $\lambda_0=0$. When we express the pathwise derivative with respect to the first-stage estimation according to the following expression:
\begin{equation}
	\dot{\Psi}_{\zeta_j}(\theta_0,\zeta_0)[\bm{h}]=\int \bm{h}(\cdot)g_j(\cdot,\theta)dP_0.
\end{equation}
We can regroup various terms together corresponding to the multiplicative functions with respect to $(\widehat{\zeta}_1-\zeta_{01})$ and $(\widehat{\zeta}_2-\zeta_{02})$, respectively. In particular, we get
\begin{footnotesize}
	\begin{align*}
		&g_1(Z_i,\theta)\\
		=-&\frac{\mathbb{I}\{C_i=0\}}{P_0\left(C_i=0\mid W_i\right)}\left[\Phi_{11}\left(-W_{i1}^\top\theta_0,-W_{i2}^\top\theta_0;0 \right)W_{i1}\lambda_{0}+\Phi_{12}\left(-W_{i1}^\top\theta_0,-W_{i2}^\top\theta_0;0 \right)W_{2i}\lambda_{0}\right]\\
		-&\frac{\mathbb{I}\{C_i=1\}}{P_0\left(C_i=1\mid W_i\right)}\Phi_{11}\left(\frac{(W_{i1}-W_{i2})^\top\theta_0}{\sqrt{2}},W_{i1}^\top\theta_0;\frac{1}{\sqrt{2}} \right)\frac{W_{1i}-W_{2i}}{2}\lambda_{0}\\
		-&\frac{\mathbb{I}\{C_i=1\}}{P_0\left(C_i=1\mid W_i\right)}\Phi_{12}\left(\frac{(W_{i1}-W_{i2})^\top\theta_0}{\sqrt{2}},W_{i1}^\top\theta_0;\frac{1}{\sqrt{2}} \right)\frac{W_{i1}}{\sqrt{2}}\lambda_{0}\\
		-&\frac{\mathbb{I}\{C_i=1\}}{P_0\left(C_i=1\mid W_i\right)}\Phi_{21}\left(\frac{(W_{i1}-W_{i2})^\top\theta_0}{\sqrt{2}},W_{i1}^\top\theta_0;\frac{1}{\sqrt{2}} \right)\frac{W_{i1}-W_{i2}}{\sqrt{2}}\lambda_{0}\\
		-&\frac{\mathbb{I}\{C=1\}}{P_0\left(C_i=1\mid W_i\right)}\Phi_{22}\left(\frac{(W_{i1}-W_{i2})^\top\theta_0}{\sqrt{2}},W_{i1}^\top\theta_0;\frac{1}{\sqrt{2}} \right)W_{i1}\lambda_{0}\\
		+&\frac{\mathbb{I}\{C_i=2\}}{P_0\left(C_i=2\mid W_i\right)}\Phi_{11}\left(\frac{(W_{i2}-W_{i1})^\top\theta_0}{\sqrt{2}},W_{i2}^\top\theta_0;\frac{1}{\sqrt{2}} \right)\frac{W_{i2}-W_{i1}}{2}\lambda_0\\
		+&\frac{\mathbb{I}\{C_i=2\}}{P_0\left(C_i=2\mid W_i\right)}\Phi_{21}\left(\frac{(W_{i2}-W_{i1})^\top\theta_0}{\sqrt{2}},W_{i2}^\top\theta_0;\frac{1}{\sqrt{2}} \right)\frac{W_{i2}-W_{i1}}{\sqrt{2}}\lambda_{0}\\
		+	&\frac{\mathbb{I}\{C_i=0\}\partial P_0\left(C_i=0\mid W_i\right)/\partial \theta^\top }{P^2_0\left(C_i=0\mid W_i\right)}\lambda_{0}\Phi_1\left(-W_{i1}^\top\theta_0,-W_{i2}^\top\theta_0;0 \right)\\
		+&\frac{\mathbb{I}\{C_i=1\}\partial P_0\left(C_i=1\mid W_i\right)/\partial \theta^\top }{P^2_0\left(C_i=1\mid W_i\right)}\lambda_{0} \left[	\Phi_1\left(\frac{(W_{i1}-W_{i2})^\top\theta_0}{\sqrt{2}},W_{i1}^\top\theta_0;\frac{1}{\sqrt{2}} \right)\frac{1}{\sqrt{2}} +\Phi_2\left(\frac{(W_{i1}-W_{i2})^\top\theta_0}{\sqrt{2}},W_{i1}^\top\theta_0;\frac{1}{\sqrt{2}} \right)\right]\\
		+&\frac{\mathbb{I}\{C_i=2\}\partial P_0\left(C_i=2\mid W_i\right)/\partial \theta^\top }{P^2_0\left(C_i=2\mid W_i\right)} \Phi_1\left(\frac{(W_{i2}-W_{i1})^\top\theta_0}{\sqrt{2}},W_{i2}^\top\theta_0;\frac{1}{\sqrt{2}} \right)\frac{\lambda_{0}}{\sqrt{2}}. 
	\end{align*}
\end{footnotesize}
The formula for $g_2(\cdot,\theta)$ is obtained in a complete analogous way. For the more general case that allows for the unknown correlation between the latent errors, we would proceed with the following parameterization $\eta\equiv\tau^2=1-\sigma_{12}$:
\begin{align*}
	&P_0\left(C_i=0\mid W_i\right)=\Phi\left(-W_{i1}^\top\beta_0,-W_{i2}^\top\beta_0;1-\tau^2 \right),\\
	&P_0\left(C_i=1\mid W_i\right)=\Phi\left(\frac{(W_{i1}-W_{i2})^\top\beta_0}{\sqrt{2}\tau},W_{i1}^\top\beta_0;\frac{\tau}{\sqrt{2}} \right),\\
	&P_0\left(C_i=2\mid W_i\right)=\Phi\left(\frac{(W_{i2}-W_{i1})^\top\beta_0}{\sqrt{2}\tau},W_{i2}^\top\beta_0;\frac{\tau}{\sqrt{2}} \right).
\end{align*}
As it requires straightforward yet tedious calculation without much further insight, we omit the details.

\section{Appendix E: Additional Simulation Results}

In order to further highlight the novelty of our bootstrap scheme, we compare our method with an alternative proposal in this section. Our quasi-Bayesian bootstrap inference (described in Algorithm 1) is based on the bootstrap distribution of quasi-posterior means. Now we describe a different algorithm, denoted by QBB$_{1}$, which takes a single posterior draw from each bootstrap sample. One of the key reasons for the failure of the quasi-Bayesian credible set by itself is that the uncertainty from the first stage estimation is ignored. A possibility would be to approximate the joint posterior distribution based on the following decomposition: \footnote{We thank one anonymous referee for making this suggestion.}
\[
\pi(\theta|\bm{Y}^n_{2},\zeta)\hat{p}(\zeta\mid \bm{Y}^n_{1}),
\]
where $\hat{p}(\zeta\mid \bm{Y}^n_{1})$ is an approximation of the limiting distribution of $\widehat{\zeta}_n$ (i.e., the sampling distribution of the frequentist estimator). When it comes to the practical implementation, it is straightforward to approximate the aforementioned sampling distribution by its bootstrap analog. For $s=1,2,...,S$, repeat the following two steps:\\
1. Draw $\zeta^{(s)}$ from the sampling distribution of $\widehat{\zeta}_n$ via the nonparametric bootstrap. \\
2. Draw $\theta^{(s)}$ from the conditional posterior distribution $p(\theta\mid \bm Y^n_{2}, \zeta^{(s)})$.\\
In order to illustrate the connection and the subtle difference to our bootstrap scheme, we describe this algorithm as follows.

\begin{algorithm}
\caption*{ \textbf{Algorithm QBB$_{1}$:} One posterior draw per bootstrap}
\begin{algorithmic}
    \STATE \textbf{Input:} Data $\bY^n$, number of bootstrap replications $B$
   \FOR{$b=1,\ldots, B$}
    \STATE Resample data $\bY^n$ with replacement to form the $b$th-bootstrap sample $\bY_b^n$.
        \STATE  (1). Obtain a frequentist estimate $\widehat{\zeta}_{n,b}^*$ from the first-stage data $\bY^n_{1,b}$.
        \STATE (2).  Generate ${\theta}^*_{n,1,b}$ as ONE draw from the posterior of $\theta$ conditional on $\bY^n_{2,b}$ and $\widehat{\zeta}_{n,b}^*$. 
\ENDFOR
\STATE \textbf{Output:} $\left\{{\theta}^*_{n,1,b}:b=1,\ldots,B\right\}$. The $j$th coordinate of ${\theta}^*_{n,1,b}$ is denoted as ${\theta}^*_{n,1,b,j}$.\\
Let $\theta_{0,j}$ be the $j$th coordinate of the parameter $\theta_0$. The $1-\alpha$ \textbf{QBB$_{1}$} confidence interval for $\theta_{0,j}$ is formed as 
$\left[  \bar Q_{n,j}^{\ast}(\alpha/2), \bar Q_{n,j}^{\ast}(1-\alpha/2) \right], ~\text{for}~ j=1,\cdots,p$,
where $\bar Q^*_{n,j}(a)$ denotes the $a$-th quantile of the bootstrap distribution $\left\{{\theta}^*_{n,1,b,j} :b=1,\ldots,B\right\}$.
\end{algorithmic}
\end{algorithm}

\begin{figure}
			\caption{Two bootstrap methods: a simulation replication with $J+1=3$, Design I in Section 4 with $\theta = \tilde{\beta}=1$, control function $\widehat{\zeta}^*_n = \left(\hat{v}^{\dagger*}_{i1}, \hat{v}^{\dagger*}_{i2}\right),i=1,\cdots, n$. Solid vertical line in magenta = true value of the parameter, dashed vertical line in black = mean of the distribution.}\label{fig:flowchart_2}
			\vspace{2mm}
			\begin{tabular}{cc}
				(a). Quasi-posterior with bootstrap & (b). QBB$_1$: one draw each bootstrap\\
				\includegraphics[scale=0.7]{flow_chart_boot.pdf}&
				\includegraphics[scale=0.7]{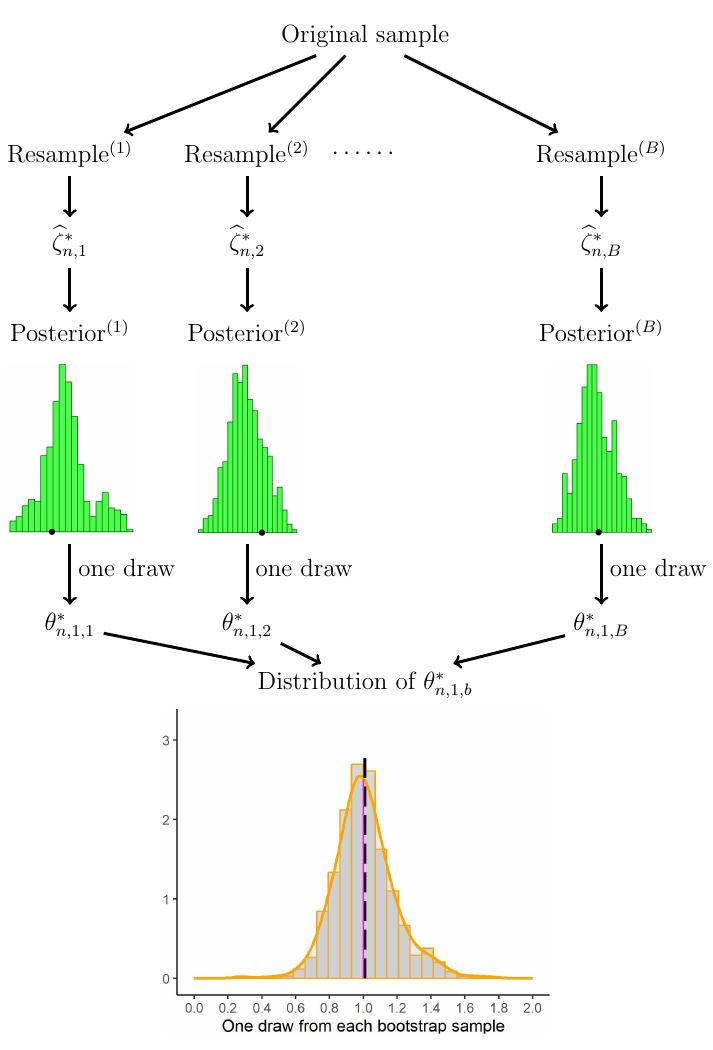}
			\end{tabular}
		\end{figure}	

Figure \ref{fig:flowchart_2} provides a graphical illustration of Algorithm QBB$_{1}$ and compares it to our proposal Algorithm 1 in Section 2.3. The histogram and density plots in the figure are based on one simulated sample in Section 4. To clarify the phrase about making ``one draw from the posterior'' in Algorithm QBB$_{1}$, we stress that this draw is taken after a burn-in period of MCMC in order to approximate the posterior. For example, if one discards the first $S$ draws (burn-out period) and uses the subsequent one draw to generate QBB$_1$, while relying on the subsequent $S$ draws to form QBB (in Table 1), then the computation time for QBB$_1$ is half of that for QBB.

Algorithm QBB$_1$ intuitively takes into account the first-stage estimation uncertainty, but its theoretical validity is not clear. Our bootstrap proposal crucially depends on the fact that the quasi-posterior mean micks the asymptotics of a frequentist two-stage estimator. And this relationship is maintained in its bootstrap analog. This can not be shown easily for the alternative approach based on Algorithm QBB$_{1}$. Table \ref{table:CR_QBB_1} presents the finite sample performance of the confidence interval produced by QBB$_1$, for the same simulation setup as in Section 4. The results of QBB in Table 1 are included for comparison. Table \ref{table:CR_QBB_1} shows that QBB$_1$ over-covers, and yields noticeably longer confidence intervals than QBB. This is not surprising, since QBB$_1$ takes one posterior draw for each bootstrap sample, it is expected to be more volatile than QBB which uses the posterior mean for each bootstrap. Figure \ref{fig:flowchart_2} also illustrates that the distribution of ${\theta}^*_{n,b,1}$ (Panel (b)) is more spread-out than that of ${\theta}^*_{n,b}$ (Panel (a)).

\begin{table}
\caption{Finite sample performance of the inference procedures \textbf{QBB$_1$} for the coefficient $\widetilde\beta$, $J+1$= number of choices, $n=1000$, nominal coverage $1-\alpha=0.90$, $0.95$, and $0.99$.}\label{table:CR_QBB_1}
\begin{tabular}{cclcccccccccccccc}\toprule
\multicolumn{1}{c}{\multirow{1}{*}{}}&\multicolumn{1}{c}{\multirow{1}{*}{}}&\multicolumn{1}{c}{\multirow{1}{*}{}}&\multicolumn{1}{c}{}& \multicolumn{3}{c}{Coverage probability}&\multicolumn{1}{c}{}&\multicolumn{3}{c}{Average length}\\
\cline{5-7}\cline{9-11}
\multicolumn{1}{c}{\multirow{1}{*}{$J+1$}}&\multicolumn{1}{c}{\multirow{1}{*}{Design}}&\multicolumn{1}{c}{Methods}&\multicolumn{1}{c}{\multirow{1}{*}{}}& \multicolumn{1}{c}{$0.90$}&\multicolumn{1}{c}{$0.95$}&\multicolumn{1}{c}{$0.99$}&\multicolumn{1}{c}{}&\multicolumn{1}{c}{$0.90$}&\multicolumn{1}{c}{$0.95$}&\multicolumn{1}{c}{$0.99$}\\
\toprule
3 & I & QBB$_1$ && 0.961 & 0.984 & 0.997 && 0.527 & 0.657 & 0.953\\
& & QBB && 0.900 & 0.951 & 0.992 && 0.414 & 0.511 & 0.738\\
\cline{2-11}
 & II & QBB$_1$ && 0.967 & 0.988 & 0.998 && 0.496 & 0.617 & 0.888\\
& & QBB && 0.897 & 0.950 & 0.991 && 0.391 & 0.480 & 0.676\\
\midrule
4 &  I &  QBB$_1$ && 0.960 & 0.986 & 0.998 && 0.615 & 0.792 & 1.196 \\
& & QBB && 0.915 & 0.958 & 0.992 && 0.482 & 0.602 & 0.886 \\
\cline{2-11}
& II &QBB$_1$ &&  0.974 & 0.993 & 0.999 && 0.617 & 0.808 & 1.217\\
& & QBB && 0.939 & 0.971 & 0.997 && 0.483 & 0.606 & 0.896\\
\bottomrule
\end{tabular}
\end{table}
\bibliographystyle{econometrica}
\bibliography{Bayes_bib}